\documentclass[12pt,english]{article}
\usepackage[T1]{fontenc}
\usepackage[latin1]{inputenc}
\usepackage{subfigure}
\usepackage{graphicx}
\usepackage{rotating}
\providecommand{\tabularnewline}{\\}

\makeatletter

\providecommand{\tabularnewline}{\\}

\usepackage{babel}
\makeatother
\usepackage{geometry}

\usepackage{babel}
\usepackage{graphics}
\def\beq{\begin{equation}}
\def\eeq{\end{equation}}

\def\beqn{\begin{eqnarray}}
\def\eeqn{\end{eqnarray}}
\def\ba{\begin{eqnarray}}
\def\ea{\end{eqnarray}}

\def\cc{Corian\`{o}}
\newcommand{\beqa}{\begin{eqnarray}}
\newcommand{\eeqa}{\end{eqnarray}}

\makeatletter

\providecommand{\LyX}{L\kern-.1667em\lower.25em\hbox{Y}\kern-.125emX\@}

\makeatother
\begin{document}
\begin{flushright}
YITP-SB-05-38
\end{flushright}
\begin{center}
\vspace{1.cm}
{\bf \Large On the Scale Variation of the Total  Cross Section for \\
 Higgs Production at the LHC and at the Tevatron}

\vspace{2cm}
 {\large $^{a, b}$Alessandro Cafarella, $^a$Claudio Corian\`{o}, $^a$Marco Guzzi and $^c$J. Smith\\}
\vspace{0.5cm}
{\it $^a$ Dipartimento di Fisica, Università di Lecce and INFN sezione
di Lecce\\
Via per Arnesano, 73100 Lecce, Italy\footnote
{\it e-mail: alessandro.cafarella@le.infn.it, claudio.coriano@le.infn.it, marco.guzzi@le.infn.it,\\ smith@insti.physics.sunysb.edu}\\}
\vspace{0.5cm}
{\it $^b$Department of Physics, University of Crete, 71003 Heraklion, Greece\\}
\vspace{0.5cm}
{\it $^c$ C.N. Yang Institute for Theoretical Physics\\
State University of New York at Stony Brook, New York 11794-3840, U.S.A.\\
and\\
NIKHEF, Postbus 41882, 1009 DB, Amsterdam, The Netherlands\\}
\end{center}

\vspace{1.cm}
\begin{abstract}
We present a detailed study of the total $p\,p$ cross section for scalar 
Higgs production to next-to-next-to leading order in $\alpha_s$ at LHC 
energies, and of the $p\bar p$ cross section at the Tevatron, combining 
an implementation of the solutions of the parton evolution equations 
at the 3-loop order with the corresponding hard scatterings, evaluated at 
the same perturbative order. Our solutions of the DGLAP equations are
implemented directly in $x$-space and allow the study of the dependence 
of the results on the factorization ($\mu_F$) and renormalization scales 
$(\mu_R)$ typical of a given process, together with the stability of the 
perturbative expansion.  The input sets for the parton evolutions 
are those given by Martin, Roberts, Stirling and Thorne and by Alekhin.
Results for K-factors are also presented. The
NNLO corrections can be quite sizeable at typical collider energies. 
The stability region of the perturbative expansion is found 
when $\mu_R\,> \,m_H\sim\mu_F$.

\end{abstract}
\newpage
\section{Introduction}

The validity of the mechanism of mass generation in the Standard Model
will be tested at the new collider, the LHC.  For this we require precision 
studies in the Higgs sector to confirm its existence. 
This program involves a rather complex analysis of the QCD backgrounds 
with the corresponding radiative corrections fully taken into account.
Studies of these corrections for specific processes have been performed 
by various groups, to an accuracy which has reached the 
next-to-next-to-leading order (NNLO) level in $\alpha_s$, 
the QCD coupling constant. The quantification of the impact of these 
corrections requires the determination of the hard scattering partonic 
cross sections up to order $\alpha_s^3$, together with the DGLAP kernels 
controlling the evolution of the parton distributions 
determined at the same perturbative order. 
Therefore, the study of the evolution of the parton distributions, 
using the three-loop results on the anomalous dimensions \cite{vogt1}, is  
critical for the success of this program. Originally NNLO predictions 
for some particular processes such as total cross sections 
\cite{RSVN} have been obtained using the approximate expressions for 
these kernels \cite{VN1}. The completion of the exact computation of the 
NNLO DGLAP kernels motivates more detailed studies of the same 
observables based on these exact kernels and the investigation of the 
factorization and renormalization scale dependences of the result, 
which are still missing. In this work we are going to reanalyse 
these issues from a broader perspective.  Our analysis is here exemplified 
in the case of the total cross sections at the LHC ($pp$) and at the 
Tevatron ($p \bar{p}$) for Higgs production using the hard scatterings 
computed in \cite{RSVN} and their dependence on the factorization and 
renormalization scales. The DGLAP equation is solved directly in $x$-space 
using a method which is briefly illustrated below 
and which is accurate up to order $\alpha_s^2$. Our input distributions at 
a small scale will be specified below. We also analyse the corresponding 
K-factors and the region of stability of the perturbative expansion 
by studying their variation under changes in all the relevant 
scales. It is shown that the NNLO corrections are sizeable while the region 
of reduced scale dependence is near the value $m_H=\mu_F$ 
with $\mu_R$ around the same value but slightly higher. 

\section{Higgs production at LHC}

\begin{figure}
{\centering \resizebox*{7cm}{!}{\rotatebox{0}
{\includegraphics{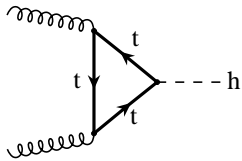}}}\par}
\caption{The leading order diagram for Higgs production by gluon fusion}
\label{g-fusion}
\end{figure}

\begin{figure}
{\centering \resizebox*{7cm}{!}{\rotatebox{0}
{\includegraphics{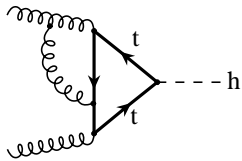}}}\par}
\caption{A typical NLO diagram for Higgs production by gluon fusion}
\label{g-fusion1}
\end{figure}
The Higgs field, being responsible for the mechanism of mass generation, 
can be radiated off by any massive state and its coupling is proportional 
to the mass of the same state. At the LHC one of the golden plated modes 
to search for the Higgs is its production via the mechanism of gluon fusion.
The leading order contribution is shown in Fig.~(\ref{g-fusion}) which 
shows that dependence of the amplitude is through the quark loop. Most of the 
contribution comes from the top quark, since this is the heaviest quark and has 
the largest coupling to the Higgs field. NLO and NNLO corrections
have been computed in the last few years by various groups. 
A typical NLO correction is shown in Fig.~(\ref{g-fusion1}). 
In the infinite mass limit of the quark mass in the loop (see \cite{dawson} 
for a review), an effective description of the process is obtained in 
leading order by the Lagrangian density
\beqn
{\cal L}_{\rm eff}&=&{\alpha_s \over 12 \pi} G_{\mu\nu}^A G^{A~\mu\nu}
\biggl({H\over v}\biggr)\nonumber \\
&=&{\beta_F \over g_s} G_{\mu\nu}^A G^{A~\mu\nu}
\biggl({H\over 2v}\biggr)(1-2 \alpha_s/\pi),\nonumber
\label{effth}
\eeqn
with
\beq
\beta_F={g_s^3 N_H\over 24 \pi^2}
\eeq
being the contribution of $N_H$ heavy fermion loops to the QCD beta function.
This effective Lagrangian can be used to compute the radiative corrections 
in the gluon sector. A discussion of the NNLO approach to the computation 
of the gluon fusion contributions to Higgs production has been presented 
in \cite{RSVN}, to which we refer for more details. We recall that in 
this paper the authors presented a study for both scalar and pseudoscalar Higgs
production, the pseudoscalar appearing in 2-Higgs doublets models. 
The diagonalization of the mass matrix for the Higgs at the minimum 
introduces scalar and pseudoscalar interactions between the various Higgs 
and the quarks, as shown from the structure of the operator $O_2$ below 
in eq. (\ref{eqn2.2}). 
In the large top-quark mass limit the Feynman rules
for scalar Higgs  production (${\rm H}$) can be derived from the
effective Lagrangian density \cite{RSVN1}
\begin{eqnarray}
\label{eqn2.1}
{\cal L}^{\rm H}_{eff}=G_{\rm H}\,\Phi^{\rm H}(x)\,O(x) \quad
\mbox{with} \quad O(x)=-\frac{1}{4}\,G_{\mu\nu}^a(x)\,G^{a,\mu\nu}(x)\,,
\end{eqnarray}
whereas the production of a pseudo-scalar Higgs (${\rm A}$) is obtained from
\begin{eqnarray}
\label{eqn2.2}
&&{\cal L}_{eff}^{\rm A}=\Phi^{\rm A}(x)\Bigg [G_{\rm A}\,O_1(x)+
\tilde G_{\rm A}\,O_2(x)\Bigg ] \quad \mbox{with} \quad
\nonumber\\[2ex]
&&O_1(x)=-\frac{1}{8}\,\epsilon_{\mu\nu\lambda\sigma}\,G_a^{\mu\nu}\,
G_a^{\lambda\sigma}(x) \,,
\nonumber\\[2ex]
&&O_2(x) =-\frac{1}{2}\,\partial^{\mu}\,\sum_{i=1}^{n_f}
\bar q_i(x)\,\gamma_{\mu}\,\gamma_5\,q_i(x)\,,
\end{eqnarray}
where $\Phi^{\rm H}(x)$ and  $\Phi^{\rm A}(x)$ represent the scalar and
pseudo-scalar fields respectively and $n_f$ denotes the number of light
flavours.
$G_a^{\mu\nu}$ is the field strength of QCD
and the quark fields are denoted by $q_i$. 
We refer the reader to \cite{RSVN} for further details.

Using the effective Lagrangian one can calculate the
total cross section of the reaction

\begin{eqnarray}
\label{eqn2.10}
H_1(P_1)+H_2(P_2)\rightarrow {\rm H} + X\,,
\end{eqnarray}
where $H_1$ and $H_2$ denote the incoming hadrons and $X$ represents an
inclusive hadronic state and ${\rm B}$ denotes the scalar or the psudoscalar
particle produced in the reaction.
The total cross section is given by
\begin{eqnarray}
\label{eqn2.11}
&&\sigma_{\rm tot}=\frac{\pi\,G_{\rm B}^2}{8\,(N^2-1)}\,\sum_{a,b=q,\bar q,g}\,
\int_x^1 dx_1\, \int_{x/x_1}^1dx_2\,f_a(x_1,\mu^2)\,f_b(x_2,\mu^2)\,
\nonumber\\[2ex] && \qquad\qquad\times
\Delta_{ab,{\rm B}}\left ( \frac{x}{x_1\,x_2},\frac{m^2}{\mu^2} \right ) \,,
\nonumber\\[2ex]
&&\mbox{with}\quad x=\frac{m^2}{S} \quad\,,\quad S=(P_1+P_2)^2\quad
\,,
\end{eqnarray}
where the factor $1/(N^2-1)$ is due to the average over colour.
The parton distributions $f_a(y,\mu^2)$ ($a,b=q,\bar q,g$)
depend on the mass factorization/renormalization scale $\mu$.
$\Delta_{ab,{\rm B}}$ denotes the partonic hard scattering coefficient 
computed with NNLO accuracy.

\section{The NNLO Evolution}
We summarize the main features of the NNLO DGLAP evolution. 
As usual we introduce singlet $(+)$ and non-singlet $(-)$ parton distributions

\begin{equation}
q_{i}^{(\pm)}=q_{i}\pm\overline{q}_{i},\qquad q^{(\pm)}=\sum_{i=1}^{n_{f}}q_{i}^{(\pm)}
\label{eq:definizioni}
\end{equation}
whose evolution is determined by the corresponding equations 

\begin{equation}
\frac{\textrm{d}}{\textrm{d}\log Q^{2}}\left(\begin{array}{c}
q^{(+)}(x,Q^{2})\\
g(x,Q^{2})\end{array}\right)=\left(\begin{array}{cc}
P_{qq}(x,\alpha_{s}(Q^{2})) & P_{qg}(x,\alpha_{s}(Q^{2}))\\
P_{gq}(x,\alpha_{s}(Q^{2})) & P_{gg}(x,\alpha_{s}(Q^{2}))\end{array}\right)\otimes\left(\begin{array}{c}
q^{(+)}(x,Q^{2})\\
g(x,Q^{2})\end{array}\right)
\label{eq:singlet}
\end{equation}
for the singlet combination 
and a scalar one for the non-singlet case

\begin{equation}
\frac{\textrm{d}}{\textrm{d}\log Q^{2}}q_i^{(-)}(x,Q^{2})=P_{NS}(x,\alpha_{s}(Q^{2}))\otimes q_i^-(x,Q^{2}).
\label{eq:DGLAP}
\end{equation}

The convolution product is defined by
\begin{equation}
\left[a\otimes b\right](x)=\int_{x}^{1}\frac{\textrm{d}y}{y}a\left(\frac{x}{y}\right)b(y)=
\int_{x}^{1}\frac{\textrm{d}y}{y}a(y)b\left(\frac{x}{y}\right).
\end{equation}

We recall that the perturbative expansion, up to NNLO, of the kernels
is
\begin{equation}
P(x,a_{s})= a_s P^{(0)}(x)+ a_s^2 P^{(1)}(x)+ a_s^3 P^{(2)}(x)+\ldots.
\label{kern1}
\end{equation}
where $a_s\equiv \alpha_s/(4 \pi)$.
In order to solve the evolution equations directly in $x$-space 
(see \cite{cafacor} for an NLO implementation of the method), we
assume solutions of the form \cite{CCG}
\begin{eqnarray}
f(x,Q^{2}) & = & \sum_{n=0}^{\infty}\frac{A_{n}(x)}{n!}
\log^{n}\frac{a_{s}(Q^{2})}{a_{s}(Q_{0}^{2})}+a_s (Q^{2})
\sum_{n=0}^{\infty}\frac{B_{n}(x)}{n!}\log^{n}
\frac{a_{s}(Q^{2})}{a_{s}(Q_{0}^{2})}\nonumber \\
&&+ a_{s}^2(Q^{2})\sum_{n=0}^{\infty}
\frac{C_{n}(x)}{n!}\log^{n}\frac{a_{s}(Q^{2})}{a_{s}(Q_{0}^{2})}
\label{eq:ansatz}
\end{eqnarray}
for each parton distribution $f$, where $Q_{0}$ defines the initial
evolution scale. The ansatz is introduced into the evolution equations and 
used to derive recurrence relations for its unknown coefficients 
$A_n,B_n,C_n$, involving polylogarithmic functions 
\cite{remiddi1,remiddi2} which are then implemented numerically.

This ansatz corresponds to a solution of the DGLAP equation accurate up to 
order $a_s^2$ (truncated solution).
It can be shown that \cite{CCG} this ansatz reproduces the solution of the 
DGLAP equation in (Mellin) moment space obtained with the same accuracy 
in $a_s$.  Modifications of this ansatz also allow to obtain the 
so-called ``exact'' solutions of the equations for the moments \cite{Vogt}. 
These second solutions include higher order terms in $a_s$ and can be 
identified only in the non-singlet case. Exact approaches also include 
an exact solution of the renormalization group equation 
for the $\beta$-function, which embodies the effects of the coefficients 
$\beta_0, \beta_1$ and $\beta_2$ to higher order in $a_s$. 
The term ``exact'' is, however, a misnomer since the accuracy of the solution 
is limited to the knowledge of the first three contributions to the 
expansion in the beta function and in the kernels.
It can be shown both for exact and for the truncated solutions that solving 
the equations by an ansatz in $x$-space is completely equivalent to 
searching for the solution in moment space, since in moment space the 
recursion relations can be solved exactly \cite{CCG}.

\section{Renormalization scale dependence}

For a better determination of the dependence of the perturbative cross 
section on the scales of a certain process it is important 
to keep these scales independent and study the behaviour of the corresponding 
hadronic cross section under their variation. In our case the two relevant 
scales are the factorization scale $\mu_F$ and the renormalization scale
$\mu_R$ which can be both included in the evolution by a rearrangement 
of the evolution kernels up to NNLO.

The study of the dependence of the solution upon the various scales is then 
performed in great generality and includes also the logarithmic 
contributions $\log(\mu_F/\mu_R)$ coming from the hard scatterings given 
in \cite{RSVN}, where, however, only the specific 
point $\mu_F=\mu_R=m_H$ was considered. The separation of the scales 
should then appear not only in the hard scatterings but also in the evolution 
equations. This issue has been addressed in \cite{Vogt} and can be 
reconsidered also from $x$-space \cite{CCG} using the $x$-space logarithmic 
ansatz (\ref{eq:ansatz}).

The scale dependence of the parton distribution
functions is then expressed by a generalized DGLAP equation
\ba
\label{evolution}
\frac{\partial}{\partial \ln \mu_F^2}\, f_i(x,\mu_F^2,\mu_R^2)=
P_{ij}(x,\mu_F^2,\mu_R^2) \otimes f_j(x,\mu_F^2,\mu_R^2)\,,
\label{scales}
\ea
where $\mu_F$ is now a generic factorization scale.

Generally speaking, both the kernels and the PDF's have a dependence on 
the scales $\mu_F$ and $\mu_R$, and formally, a comparison between these
scale is always possible up to a fixed order by using the renormalization 
group equations for the running coupling $\alpha_s$.

The renormalization scale dependence of the ansatz (\ref{eq:ansatz}) that 
solves (\ref{scales}) is obtained quite straighforwardly 
by a Taylor expansion of the running coupling $\alpha_s(\mu_F^2)$ in terms 
of $\alpha_s(\mu_R^2)$ \cite{CCG}
\ba
\label{alphas}
\alpha_s(\mu_F^2)=\alpha_s(\mu_R^2)-\left[\frac{\alpha_s^2(\mu_R^2)}{4\pi}
+\frac{\alpha_s^3(\mu_R^2)}{(4\pi)^2}(-\beta_0^2 L^2+\beta_1 L)\right]
\label{betaf}
\ea
where the $\mu_F^2$ dependence is included in the factor 
$L=\ln(\mu_F^2/\mu_R^2) $, and the coefficients of the $\beta$-function, 
(the $\beta_i$) are listed below
\ba
&&\beta_{0}=\frac{11}{3}N_{C}-\frac{4}{3}T_{f},\nonumber\\
&&\beta_{1}=\frac{34}{3}N_{C}^{2}-\frac{10}{3}N_{C}n_{f}-2C_{F}n_{f},\nonumber\\
&&\beta_{2}=\frac{2857}{54}N_{C}^{3}+2C_{F}^{2}T_{f}-\frac{205}{9}C_{F}N_{C}T_{f}
-\frac{1415}{27}N_{C}^{2}T_{f}+\frac{44}{9}C_{F}T_{f}^{2}+\frac{158}{27}N_{C}T_{f}^{2}.
\nonumber\\
\ea
As usual we have set 
\begin{equation}
N_{C}=3,\qquad C_{F}=\frac{N_{C}^{2}-1}{2N_{C}}=\frac{4}{3},
\qquad T_{f}=T_{R}n_{f}=\frac{1}{2}n_{f},
\end{equation}
where $N_{C}$ is the number of colors and $n_{f}$ is the number of
active flavors. This number is varied as we step into a region 
characterized by an evolution scale 
$\mu$ larger than a specific quark mass ($\mu \geq m_{q}$). Also the
NNLO matching conditions across flavor thresholds \cite{bmsn},
\cite{cs} are implemented.

Since the perturbative expansion of eq.~(\ref{kern1}) contains powers 
of $\alpha_{s}(\mu_F^2)$ which can be related to the value of 
$\alpha_s(\mu_R^2)$ by (\ref{betaf}), from 
\ba
\label{kern2}
P_{ij}^{NNLO}(x,\mu_F^2)=
\sum_{k=0}^{2} \, \left(\frac{\alpha_s(\mu_F^2)}{4\pi}\right)^{k+1}
P_{ij}^{(k)}(x)\,,
\ea
substituting eq. (\ref{alphas}) into (\ref{kern2}), we obtain the corresponding expression
of the kernels organized in powers of $\alpha_{s}(\mu_R^2)$ up to NNLO, and it reads \cite{Vogt}
\ba
\label{kern3}
&&P_{ij}(x,\mu_F^2,\mu_R^2)=\frac{\alpha_s(\mu_R^2)}{4\pi}P_{ij}^{(0)}(x) \nonumber\\
&&\hspace{2.5cm}+\frac{\alpha_s^2(\mu_R^2)}{(4\pi)^2}\left(P_{ij}^{(1)}(x)
-\beta_0P_{ij}^{(0)}(x) L\right)
\nonumber\\
&&\hspace{2.5cm}+\frac{\alpha_s^3(\mu_R^2)}{(4\pi)^3}\left[P_{ij}^{(2)}(x)-
2\beta_0 L P_{ij}^{(1)}(x)
-\left(\beta_1 L - \beta_0^2 L^2 \right) P_{ij}^{(0)}(x)\right]\,.\nonumber\\
\ea

The implementation of the method in $x$-space is quite straightforward and 
allows us to perform a separate study of the predictions in terms 
of $\mu_F$ and $\mu_R$.

\section{Numerical Results}
The use of the NNLO evolution of the parton distributions together
with the results of \cite{RSVN} allows us to provide accurate
predictions for the total cross section for Higgs production.
Here we summarize and discuss our numerical results. 

\begin{figure}
\subfigure[LHC]{\includegraphics[%
  width=6cm,
  angle=-90]{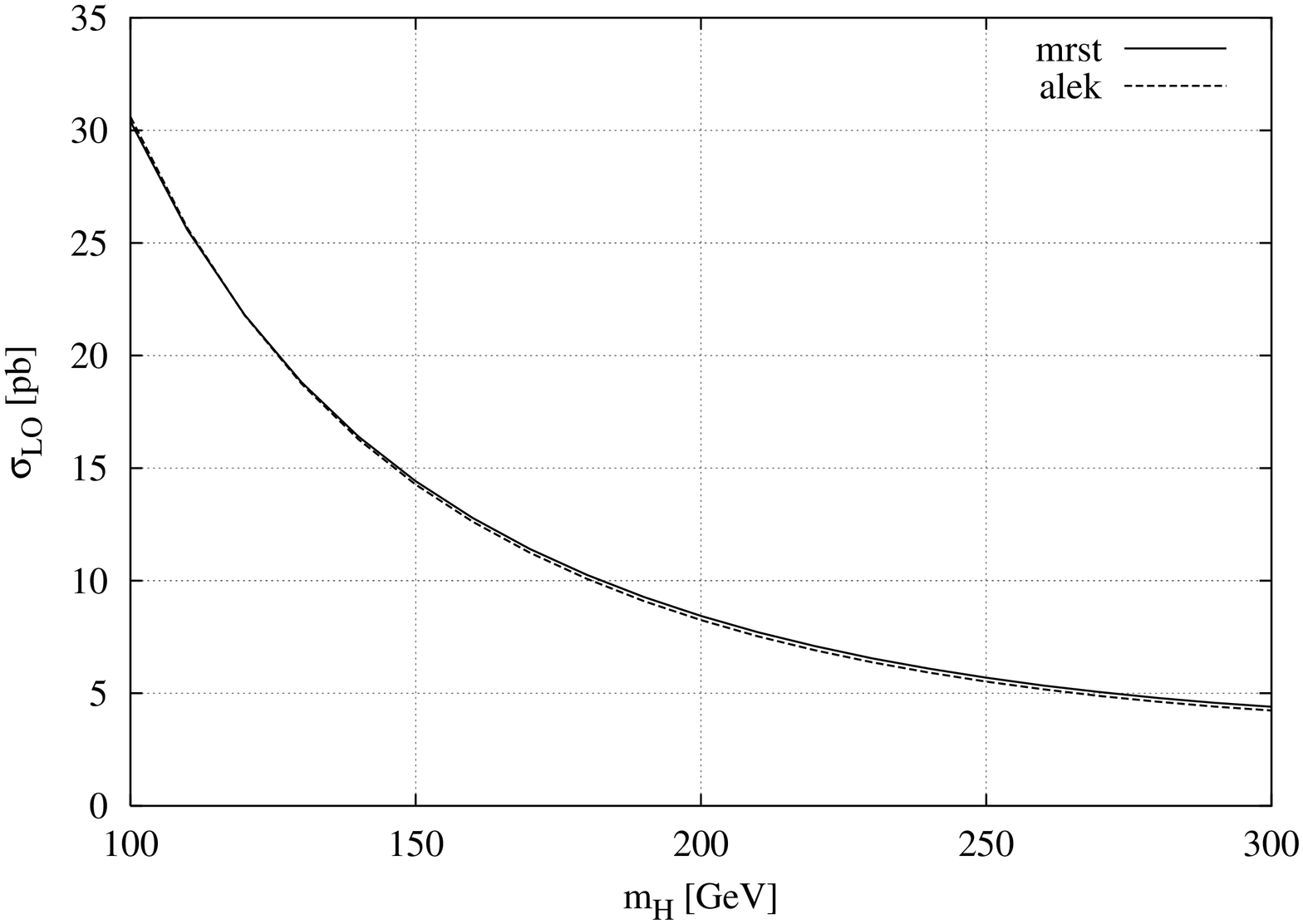}}
\subfigure[LHC]{\includegraphics[%
  width=6cm,
  angle=-90]{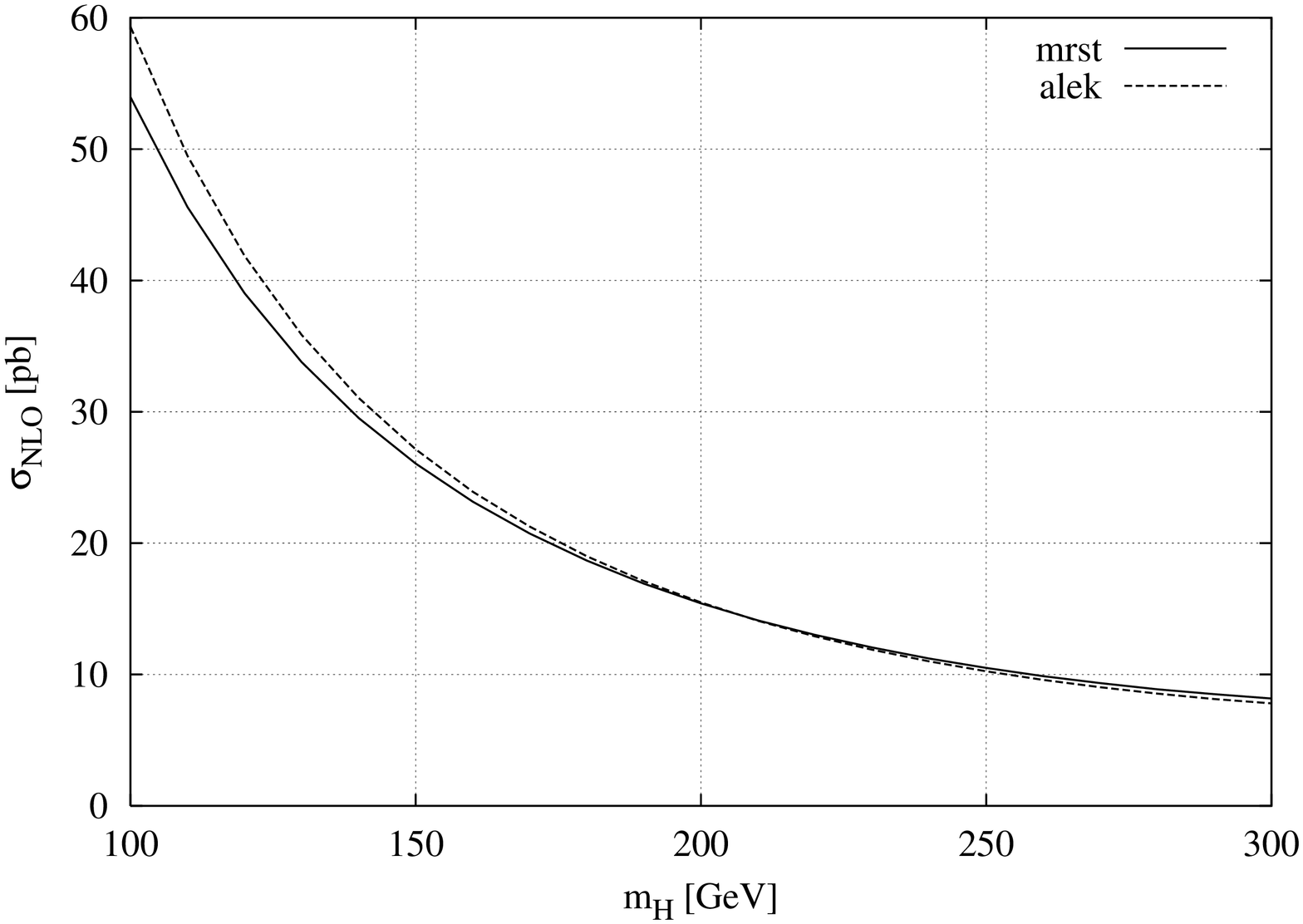}}
\subfigure[LHC]{\includegraphics[%
  width=6cm,
  angle=-90]{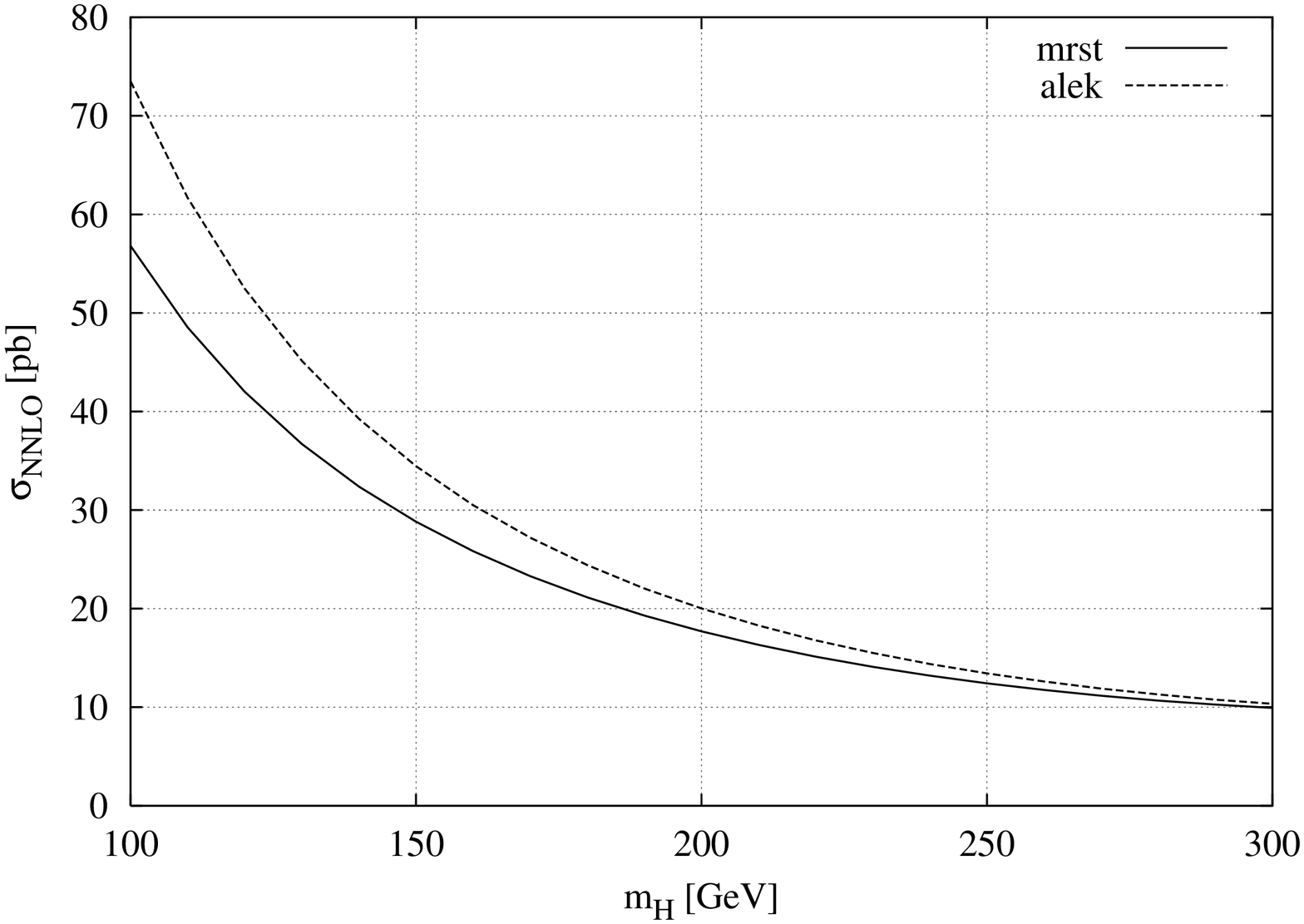}}
\subfigure[Tevatron]{\includegraphics[%
  width=6cm,
  angle=-90]{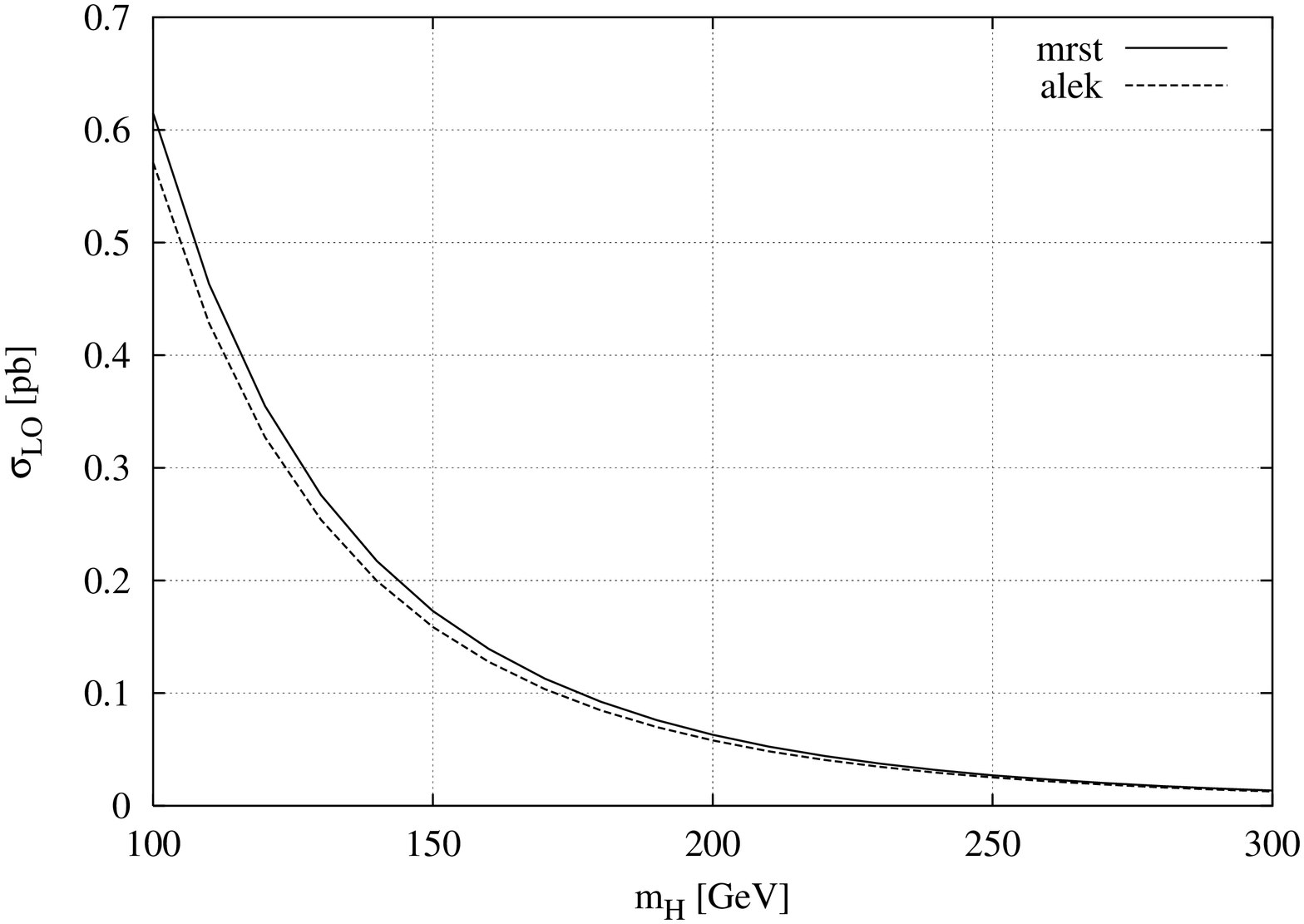}}
\subfigure[Tevatron]{\includegraphics[%
  width=6cm,
  angle=-90]{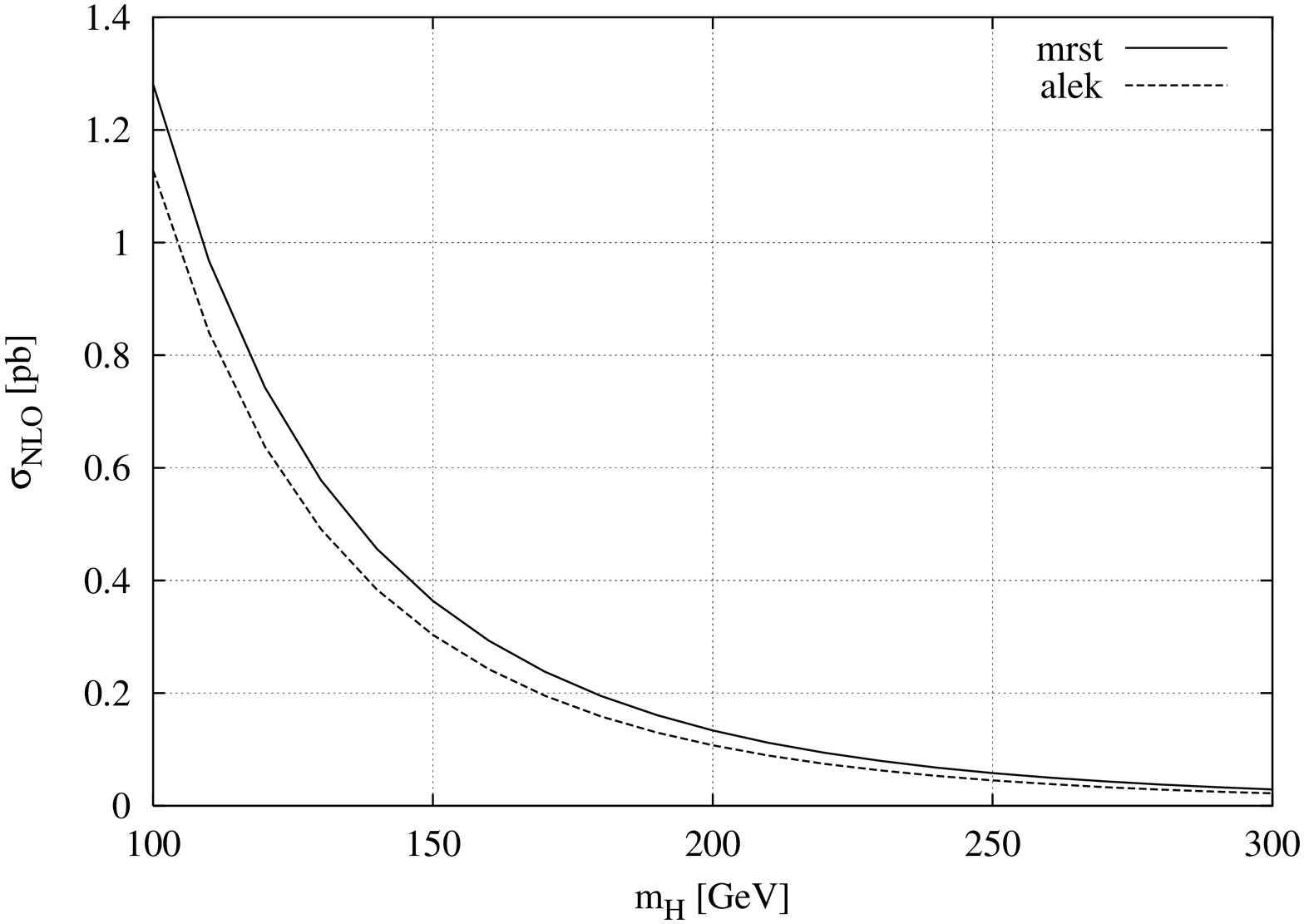}}
\subfigure[Tevatron]{\includegraphics[%
  width=6cm,
  angle=-90]{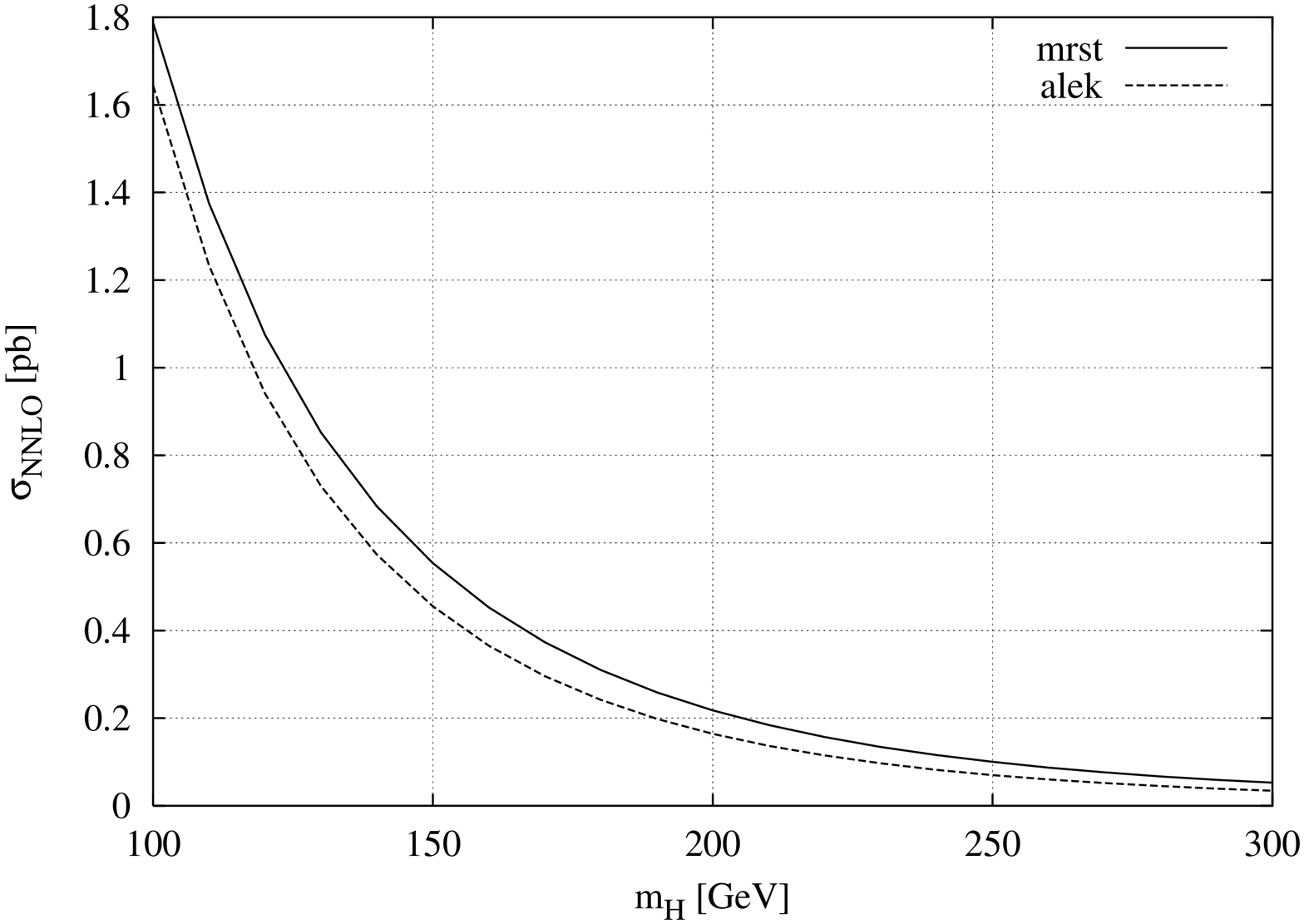}}
\caption{Cross sections for scalar Higgs production at the LHC
and at the Tevatron with $\mu_R^2=\mu_F^2=m_H^2$}
\label{S1P}
\end{figure}

We use as inital conditions at low scales the sets of distributions 
given by MRST \cite{MRST} and Alekhin \cite{Alekhin}.
Our final plots refer to center-of-mass energies which are reachable 
at the LHC, with 14 TeV being the largest one achievable in a not so 
distant future, and at the Tevatron, where we have selected the corresponding
value as 2 TeV. We have also taken the Higgs mass $m_H$ as a parameter 
in the prediction, with an interval of variability which goes from a light 
to a heavy Higgs (100 GeV to 300 GeV). Therefore $\mu_F$, $\mu_R$ and $m_H$ 
are studied chosing various combinations of their possible values in the 
determination of total cross sections at leading $(\sigma_{LO})$, 
next-to-leading $(\sigma_{NLO})$, and next-to-next-to-leading 
order $(\sigma_{NNLO})$. We present both standard two-dimensional
plots and also some three dimensional plots in order to characterize 
in detail the structure of the region of stability of the 
perturbative expansion. We have also evaluated the K-factors for the 
total Higgs cross section  at NLO, defined by 
\beq
K_1=\frac{\sigma_{NLO}}{\sigma_{LO}} 
\eeq
and at NNLO
\beq
K_2=\frac{\sigma_{NNLO}}{\sigma_{NLO}}. 
\eeq
The study of the K-factors has been performed first by keeping the three 
scales equal $(\mu_F=\mu_R=m_H)$ and then letting them vary around the 
typical value $m_H$. A second set of studies has been performed by taking 
typical values of $m_H$ and varying the value of the renormalization scale. 
We present in Figs.~\ref{S1P} LO and NLO results for the total Higgs cross 
sections  at 2 TeV and at 14 TeV. 
The LO cross sections increase by a factor of approximately 100 as 
we change the energy from 2 TeV to 14 TeV (Figs. \ref{S1P}-a and \ref{S1P}-d)  
and sharply decrease as we raise the mass of the Higgs boson. 
At 14 TeV the range of variation of $\sigma_{LO}$ 
is between 30 and 5 pb, with the highest value reached for $m_H=100$ GeV. 

In the same figure we compare LO, NLO and NNLO cross 
sections at these two typical energies. It is quite evident that the role 
of the NLO corrections is to increase by a factor of approximately 2 
the LO cross section bringing the interval of variation of $\sigma_{NLO}$ 
between 60 and 10 pb, for an increasing value of $m_H$. NNLO corrections 
at 14 TeV increase these values by an additional 10 per cent compared 
to the NLO prediction, with a growth which is more pronounced for the set 
proposed by Alekhin. The two sets give coincident predictions for values 
of $m_H > 200$ GeV.  At lower values of $m_H$ ($m_H=100$ GeV) the NNLO 
predictions for $\sigma$ are quite different in the two sets.
At the highest energy the cross sections evalutated in both models remain
below 80 pb, while at Tevatron energies they are around 2 pb.
\subsection{K-factors}
\begin{figure}
\subfigure[LHC]{\includegraphics[%
  width=6cm,
  angle=-90]{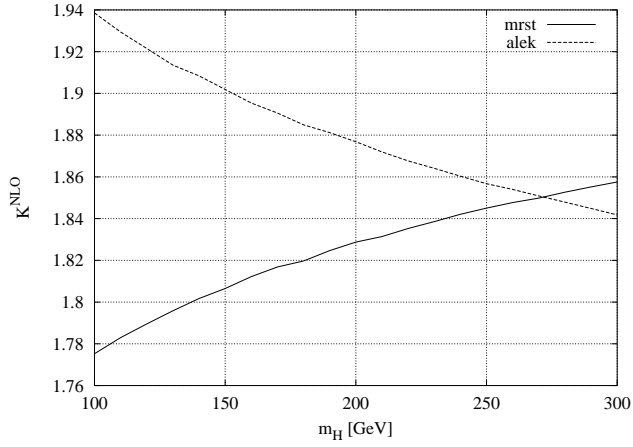}}
\subfigure[LHC]{\includegraphics[%
  width=6cm,
  angle=-90]{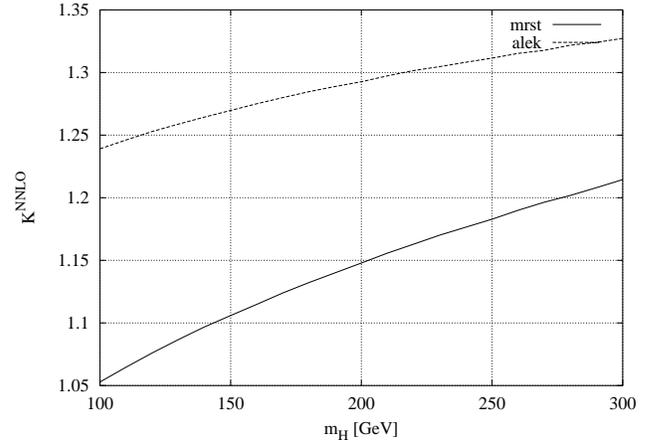}}
\subfigure[Tevatron]{\includegraphics[%
  width=6cm,
  angle=-90]{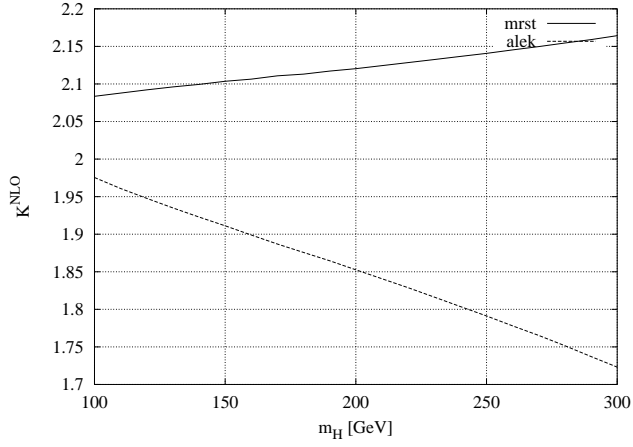}}
\subfigure[Tevatron]{\includegraphics[%
  width=6cm,
  angle=-90]{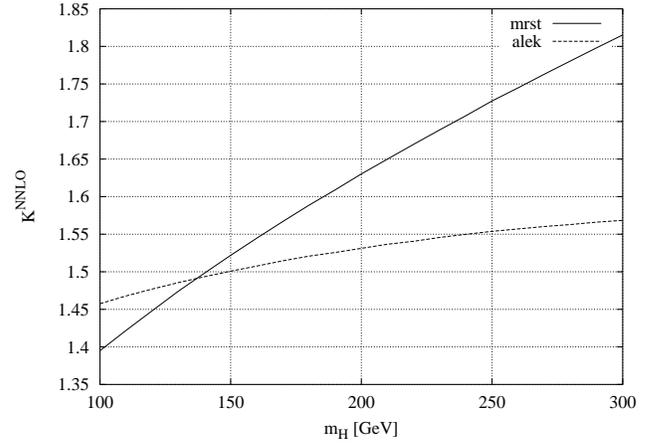}}
\caption{K-factors for scalar Higgs at NNLO/NLO
and NLO/LO with $\mu_R^2=\mu_F^2=m_H^2$}
\label{K2P}
\end{figure}

A precise indication on the impact of the NLO/NNLO corrections and the 
stability of the perturbative expansion comes from a study of the 
K-factors $K_1$ and $K_2$, defined above. From the plots in 
Figure \ref{K2P} the different behaviour of the predictions derived from 
the two models for the parton distributions is quite evident. At 14 TeV NLO 
K-factors from both models are large, as expected, since the LO dependence 
is not indicative of perturbation theory. The increase of $\sigma_{NLO}$ 
compared to the LO predictions is between 70 and 95 per cent.

 The two models also show a quite different behaviour for an increasing $m_H$: 
the impact of the NLO corrections to the LO result predicted by Alekhin 
decrease for an increasing $m_H$, while the MRST model has the opposite 
behaviour. The trend of the Alekhin set is reversed when we move to NNLO. 
The two $K_2$ factors grow between 1.05 and 1.35 at 14 TeV, with the 
corrections predicted by Alekhin being the largest ones. The evaluation of the 
overall impact of this growth on the size of these corrections should, 
however, also keep into consideration the fact that these corrections 
are enhanced in a region where the cross section is sharply 
decreasing (Figs.~\ref{S1P}). At 2 TeV $K_1$ is quite stable 
as $m_H$ increases (MRST set), varying between 2 and 2.2. 
In Alekhin's model this K-factor is smaller and its range is 
between 1.7 and 2. At the Tevatron the NNLO corrections are quite sizeable, 
with a factor $K_2$ between 1.4 and 1.8, 
as we increase the mass of the Higgs scalar. 

\subsection{Renormalization/factorization scale dependence}
\begin{figure}
\subfigure[LHC]{\includegraphics[%
  width=6cm,
  angle=-90]{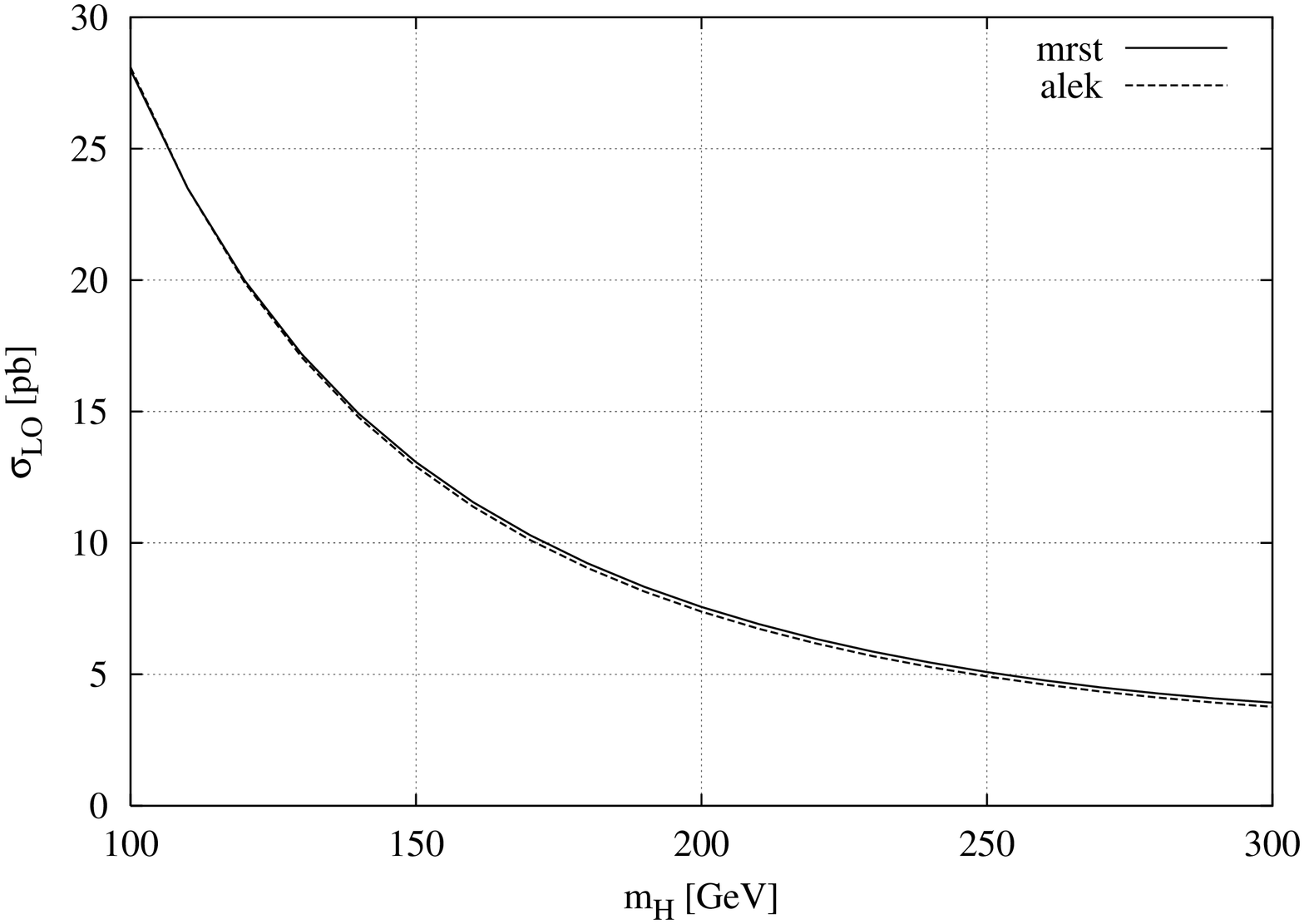}}
\subfigure[LHC]{\includegraphics[%
  width=6cm,
  angle=-90]{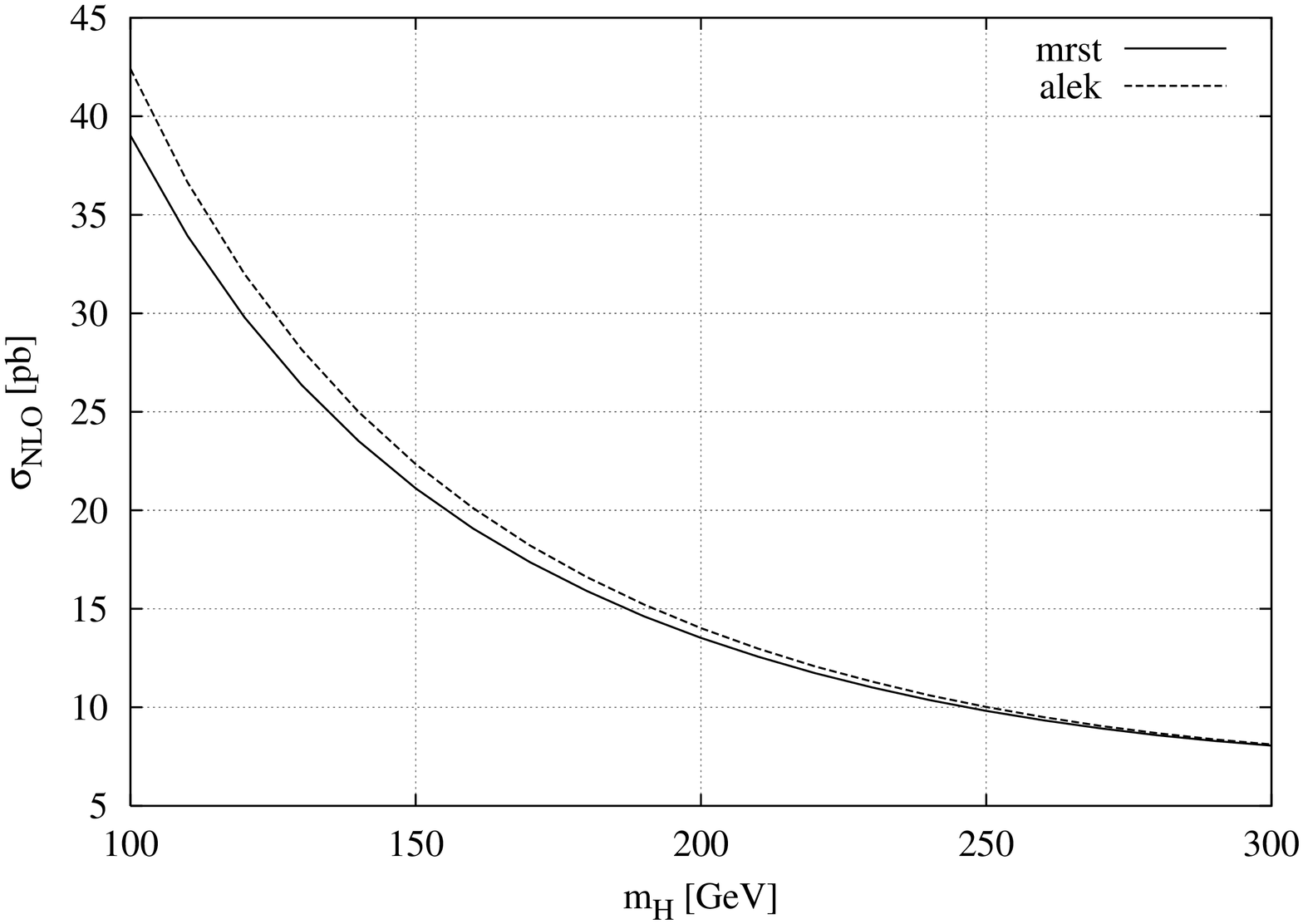}}
\subfigure[LHC]{\includegraphics[%
  width=6cm,
  angle=-90]{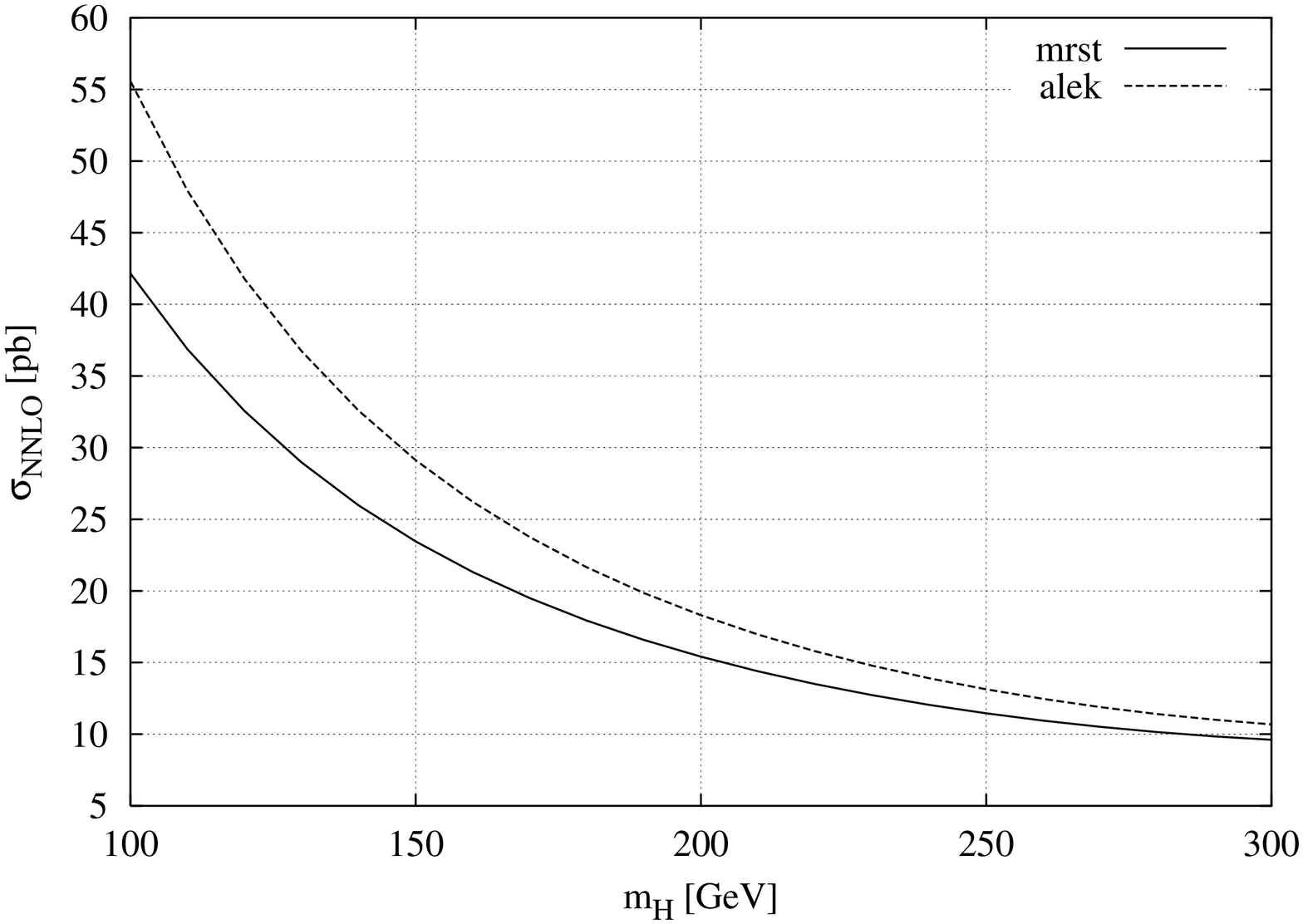}}
\subfigure[Tevatron]{\includegraphics[%
  width=6cm,
  angle=-90]{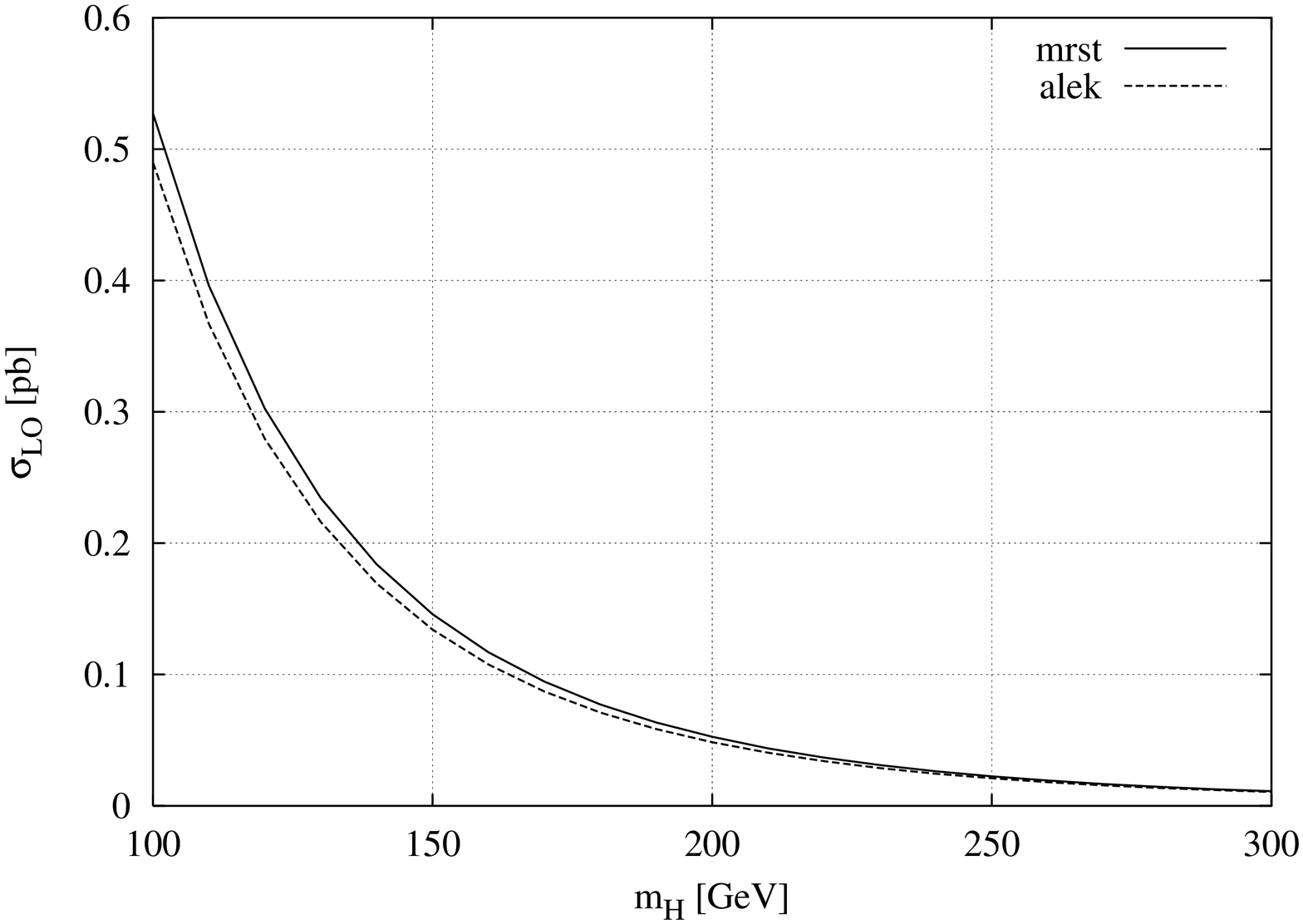}}
\subfigure[Tevatron]{\includegraphics[%
  width=6cm,
  angle=-90]{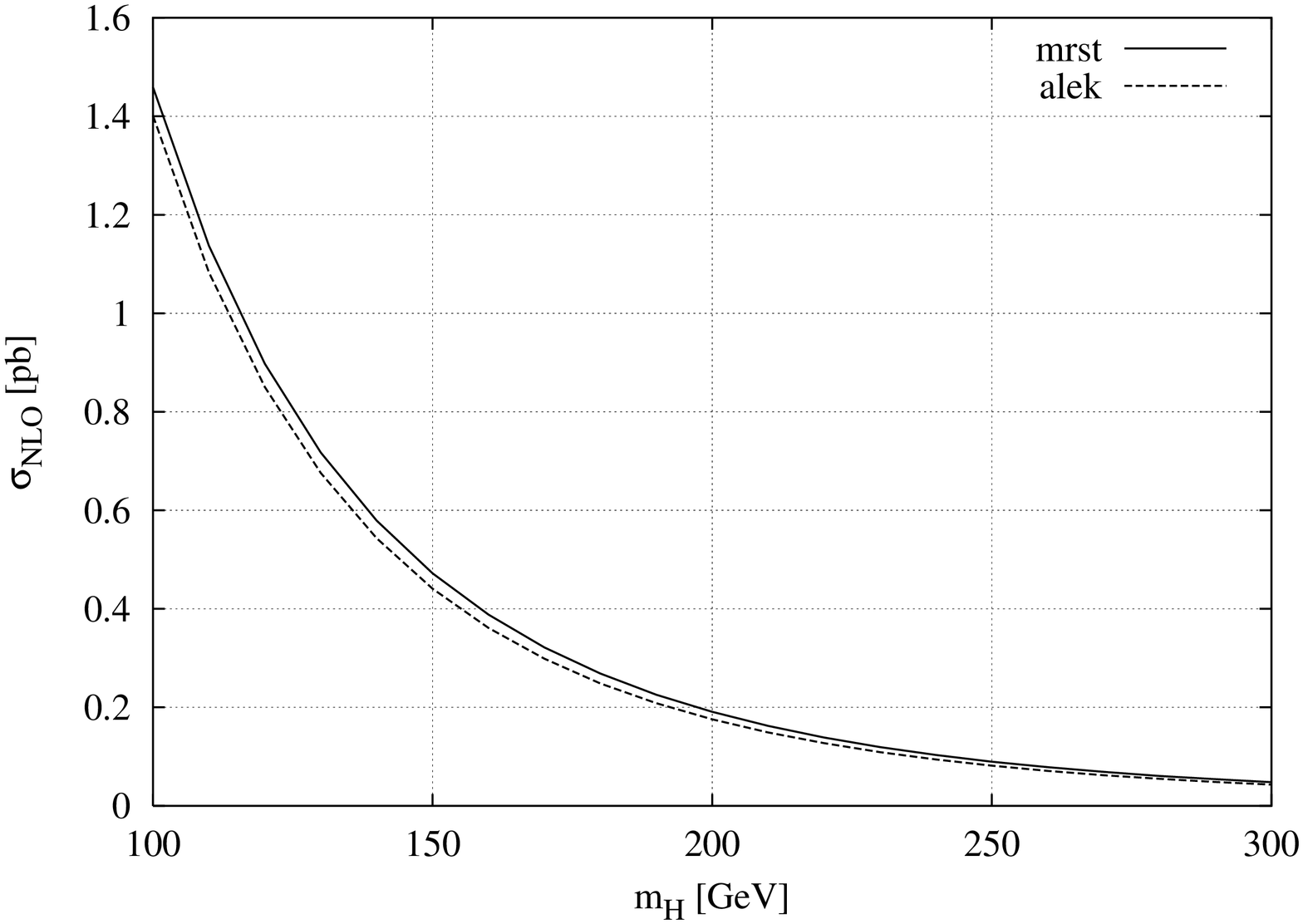}}
\subfigure[Tevatron]{\includegraphics[%
  width=6cm,
  angle=-90]{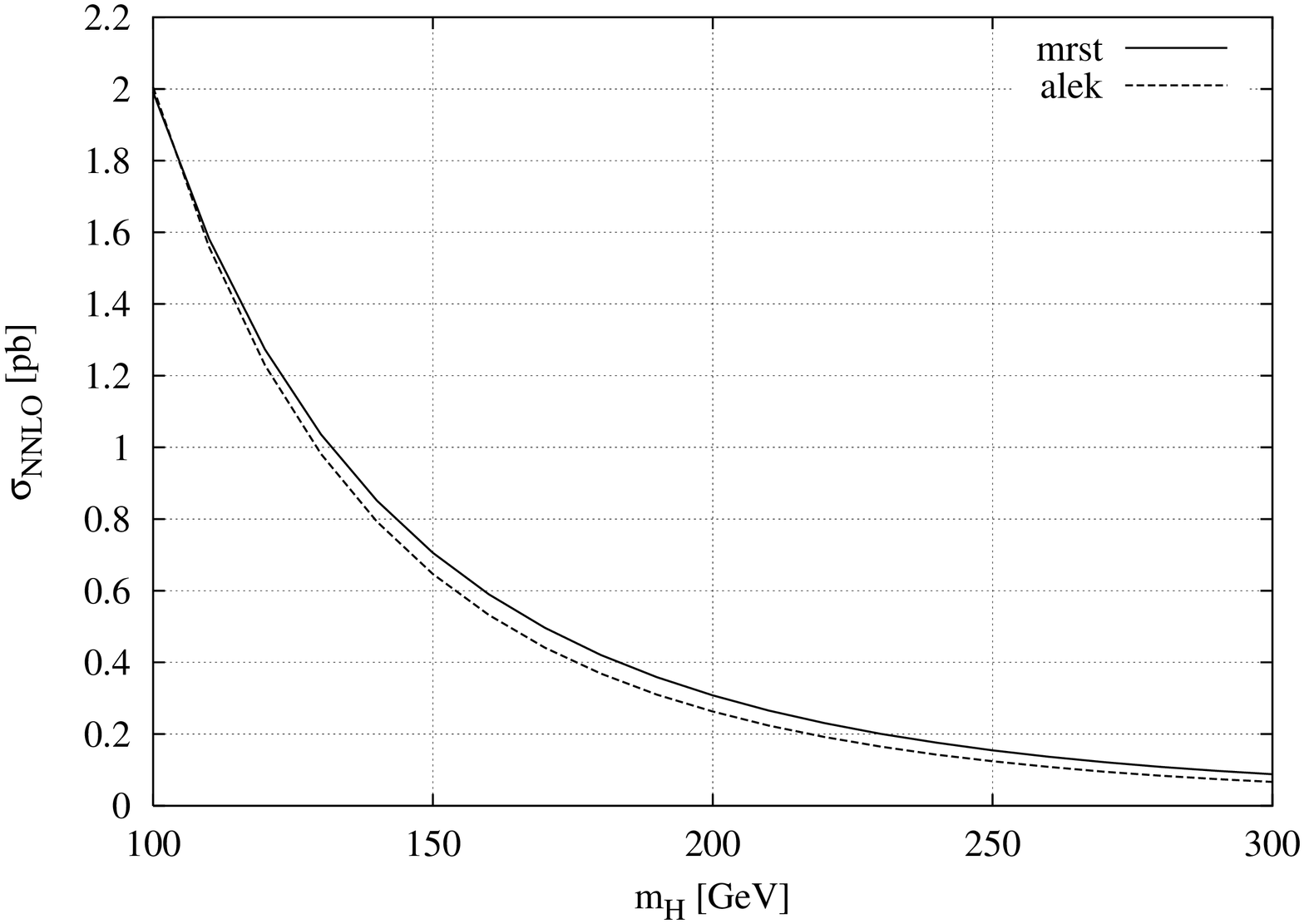}}
\caption{Cross sections for scalar Higgs production at LHC
and Tevatron with $\mu_R^2=(1/2)\mu_F^2$ and $\mu_F=2 m_H$}
\label{S3P}
\end{figure}

\begin{figure}
\subfigure[LHC]{\includegraphics[%
  width=6cm,
  angle=-90]{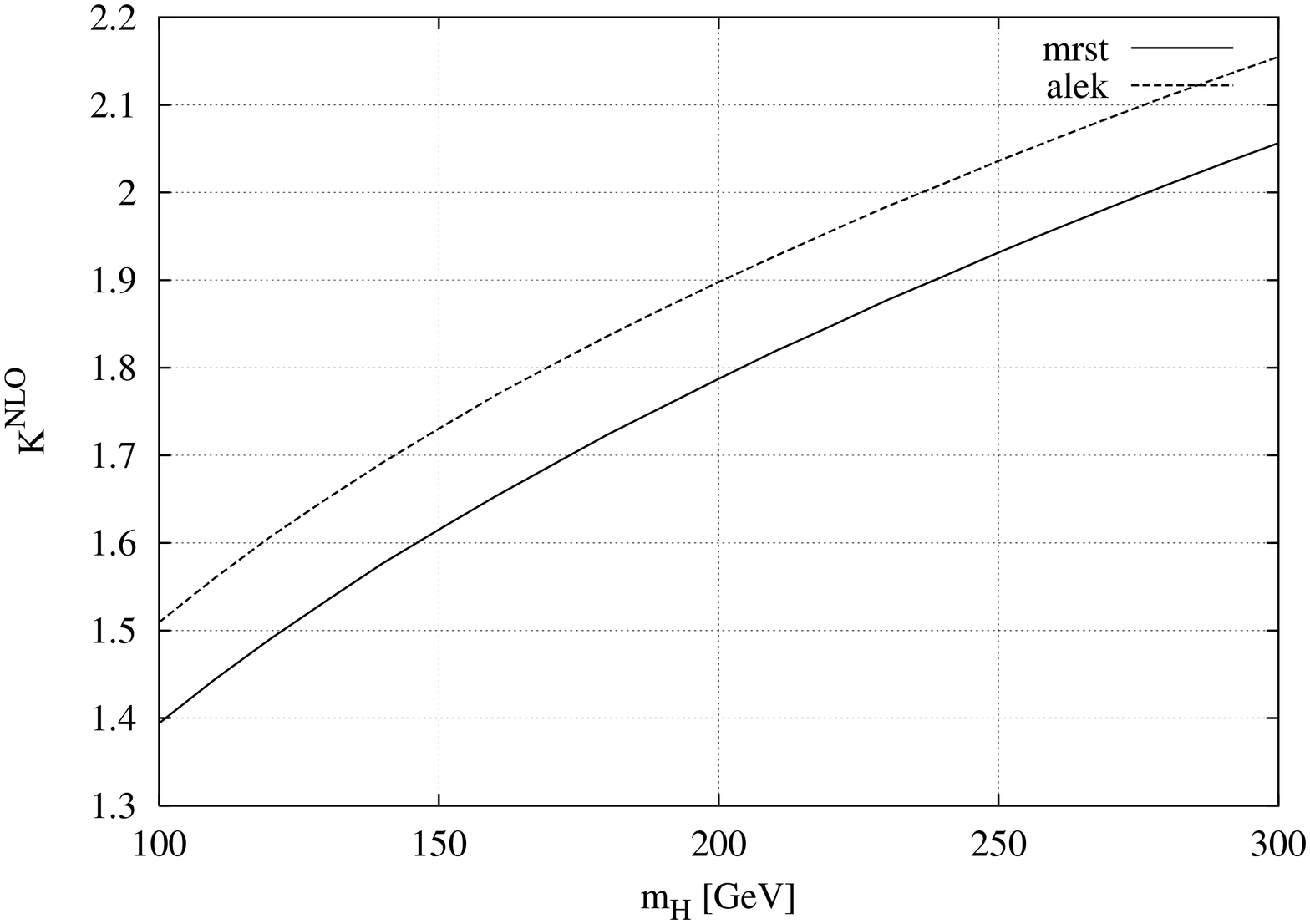}}
\subfigure[LHC]{\includegraphics[%
  width=6cm,
  angle=-90]{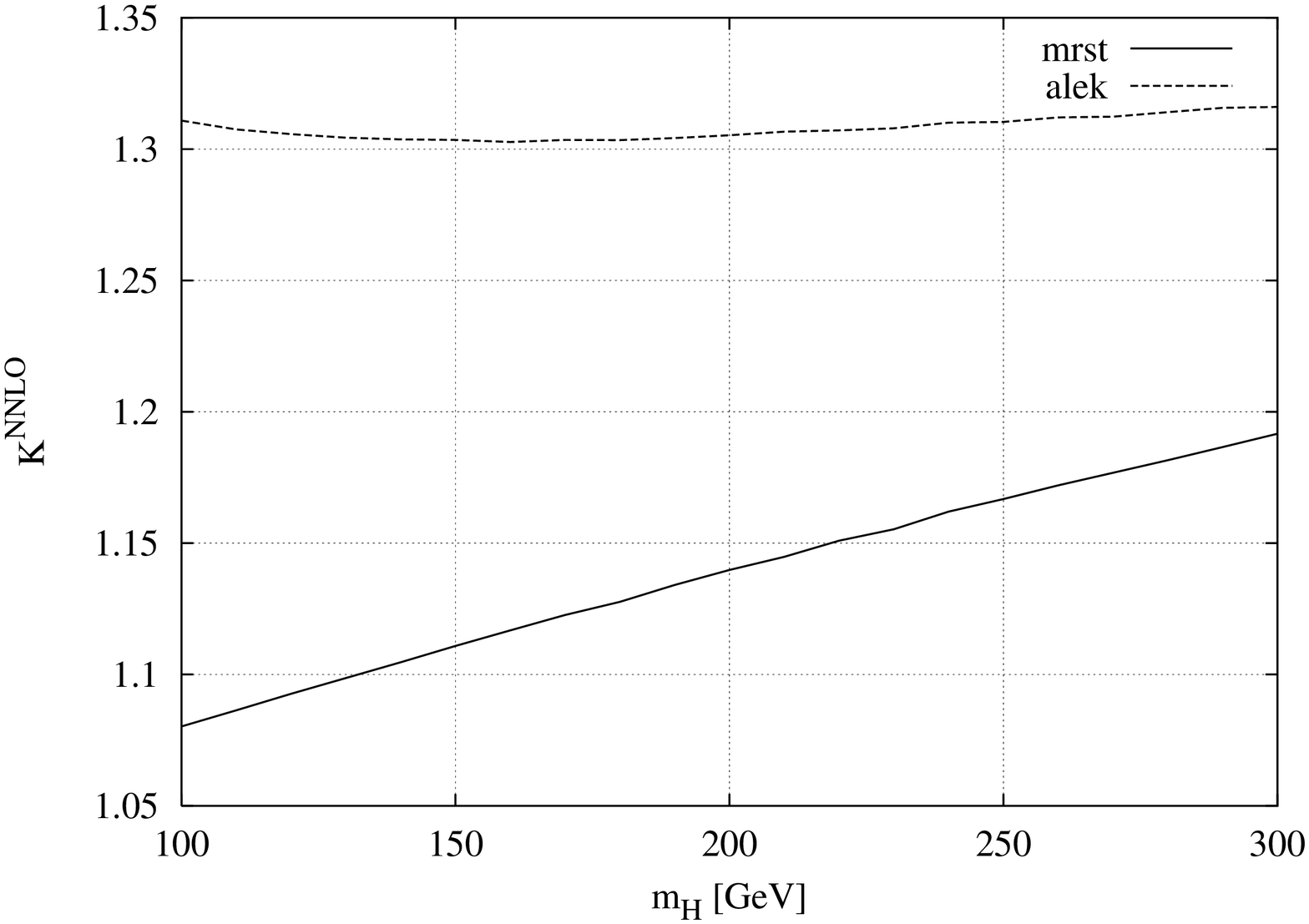}}
\subfigure[Tevatron]{\includegraphics[%
  width=6cm,
  angle=-90]{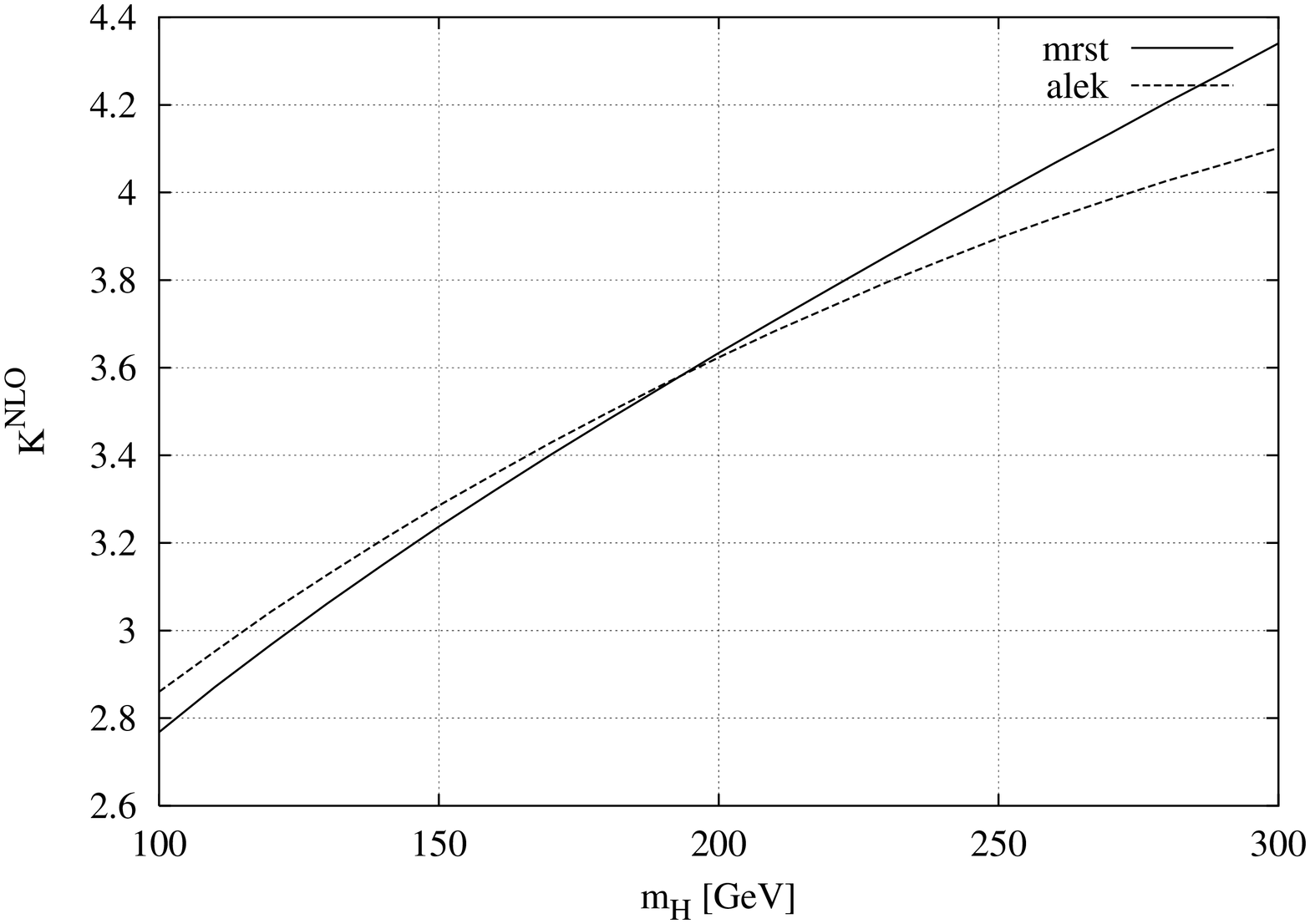}}
\subfigure[Tevatron]{\includegraphics[%
  width=6cm,
  angle=-90]{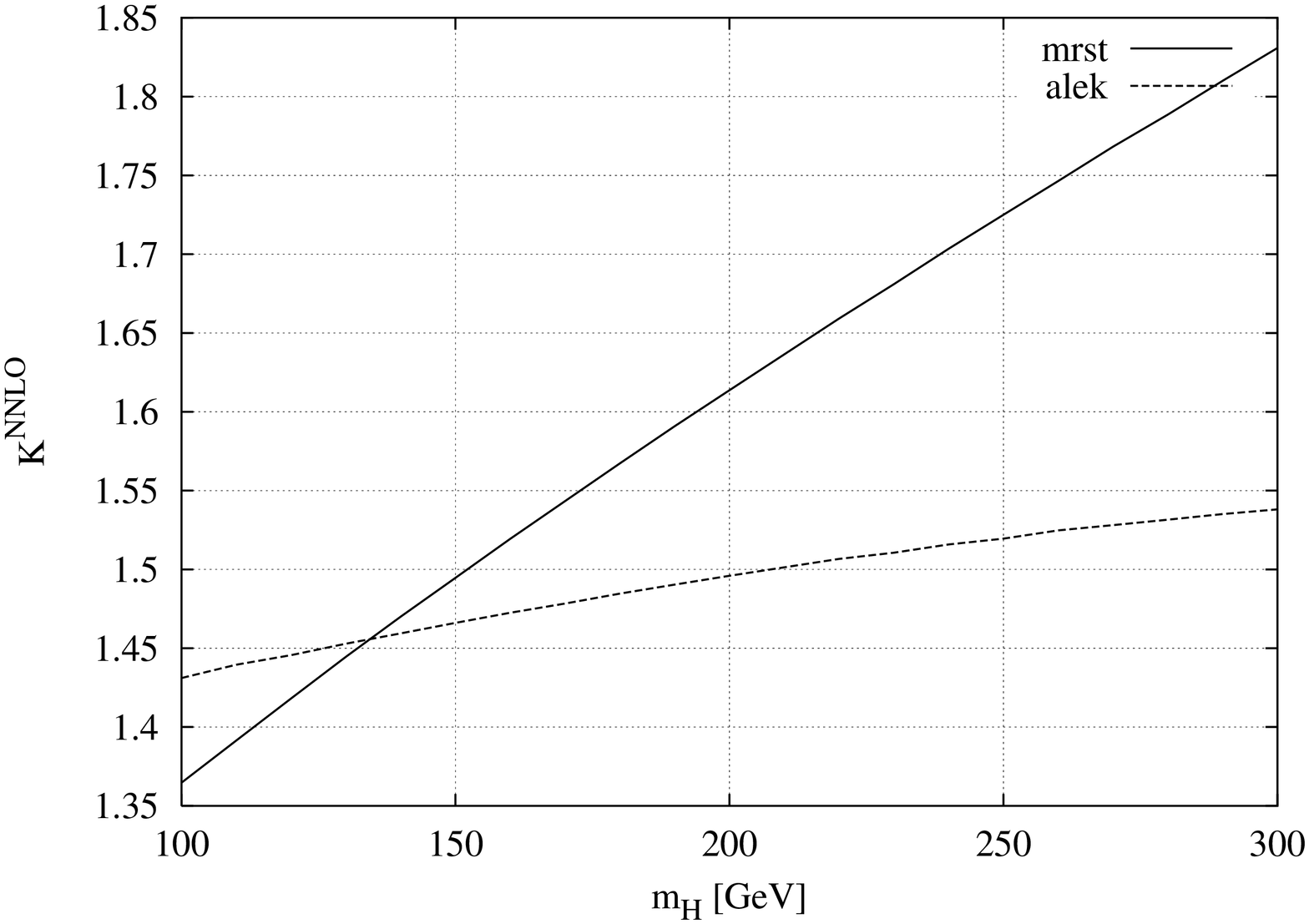}}
\caption{K-factors for scalar Higgs at NNLO/NLO and
NLO/LO with $\mu_R^2=(1/2)\mu_F^2$ and $\mu_F=2 m_H$}
\label{K4P}
\end{figure}

Now we turn to an analysis of the dependence of our results on 
$\mu_F$ and $\mu_R$.  In Figs.~\ref{S3P} we perform this study by 
computing $\sigma$ as a function of the 
Higgs mass for an incoming energy of 2 and 14 TeV and choose 
\beq
\mu_R^2=\frac{1}{2} \mu_F^2 \qquad \mu_F=2 m_H. 
\label{sel1}
\eeq
We have seen that for a typical Higgs mass around 100 GeV and 14 TeV of energy 
(for  $m_H=\mu_F=\mu_R$) the cross section doubles when 
we move from LO to NNLO, and a similar trend is also apparent if we fix 
the relations among the scales as in (\ref{sel1}). In this case, however, 
the impact of the NLO and NNLO corrections is smaller, a trend which is 
apparently uniform over the whole range of the Higgs 
mass explored. For $m_H=100$ GeV the scalar cross section $\sigma_{NNLO}$ 
is around 58 pb for coincident scales, while a different choice, 
such as (\ref{sel1}) lowers it to approximately 45 pb. At Tevatron energies 
the variations of the cross section with the changes of the various scales 
are also sizeable. In this case for $m_H=100$ GeV the LO, NLO and NNLO 
predictions ($0.6,1.2$ and 1.6 pb respectively) change approximately by 
10-20 per cent if we include variations of the other scales as well. 
A parallel view of this trend comes from the study of the 
dependence of the K-factors. This study is presented 
in Figs.~\ref{K4P}. The interval of variation of the K-factors is 
substantially the same as for coincident scales, though the trends of the 
two models \cite{MRST} and \cite{Alekhin} is structurally quite different 
at NLO and at NNLO, with several cross-overs 
among the corresponding curves taking place for $m_H$ around 200 GeV. 
Another important point is that the values of $K_2$ are, of course smaller 
than $K_1$ over all the regions explored, signaling an overall stability of 
the perturbative expansion.

\subsection{Stability and the Choice of the Scales}
\begin{figure}
\subfigure[$C=1$]{\includegraphics[%
  width=6cm,
  angle=-90]{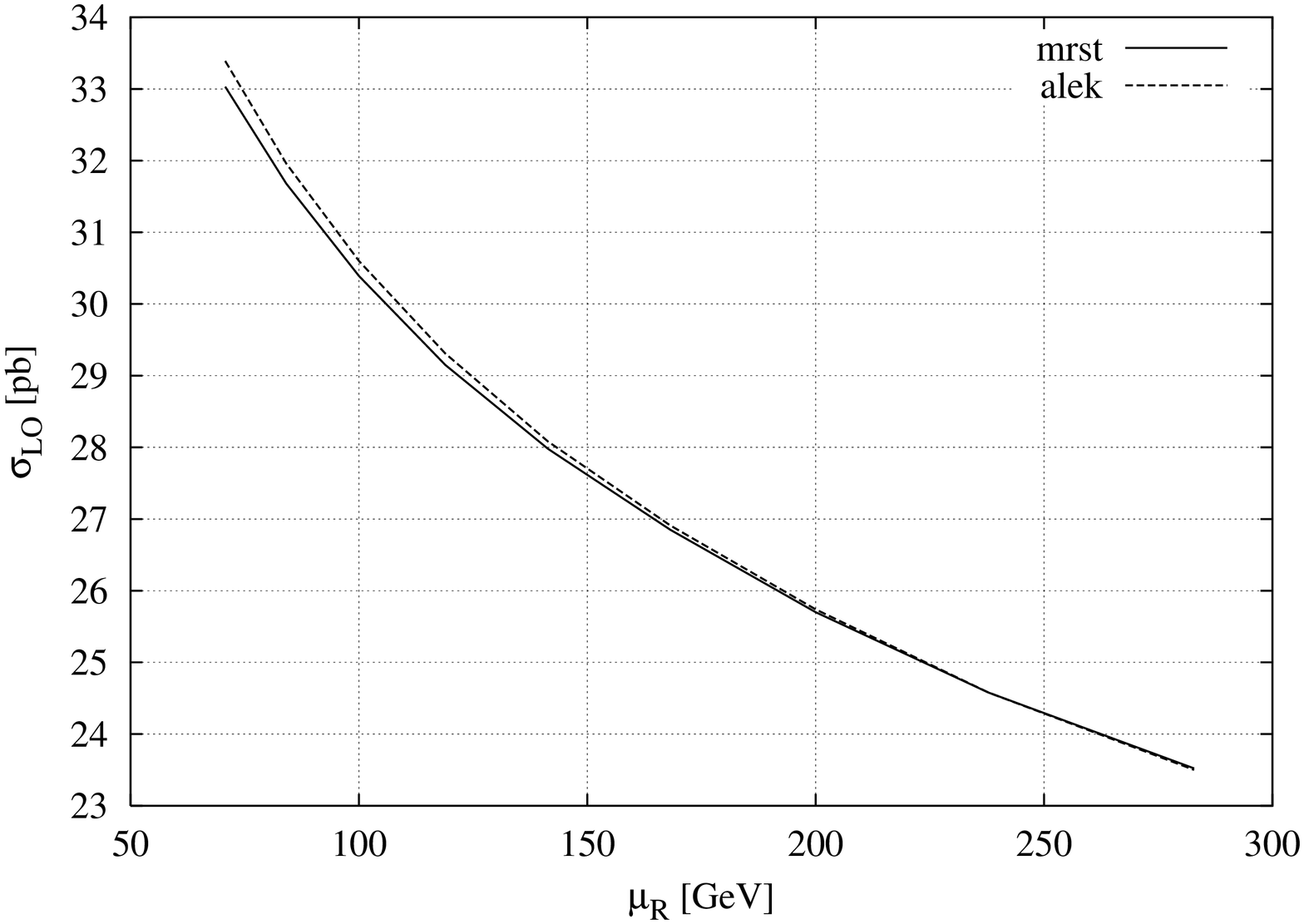}}
\subfigure[$C=1$]{\includegraphics[%
  width=6cm,
  angle=-90]{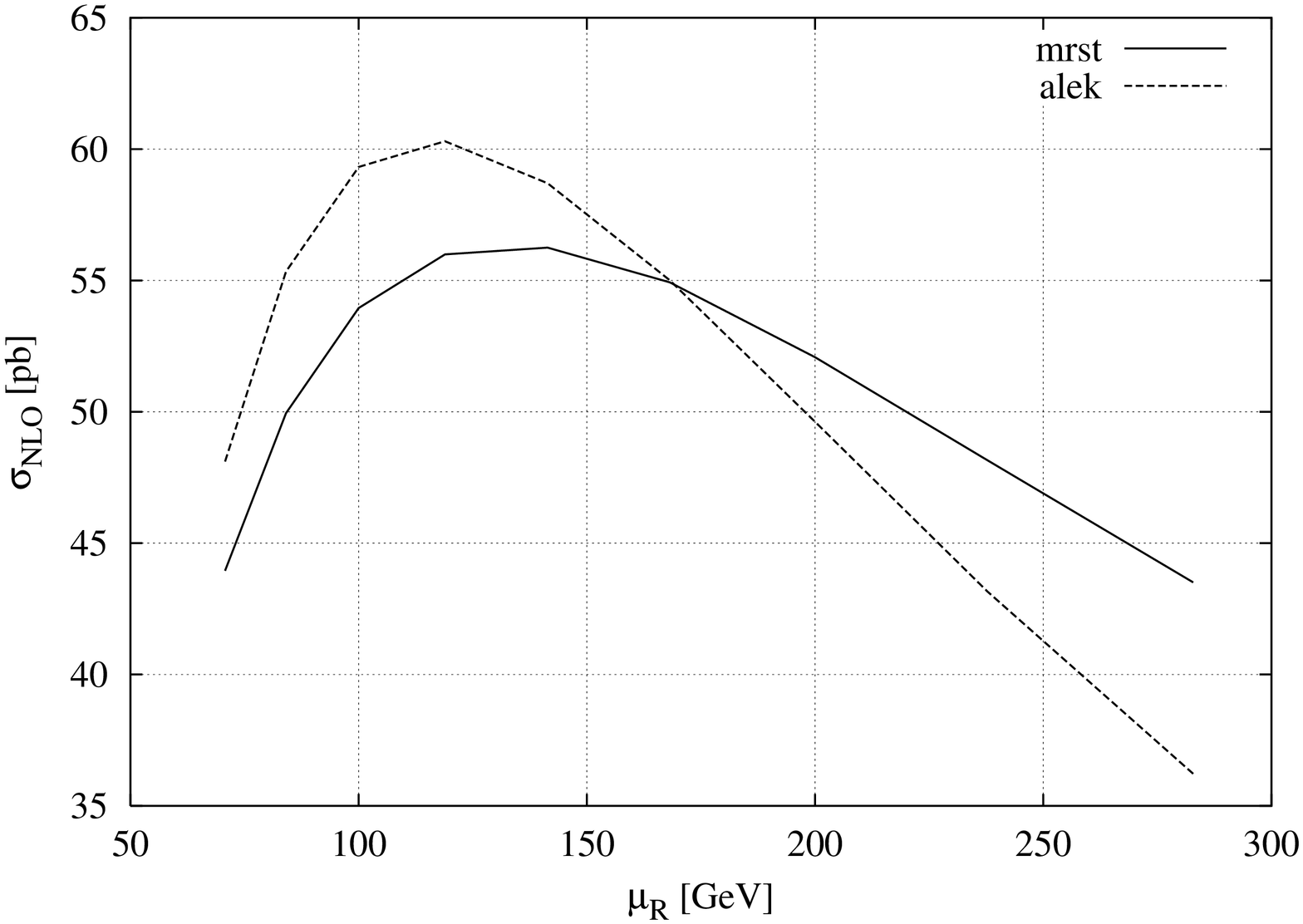}}
\subfigure[$C=1$]{\includegraphics[%
  width=6cm,
  angle=-90]{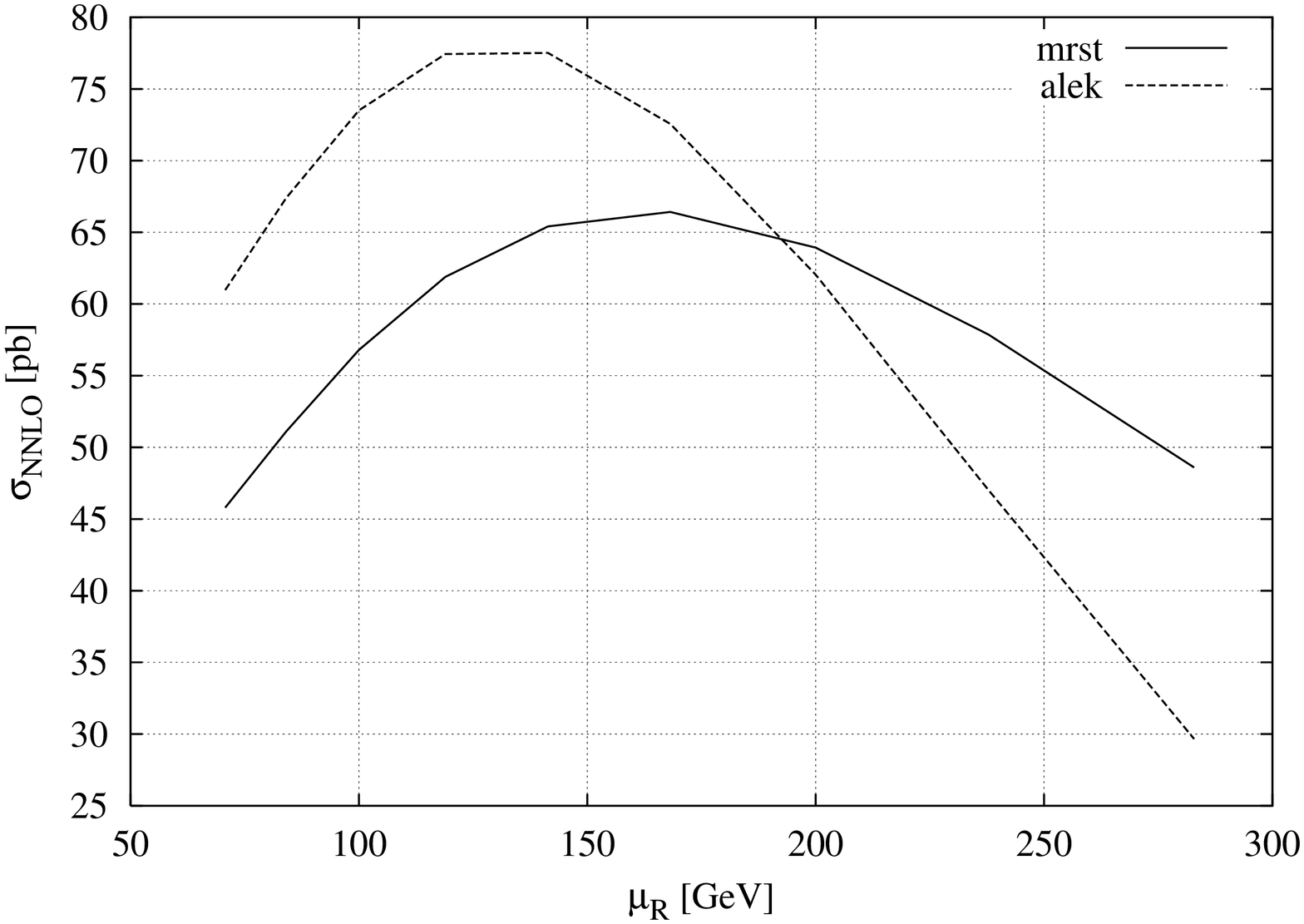}}
\subfigure[$C=1/2$]{\includegraphics[%
  width=6cm,
  angle=-90]{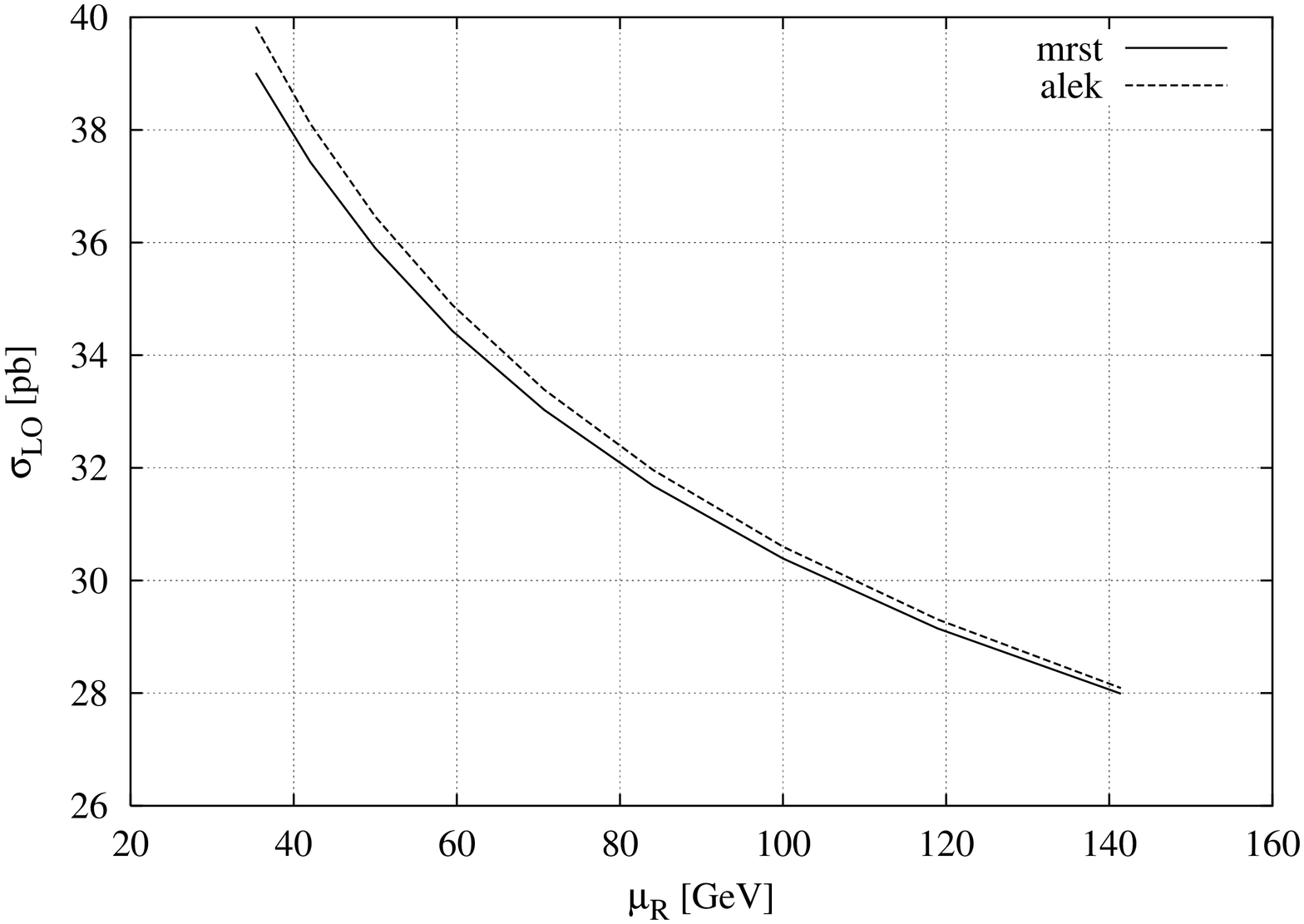}}
\subfigure[$C=1/2$]{\includegraphics[%
  width=6cm,
  angle=-90]{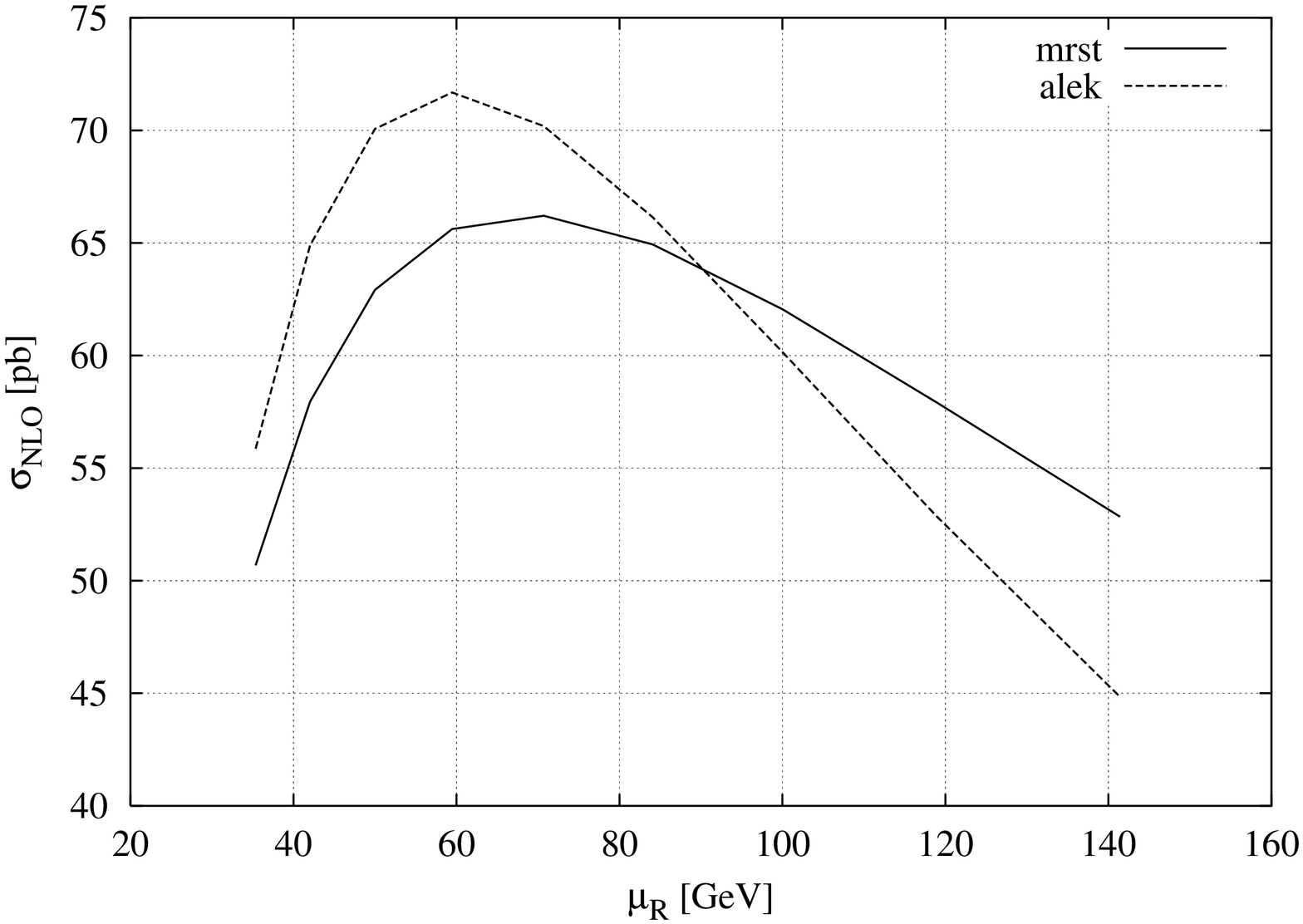}}
\subfigure[$C=1/2$]{\includegraphics[%
  width=6cm,
  angle=-90]{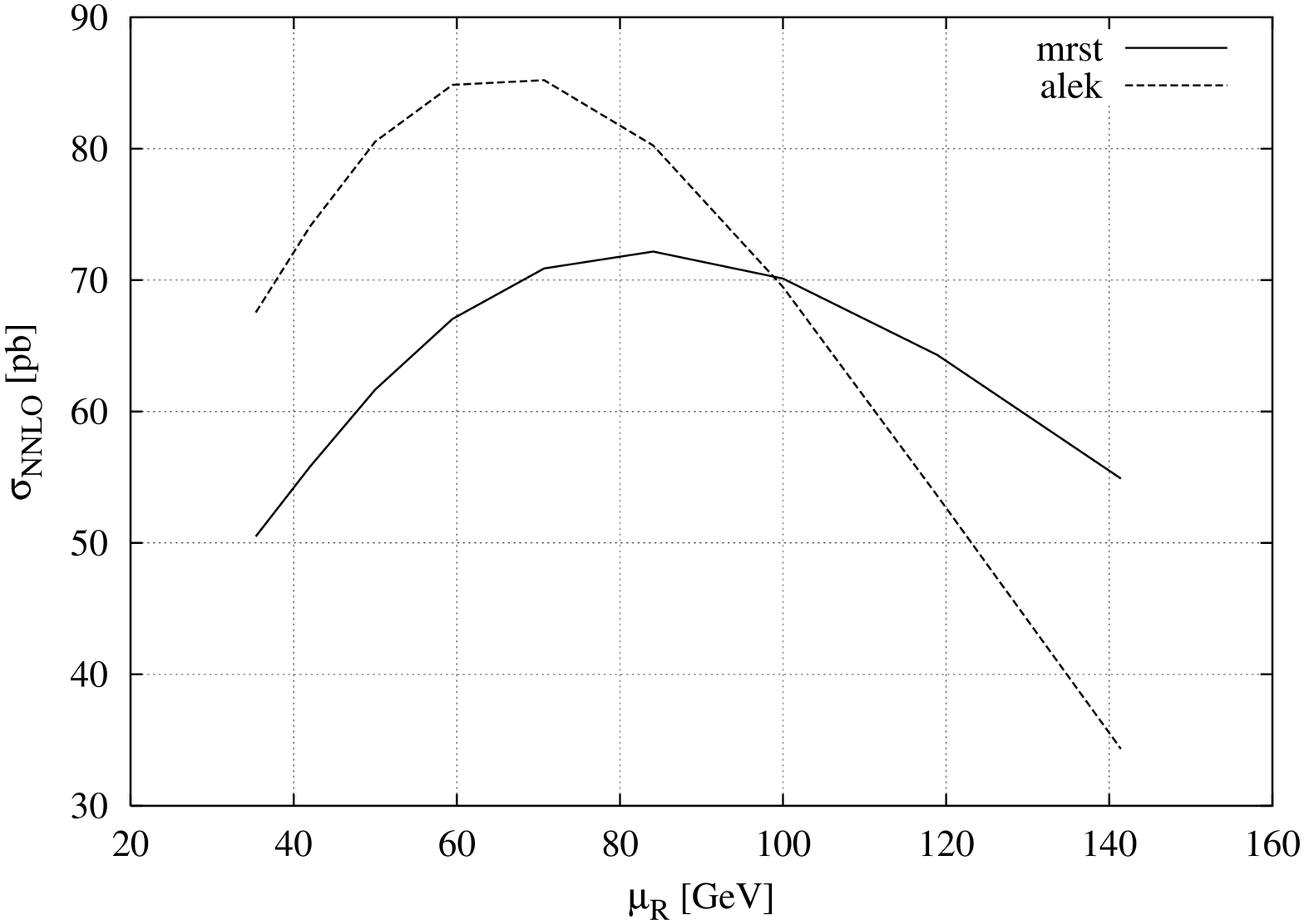}}
\caption{Cross sections for scalar Higgs production at LHC
as a function of $\mu_R$, with $\mu_F=C m_H$
and $m_H=100$ GeV}
\label{S5P}
\end{figure}

\begin{figure}
\subfigure[$C=1$]{\includegraphics[%
  width=6cm,
  angle=-90]{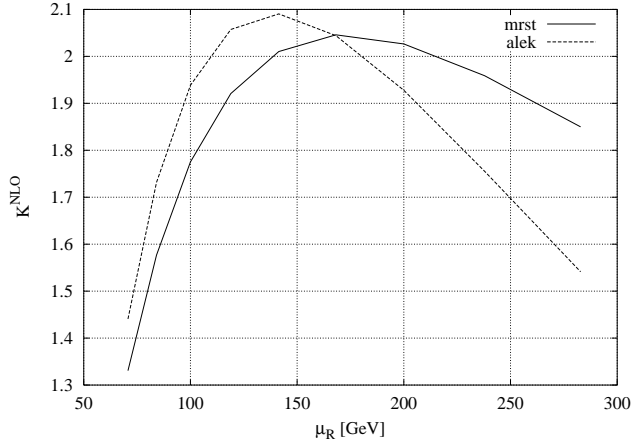}}
\subfigure[$C=1$]{\includegraphics[%
  width=6cm,
  angle=-90]{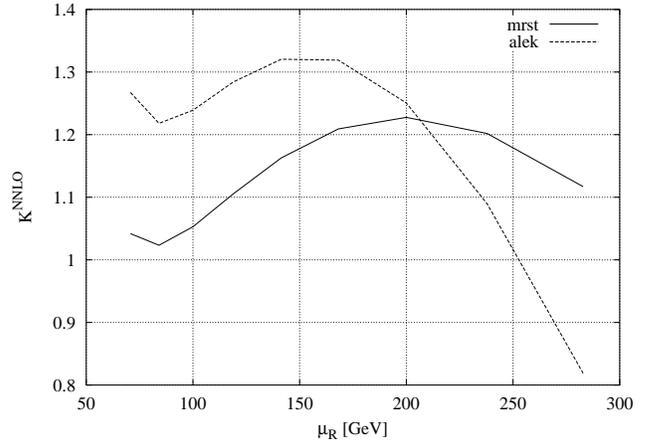}}
\subfigure[$C=1/2$]{\includegraphics[%
  width=6cm,
  angle=-90]{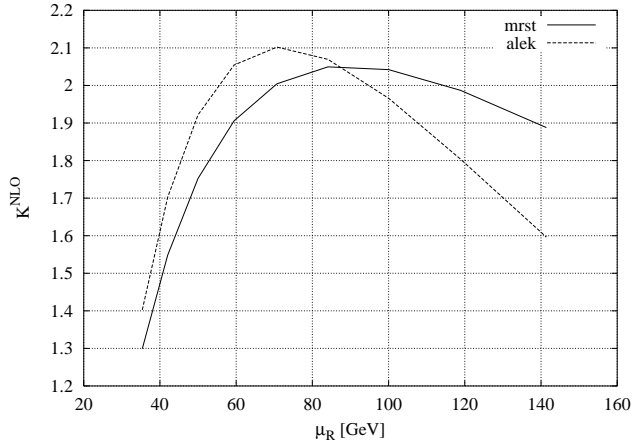}}
\subfigure[$C=1/2$]{\includegraphics[%
  width=6cm,
  angle=-90]{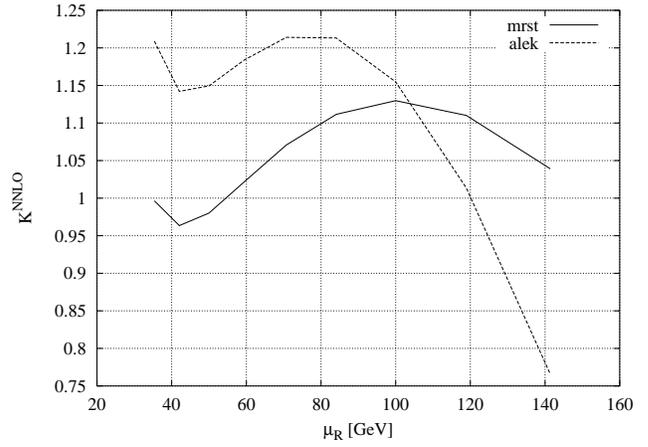}}
\caption{K-factors for scalar Higgs production at LHC, NNLO/NLO and
NLO/LO as a function of $\mu_R$, with $\mu_F=C m_H$
and $m_H=100$ GeV}
\label{K6P}
\end{figure}

 The issue of determining the best of values of $m_H$ $\mu_F$ and $\mu_R$ 
in the prediction of the total cross section is a rather important one for 
Higgs searches at LHC. We have therefore detailed in Figs.~\ref{S5P} 
and \ref{K6P} a study of the behaviour of our results varying the 
renormalization scale $\mu_R$ at a fixed value of the ratio between $\mu_F$ 
and $m_H$. In these figures we have chosen two values for the ratio between 
these two scales. Apart from the LO behaviour of the scalar cross section,
 which is clearly strongly dependent on the variation of both scales 
(see Figs. (a) and (d)), and does not show any sign of stability since the 
cross section can be drastically lowered by a different choice of $\mu_F$, 
both the NLO and the NNLO predictions show instead a clear region of 
local stability for $\mu_R > \mu_F$ but not too far away from 
the ``coincidence region'' $\mu_F=\mu_R=m_H$. This can be illustrated 
more simply using Fig. \ref{S5P}(b) as an example, where we have 
set the incoming energy of the p-p collision at 14 TeV. 
In this case, for instance, we have chosen $m_H=\mu_F=100$ GeV $(C=1)$, 
and it is clear from the plots that a plateau is present in the region 
of $\mu_R\sim 130$ GeV. Similar 
trends are also clearly visible at NNLO, though the region of the plateau 
for the scalar cross section is slightly wider. Also in this case it is 
found that the condition $\mu_R > \mu_F$ generates a reduced scale 
dependence. Away from this region the predictions 
show a systematic scale dependence, as shown also for the choice 
of $C=1/2$ in the remaining figures. In Figs.~\ref{K6P} we repeat the 
same study for the K-factors, relaxing the condition on the coincidence 
of all the scales and plotting the variations 
of $K_1$ and $K_2$ in terms of $\mu_R$. In the case $m_H=\mu_F=100$ GeV 
the plateau is reached for $\mu_R\sim 150$ GeV for $K_1$ 
and $\mu_R\sim 200$ GeV for $K_2$.  In the first case the NLO corrections 
amount to an increase by 100 per cent compared to the 
LO result, while the NNLO corrections modify the NLO estimates by about 
20 per cent (MRST).  Similar results are obtained also for $\mu_F=50$ GeV. 
In this case, at the plateau, the NLO corrections are still approximately 
100 per cent compared to the LO result and the NNLO corrections increase 
this value by around 15 per cent (MRST).  

\subsection{Energy Dependence} 
\begin{figure}
\subfigure[$C=1,\,\,k=1$]{\includegraphics[%
  width=6cm,
  angle=-90]{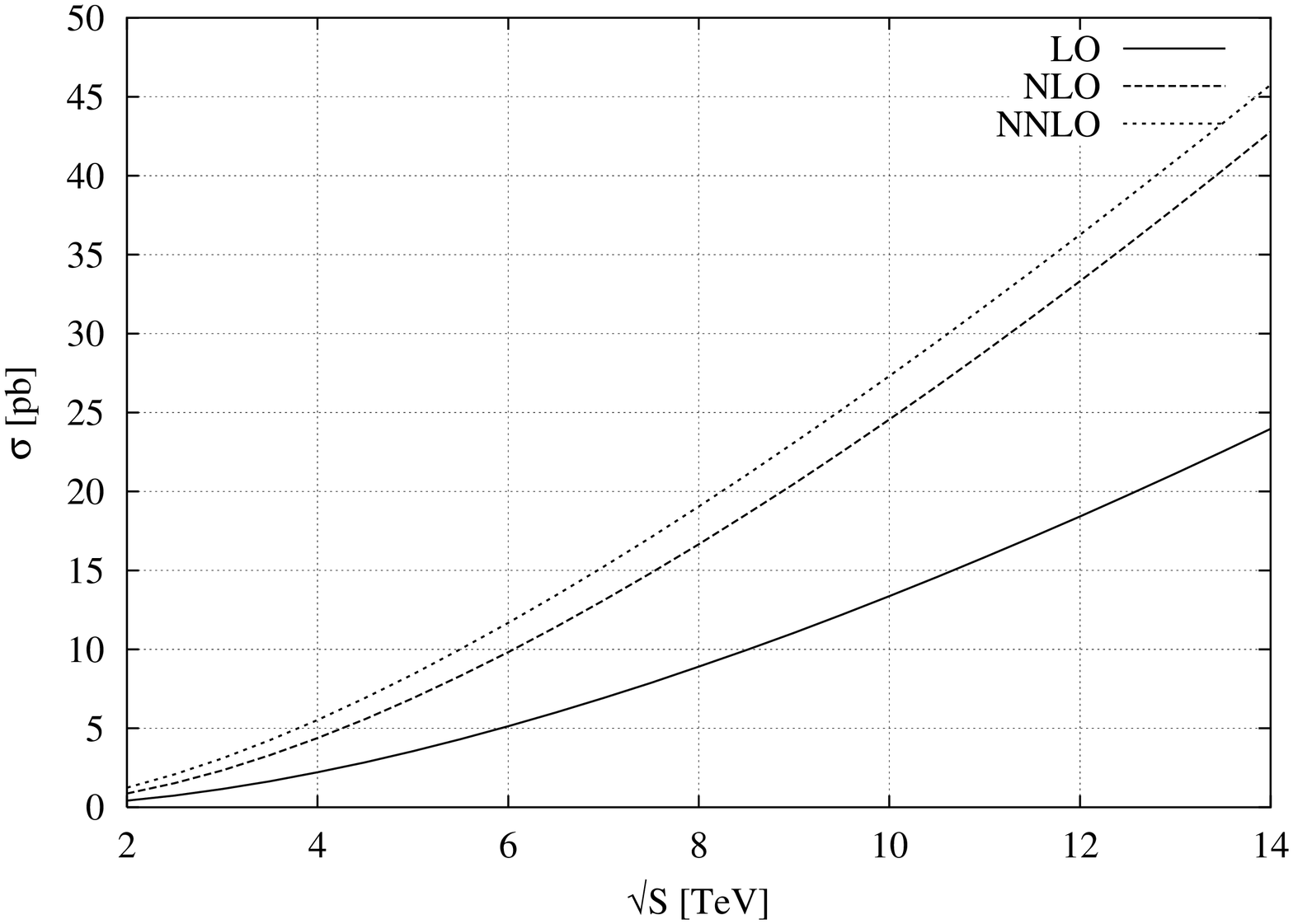}}
\subfigure[$C=1,\,\,k=1$]{\includegraphics[%
  width=6cm,
  angle=-90]{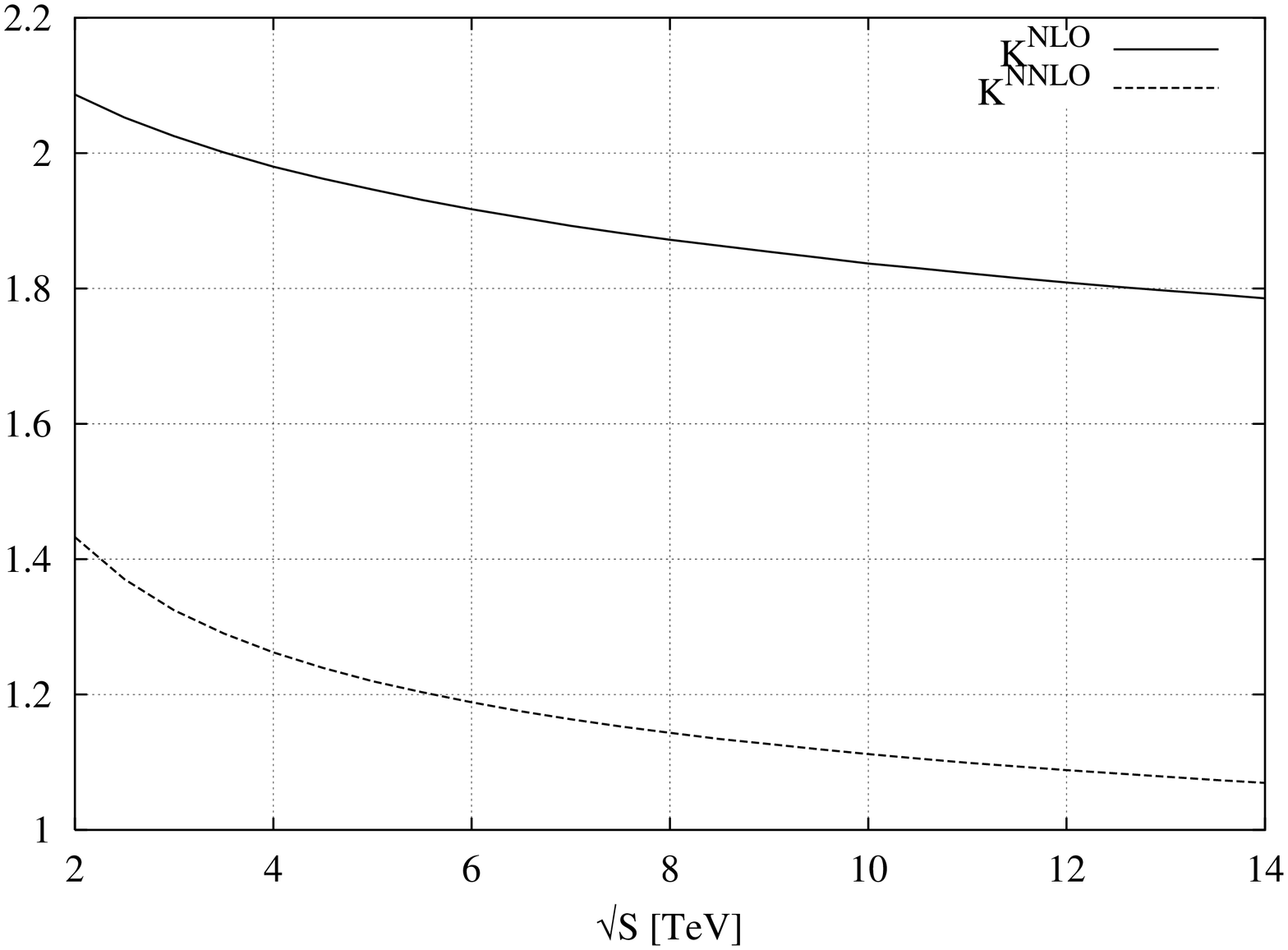}}
\subfigure[$C=2,\,\,k=1$]{\includegraphics[%
  width=6cm,
  angle=-90]{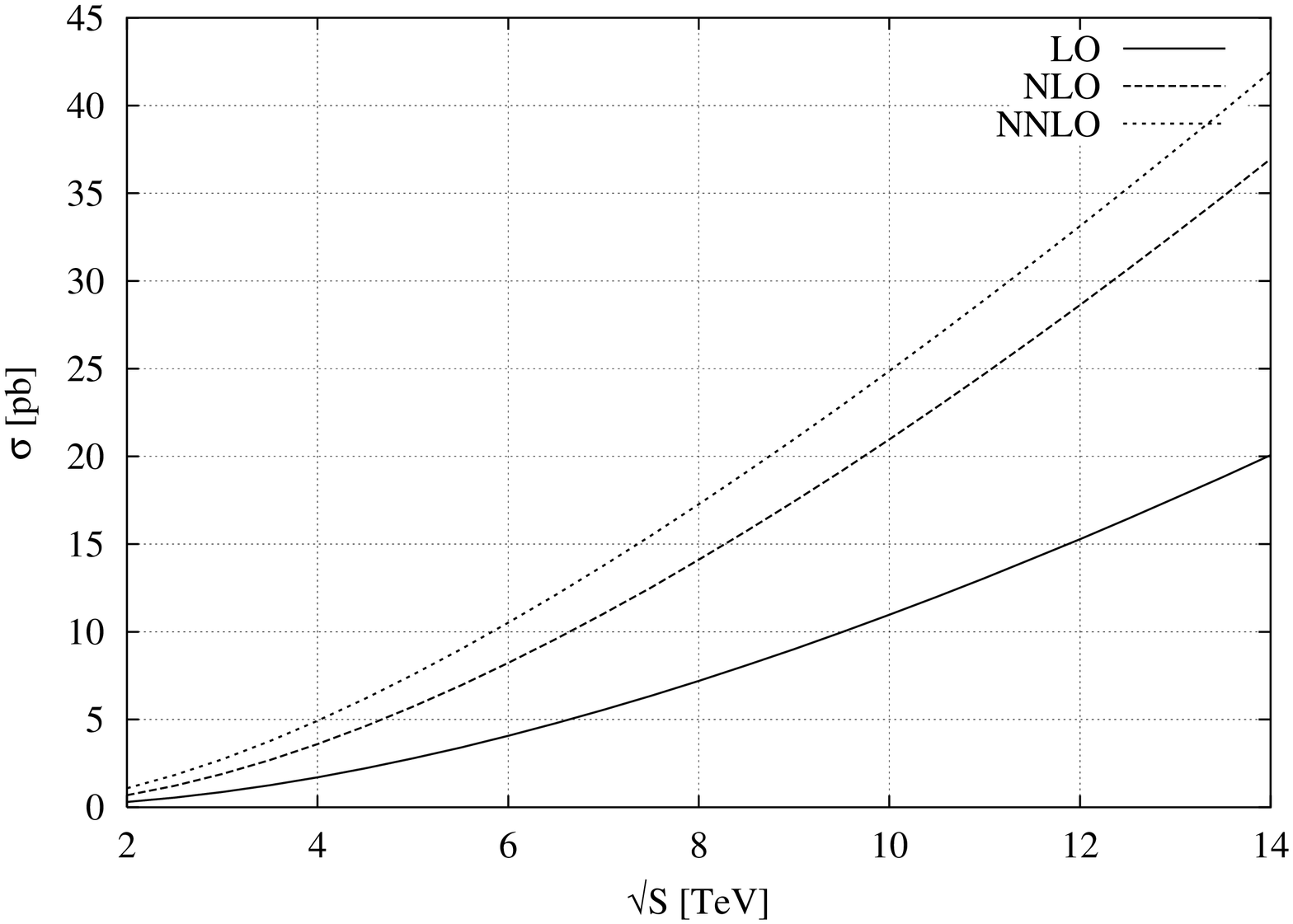}}
\subfigure[$C=2,\,\,k=1$]{\includegraphics[%
  width=6cm,
  angle=-90]{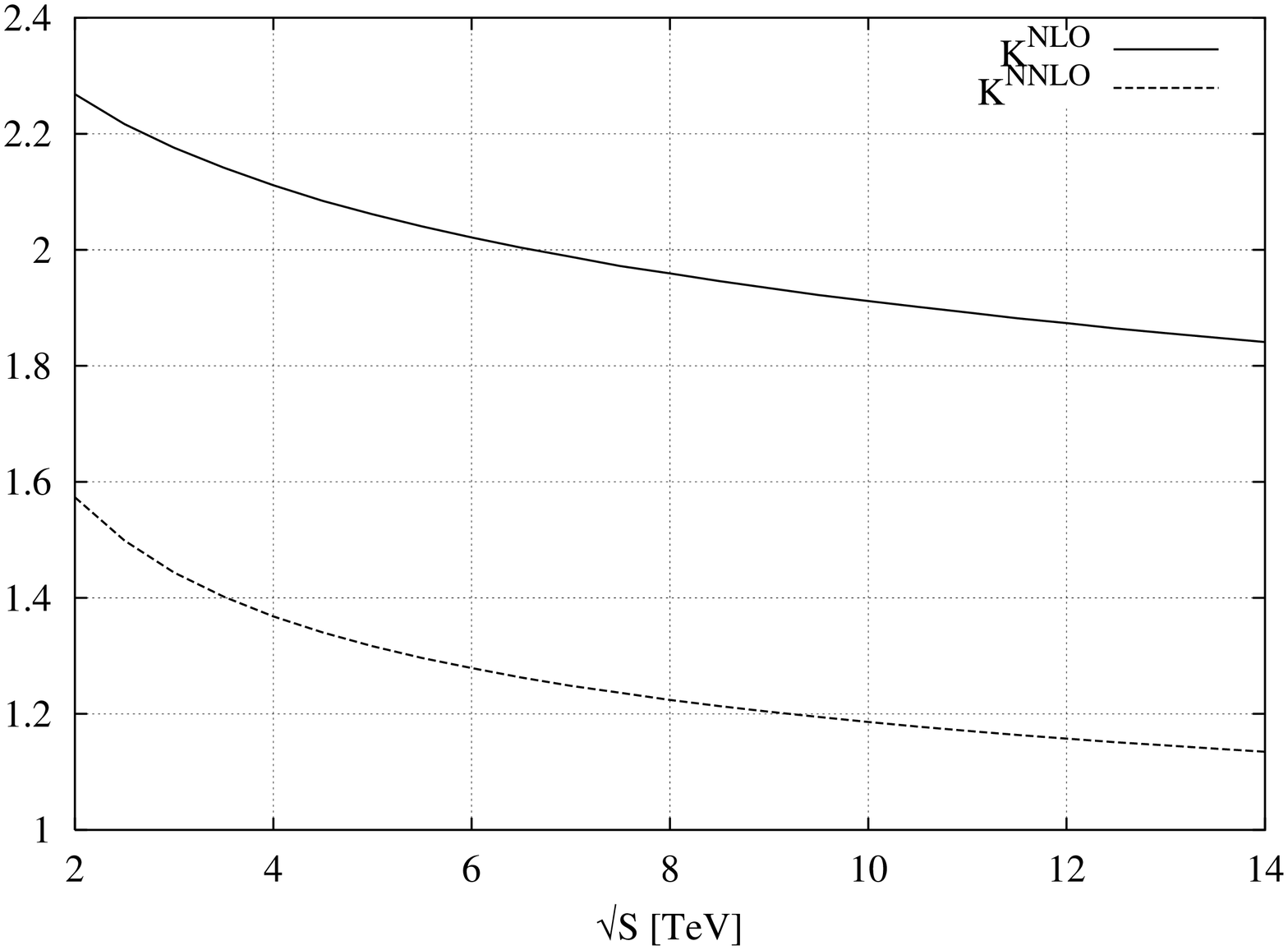}}
\subfigure[$C=1/2,\,\,k=1$]{\includegraphics[%
  width=6cm,
  angle=-90]{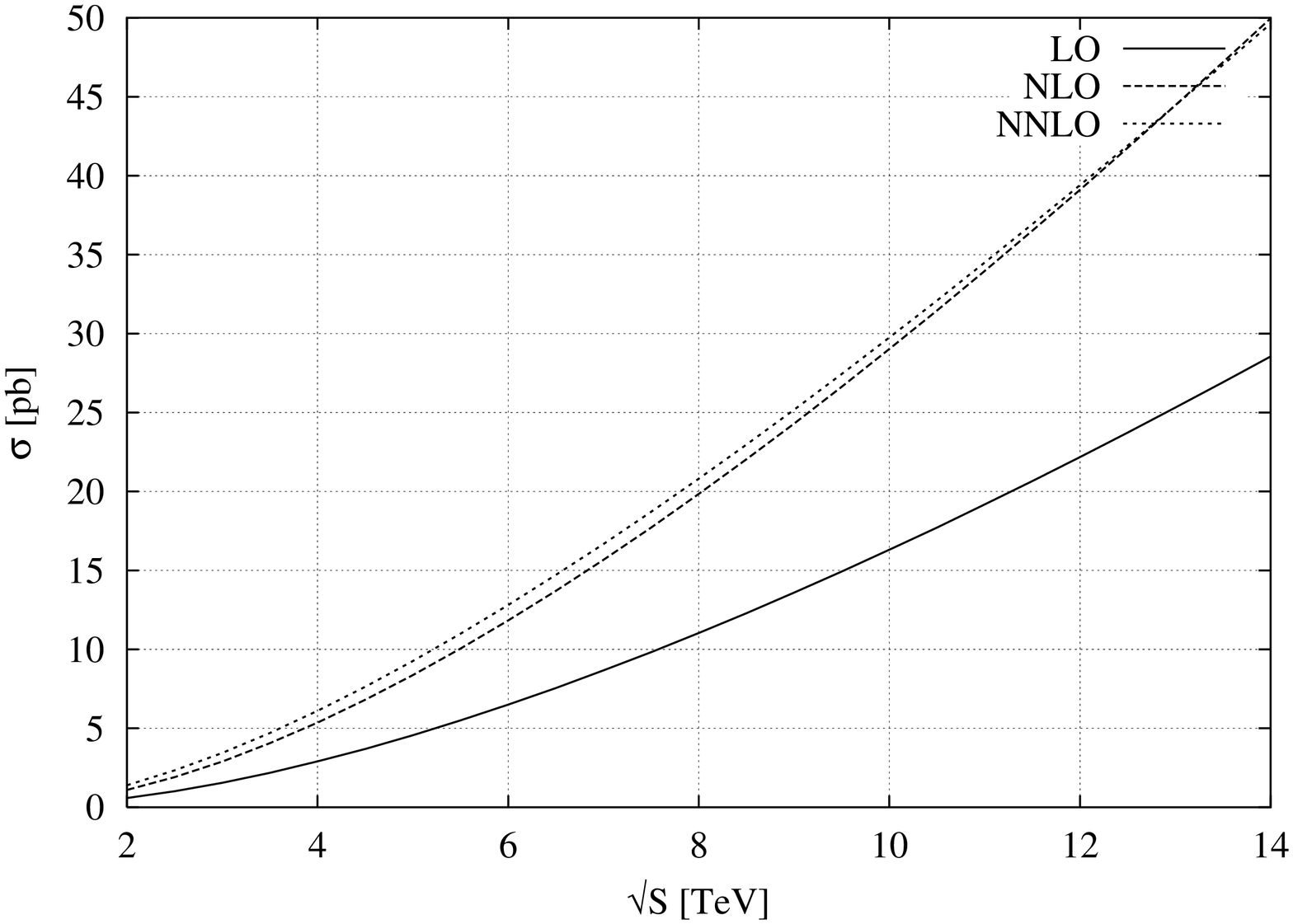}}
\subfigure[$C=1/2,\,\,k=1$]{\includegraphics[%
  width=6cm,
  angle=-90]{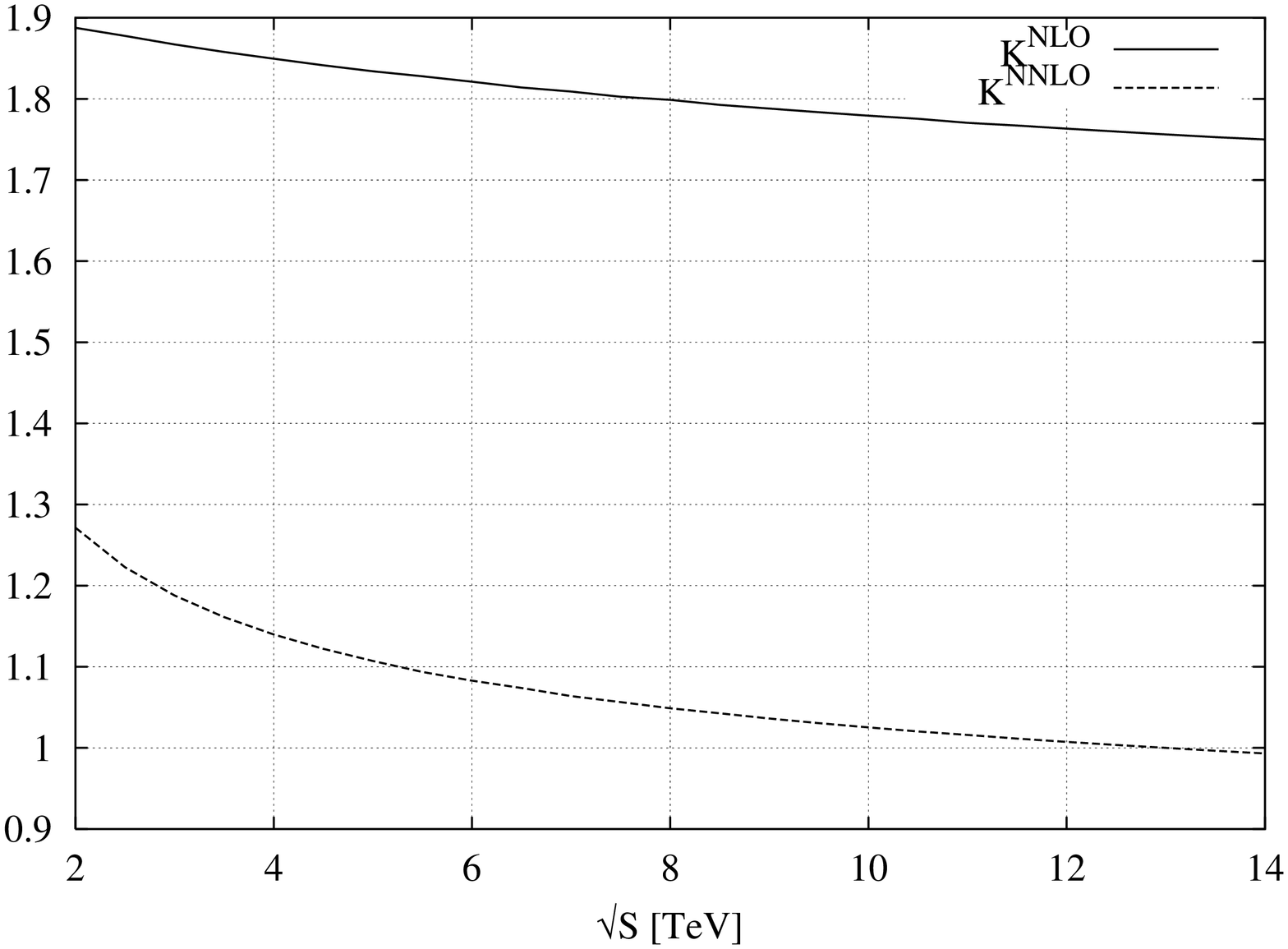}}
\caption{Cross sections and $K$-factors for scalar Higgs
production at LHC as a function of $\sqrt{S}$ with $\mu_F=C m_H$,
with $\mu_F^2=k\mu_R^2$ and $m_H=114$ GeV. MRST inputs have been used.}
\label{ener1}
\end{figure}

\begin{figure}
\subfigure[$C=1,\,\,k=2$]{\includegraphics[%
  width=6cm,
  angle=-90]{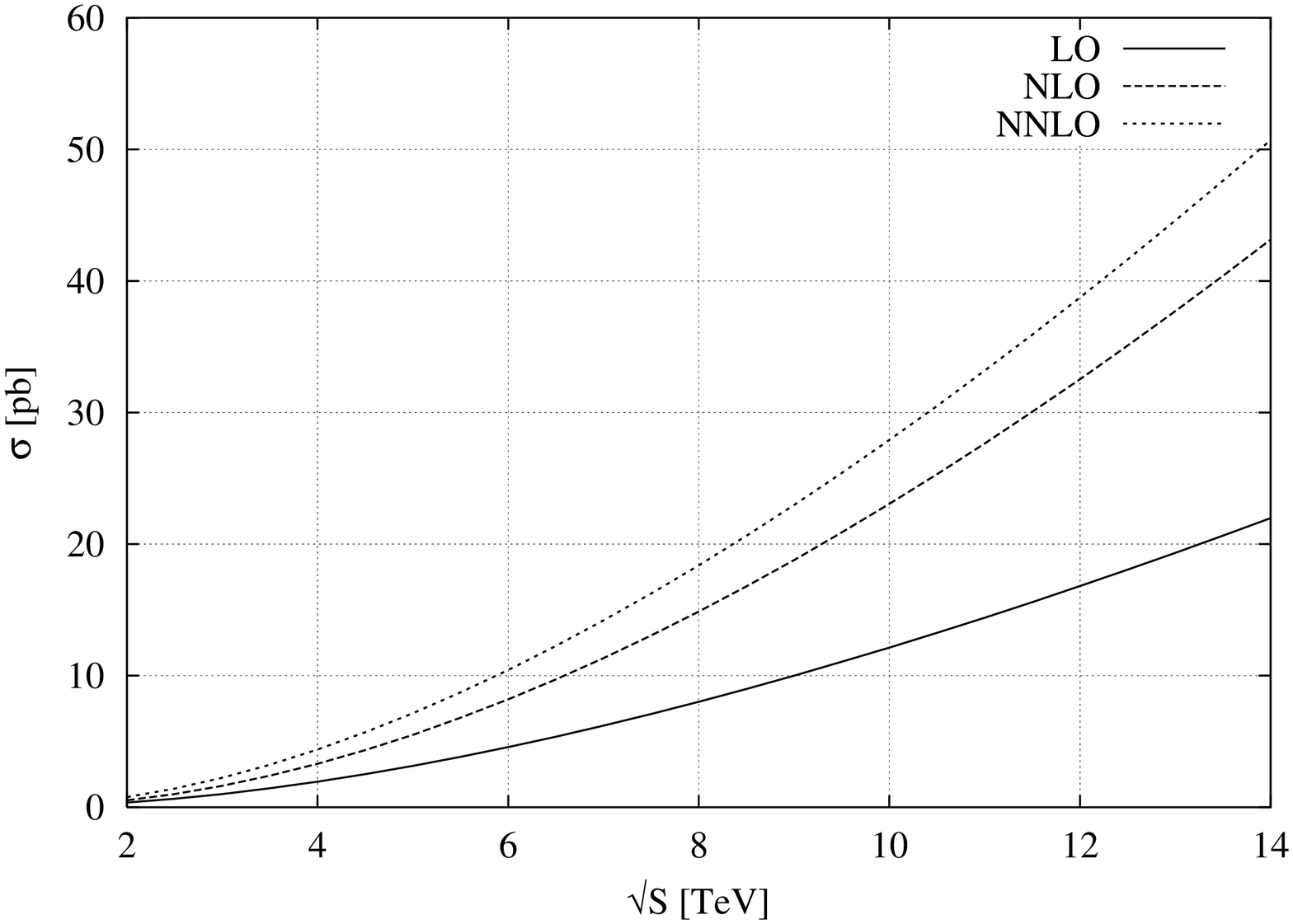}}
\subfigure[$C=1,\,\,k=2$]{\includegraphics[%
  width=6cm,
  angle=-90]{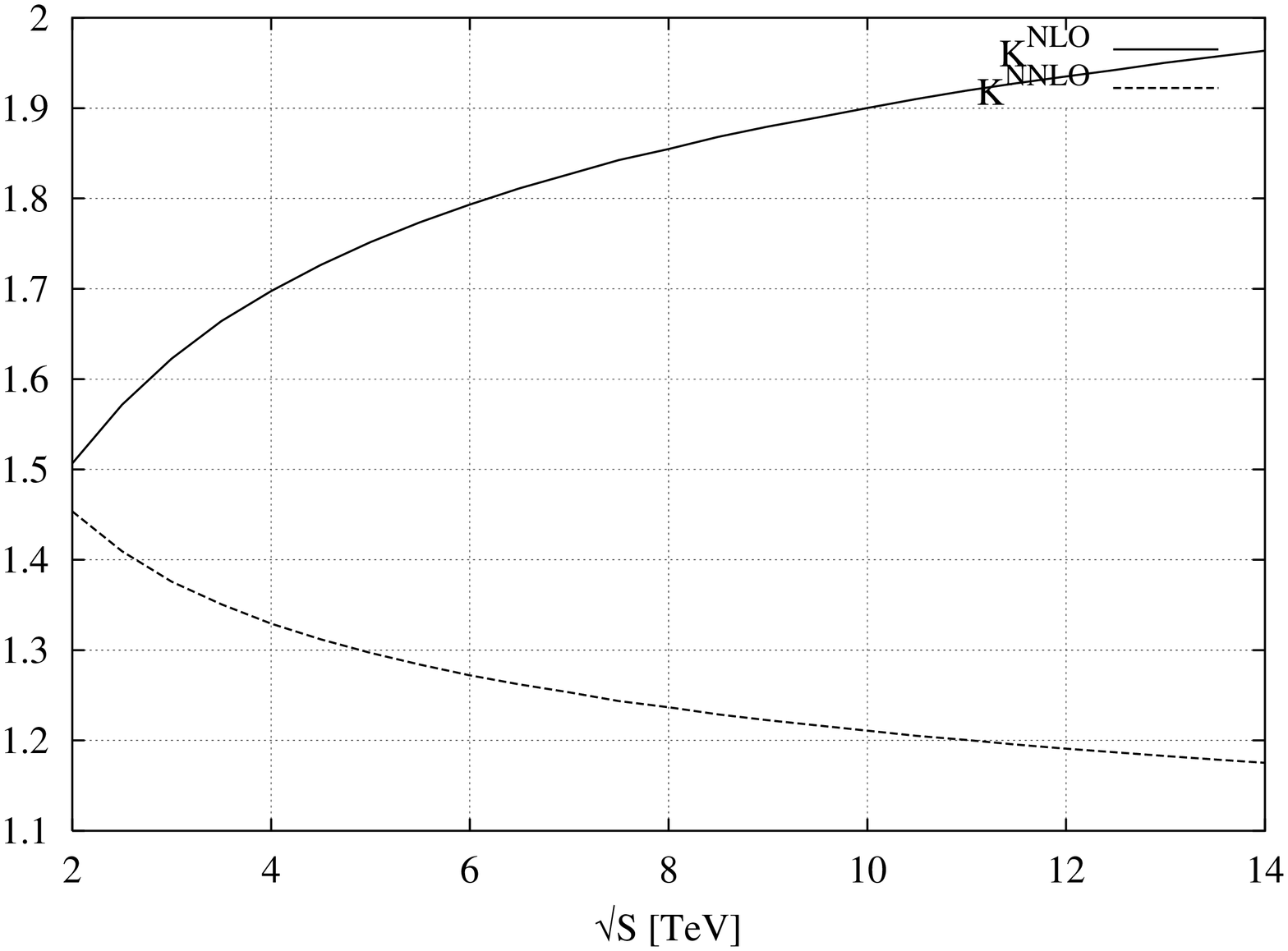}}
\subfigure[$C=2,\,\,k=2$]{\includegraphics[%
  width=6cm,
  angle=-90]{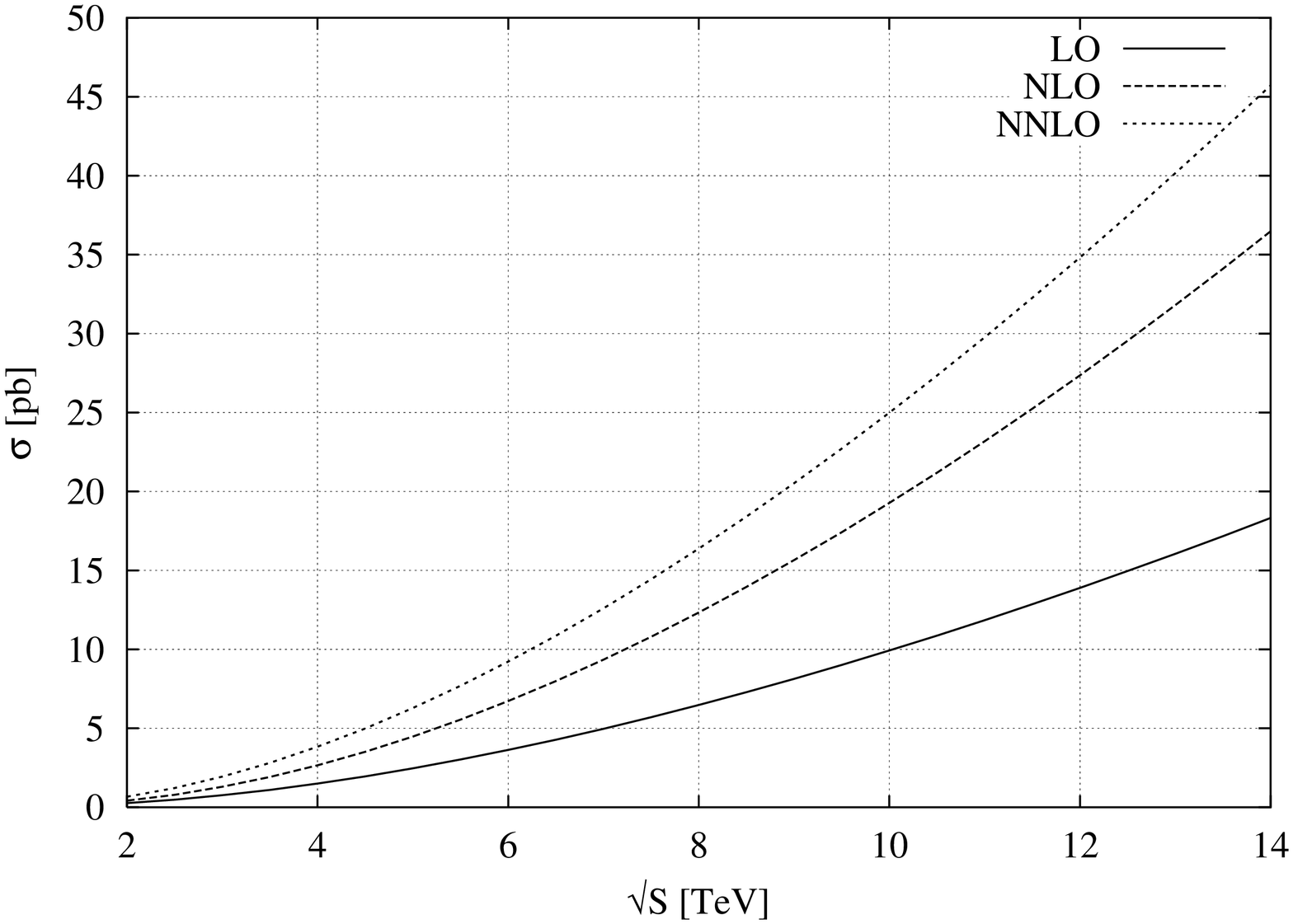}}
\subfigure[$C=2,\,\,k=2$]{\includegraphics[%
  width=6cm,
  angle=-90]{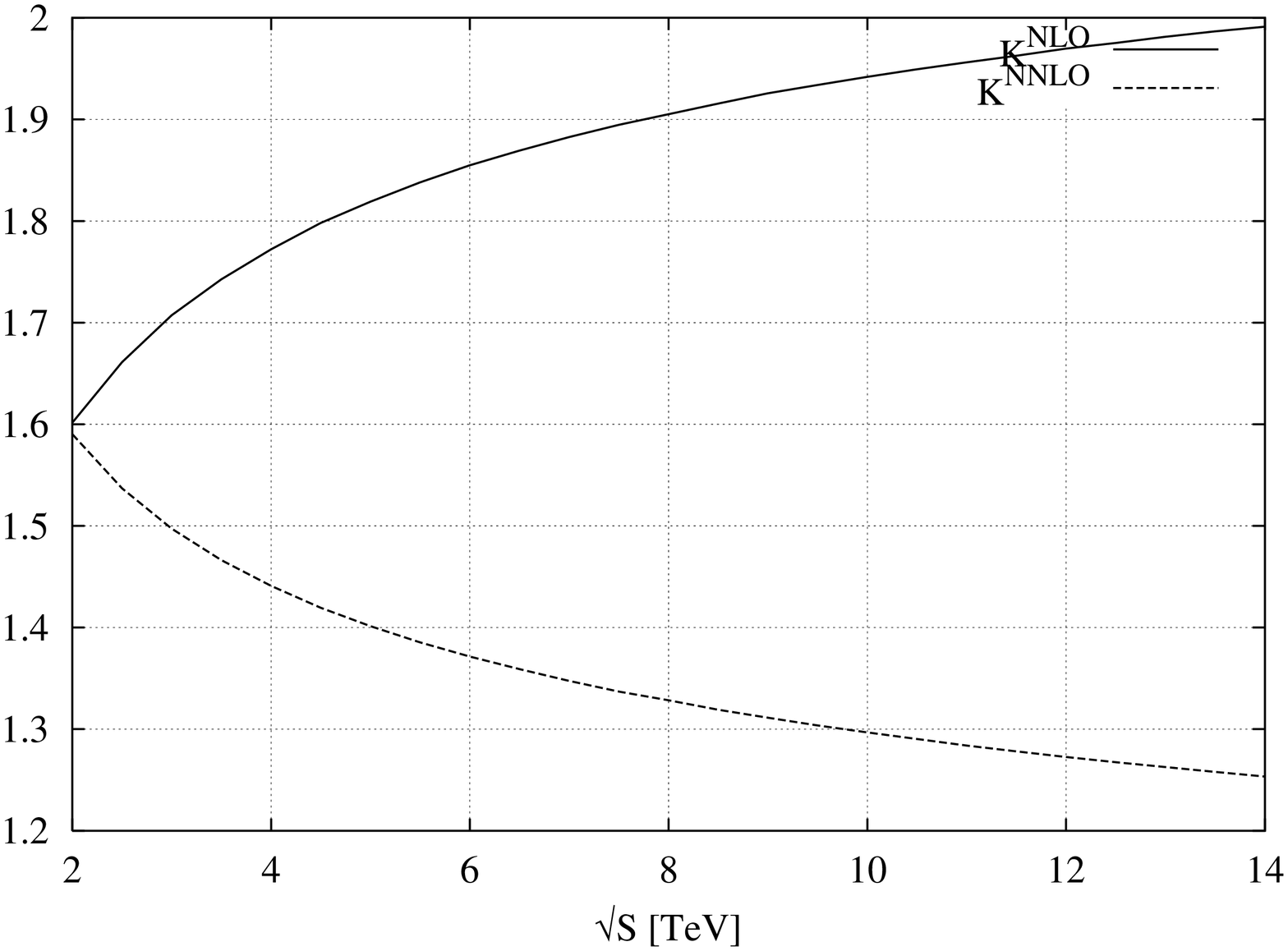}}
\subfigure[$C=1/2,\,\,k=2$]{\includegraphics[%
  width=6cm,
  angle=-90]{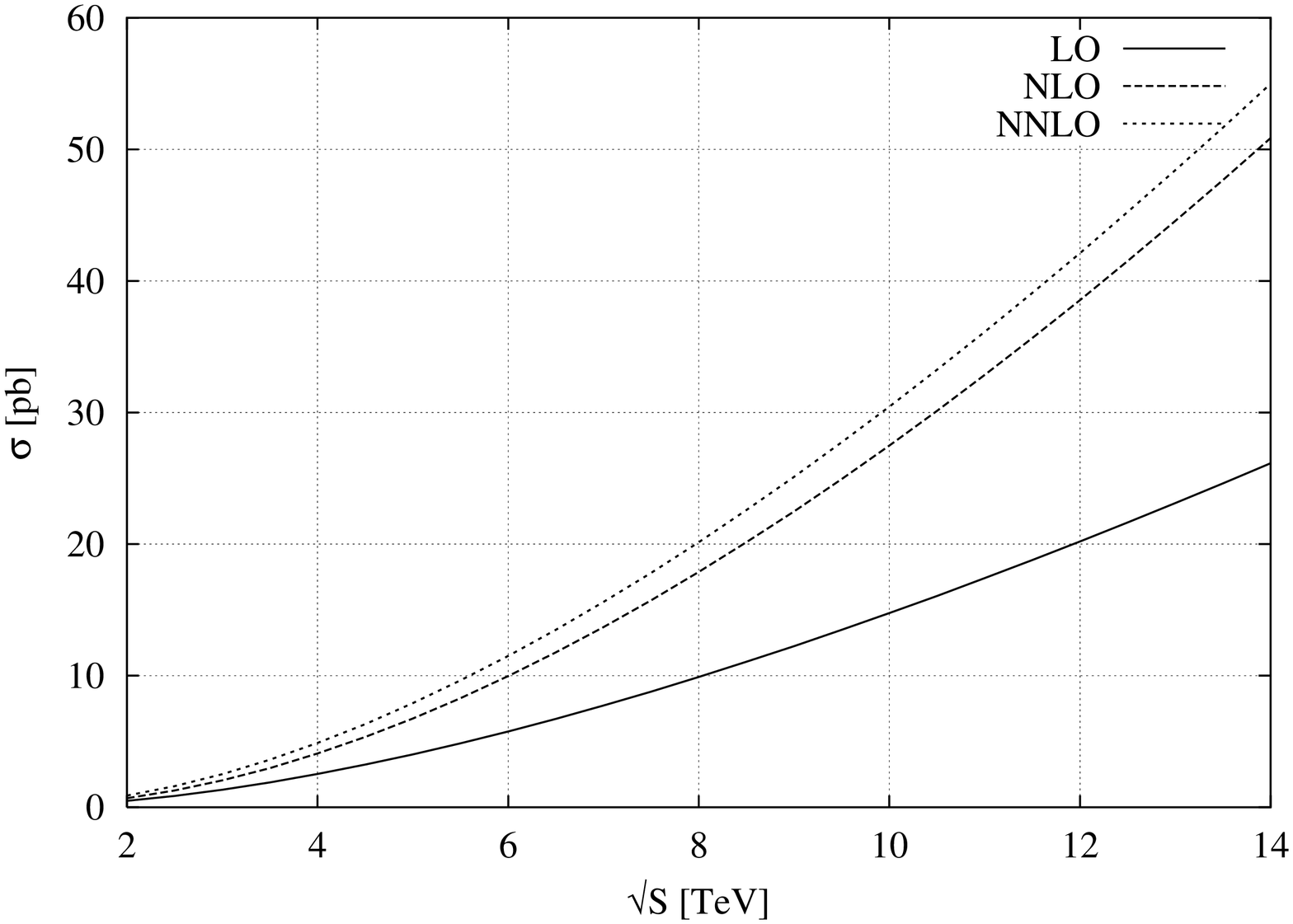}}
\subfigure[$C=1/2,\,\,k=2$]{\includegraphics[%
  width=6cm,
  angle=-90]{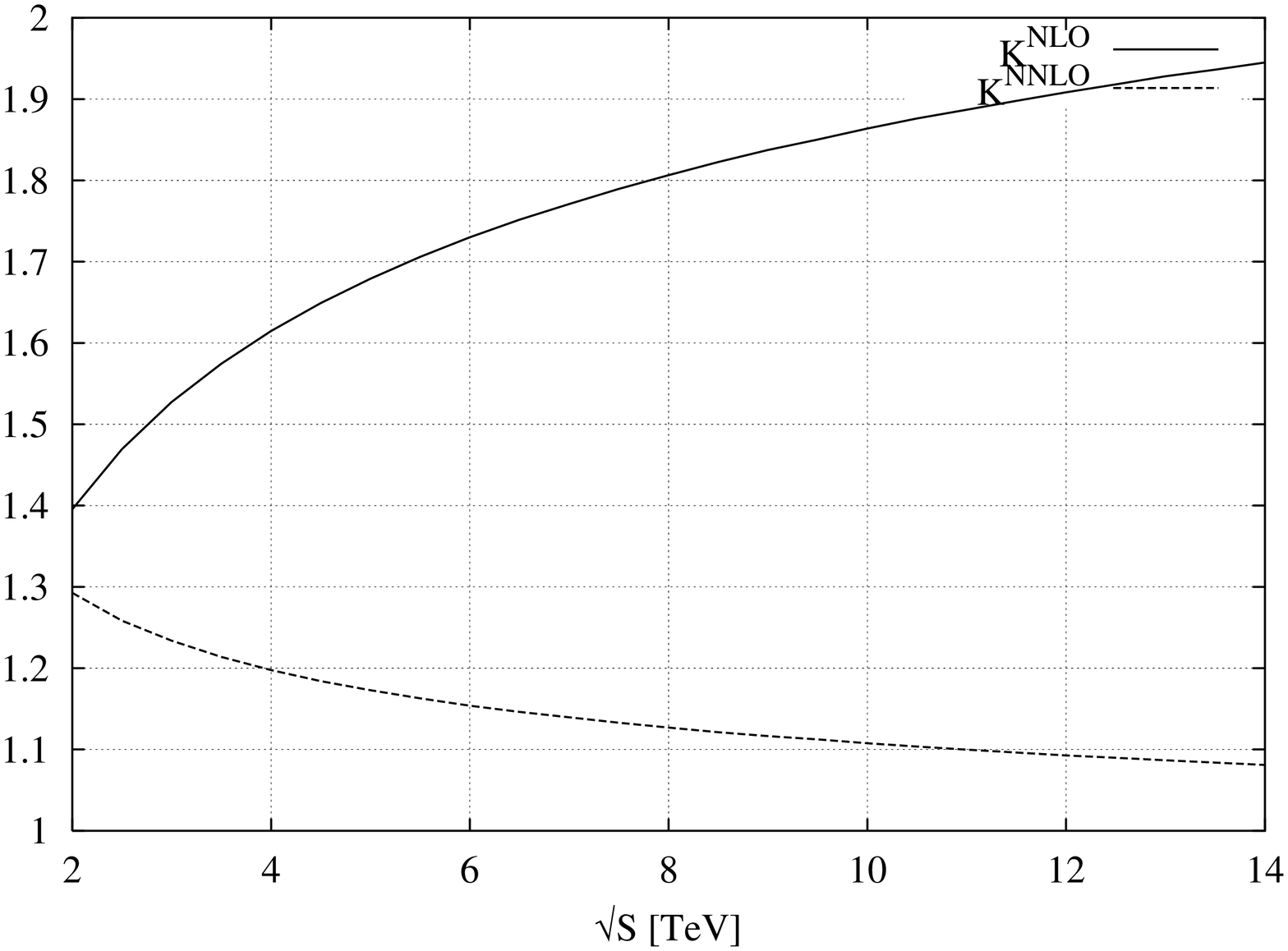}}
\caption{Cross sections and $K$-factors for the scalar Higgs
production at LHC as a function of $\sqrt{S}$ with $\mu_F=C m_H$,
with $\mu_F^2=k\mu_R^2$ and $m_H=114$ GeV. MRST inputs have been used.}
\label{ener2}
\end{figure}

\begin{figure}
\subfigure[$C=1,\,\,k=1/2$]{\includegraphics[%
  width=6cm,
  angle=-90]{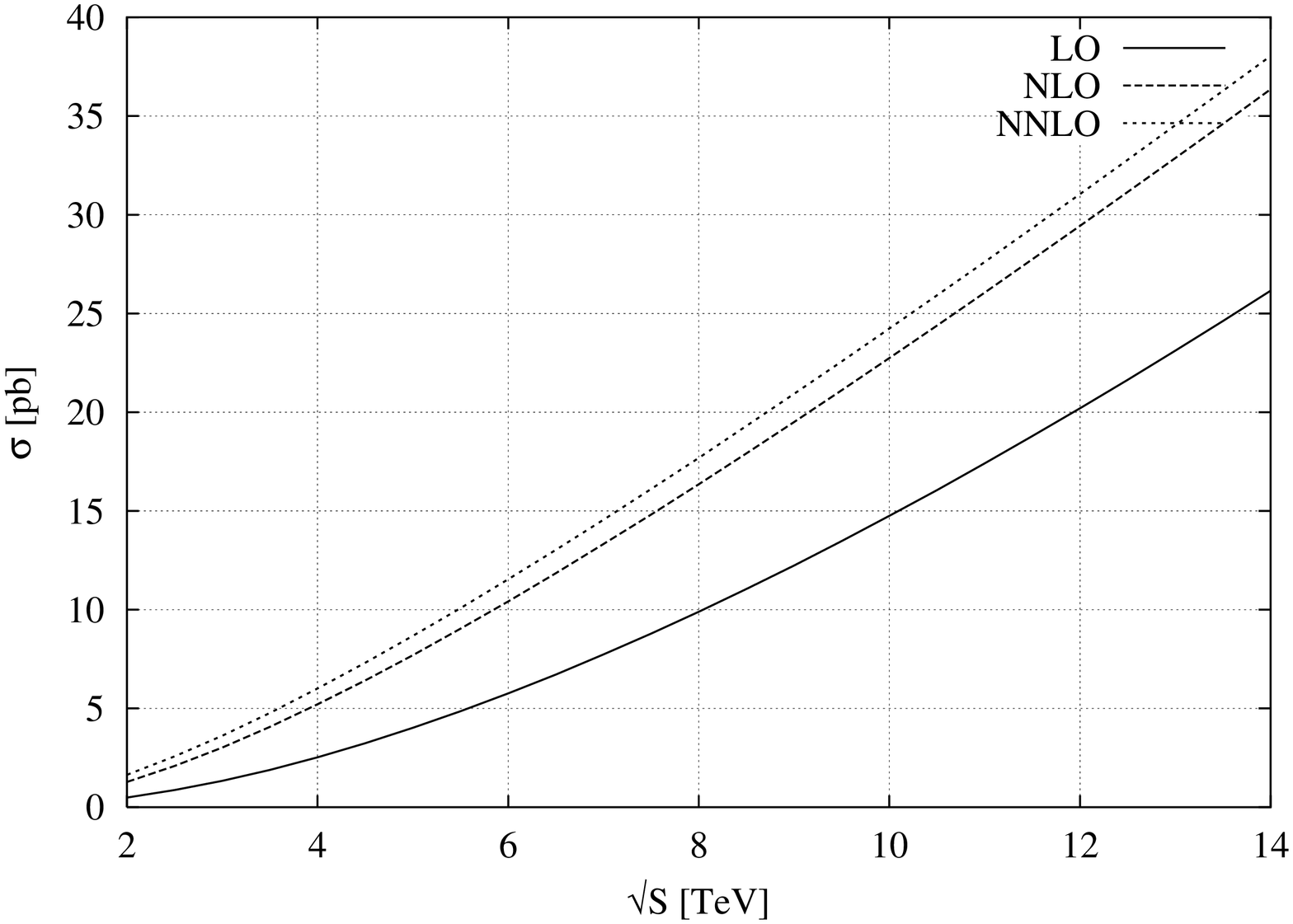}}
\subfigure[$C=1,\,\,k=1/2$]{\includegraphics[%
  width=6cm,
  angle=-90]{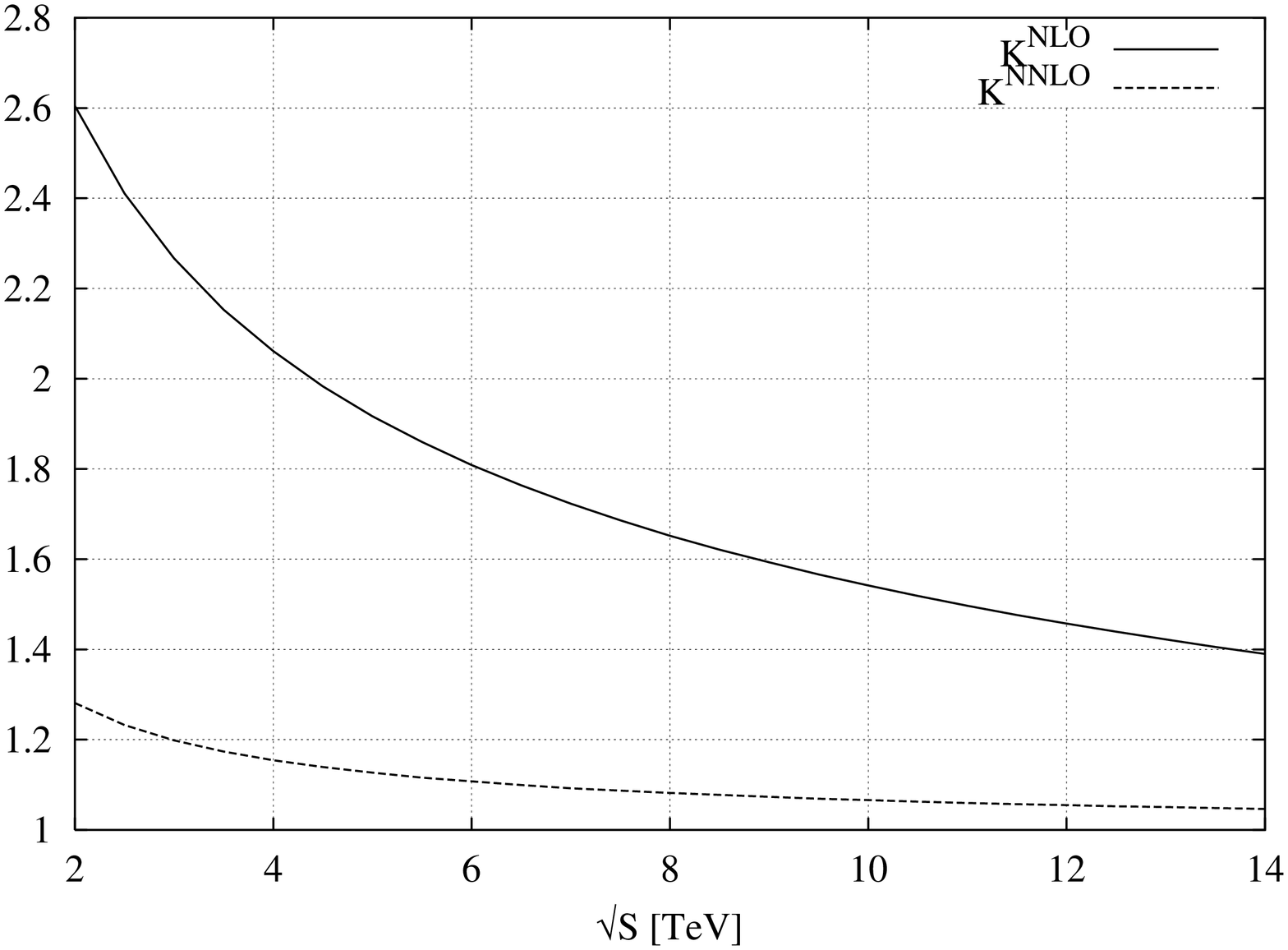}}
\subfigure[$C=2,\,\,k=1/2$]{\includegraphics[%
  width=6cm,
  angle=-90]{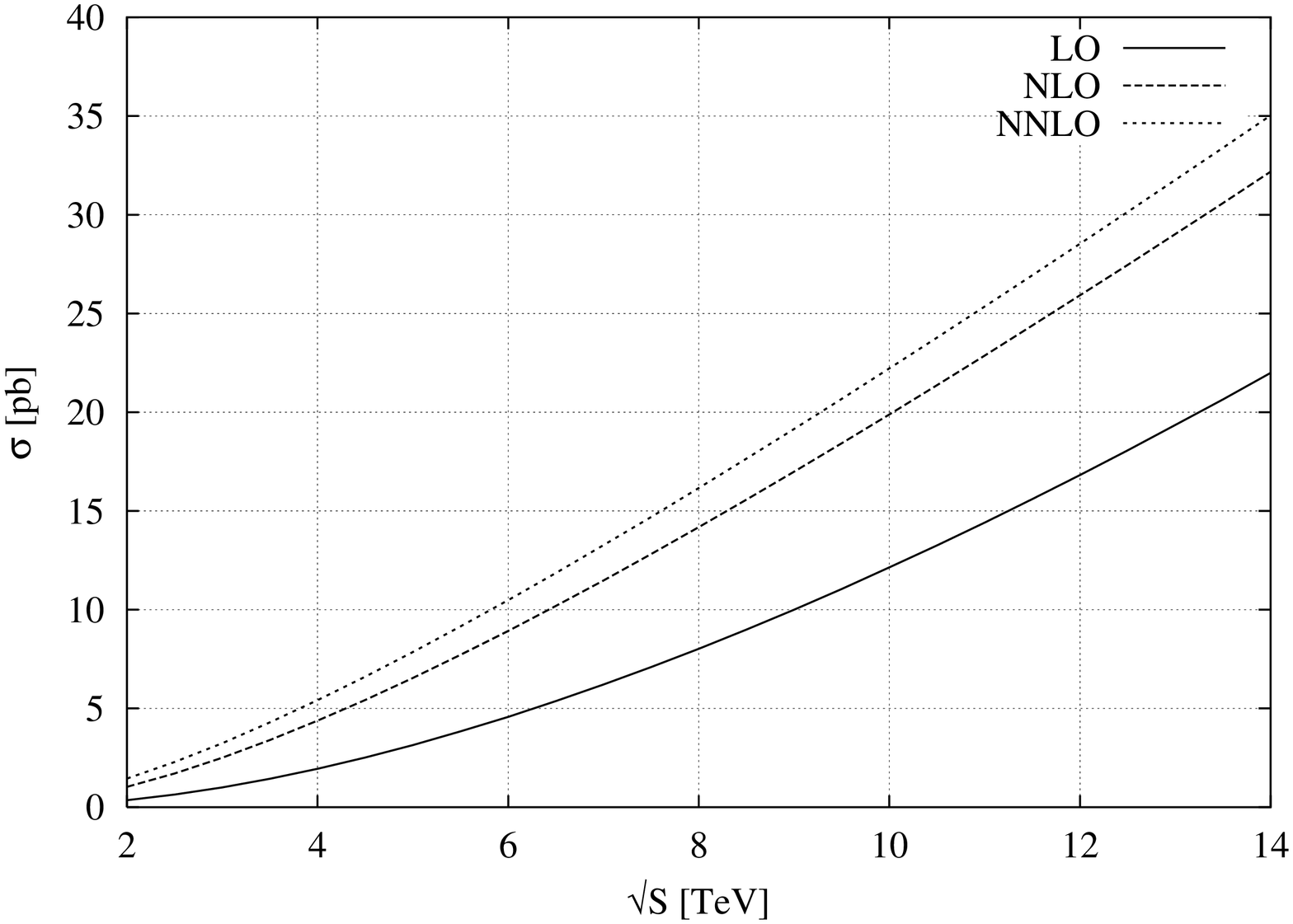}}
\subfigure[$C=2,\,\,k=1/2$]{\includegraphics[%
  width=6cm,
  angle=-90]{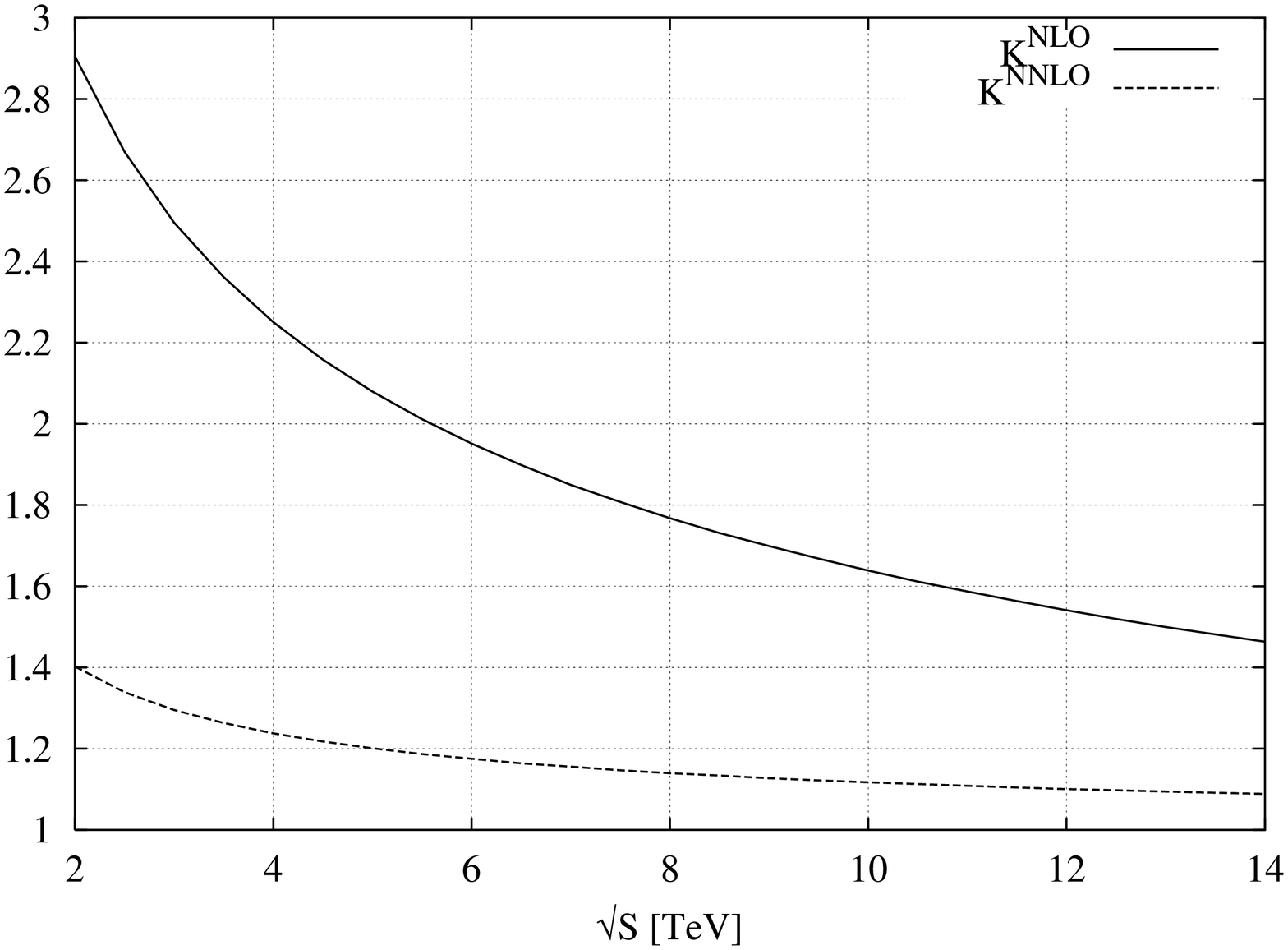}}
\subfigure[$C=1/2,\,\,k=1/2$]{\includegraphics[%
  width=6cm,
  angle=-90]{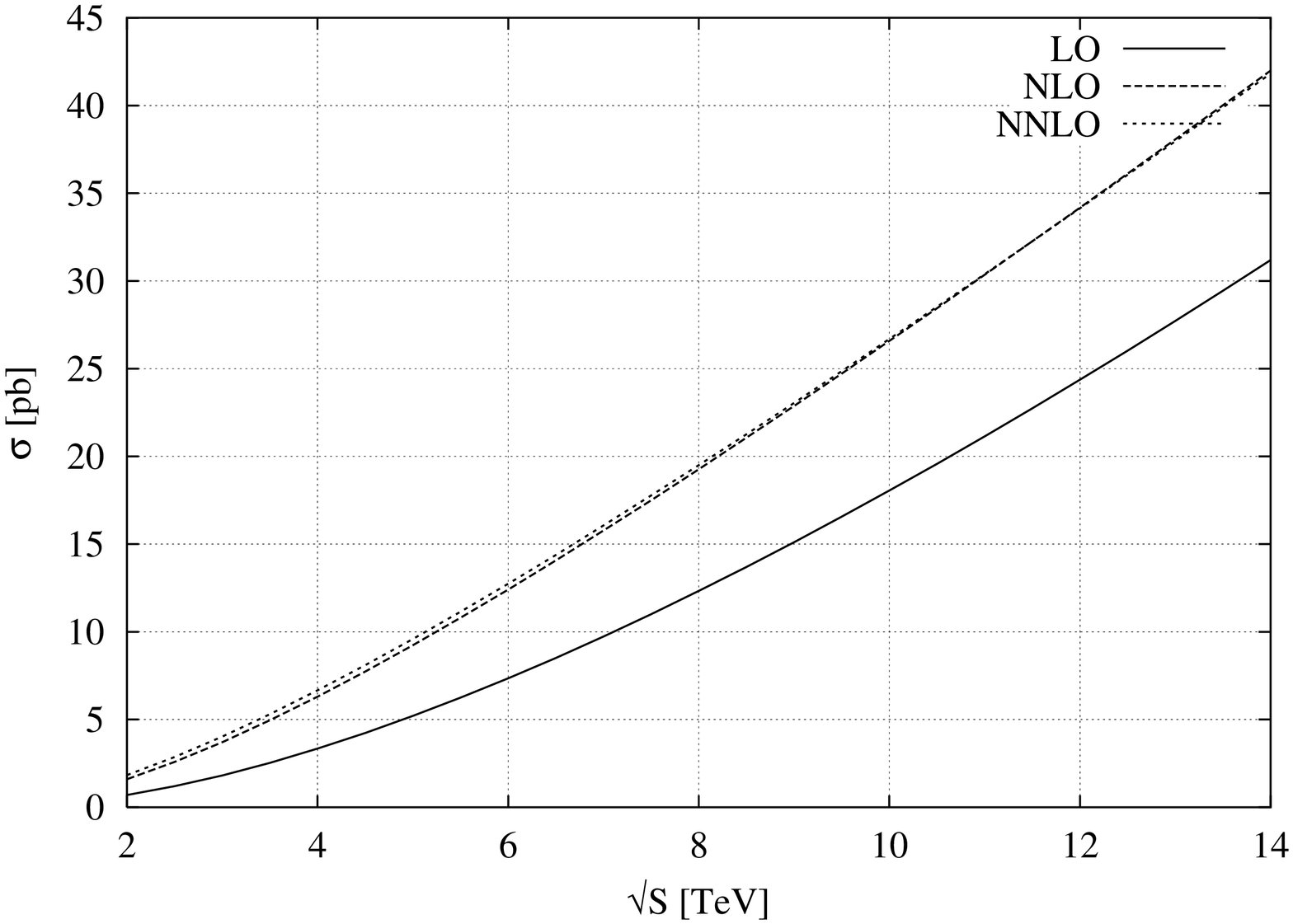}}
\subfigure[$C=1/2,\,\,k=1/2$]{\includegraphics[%
  width=6cm,
  angle=-90]{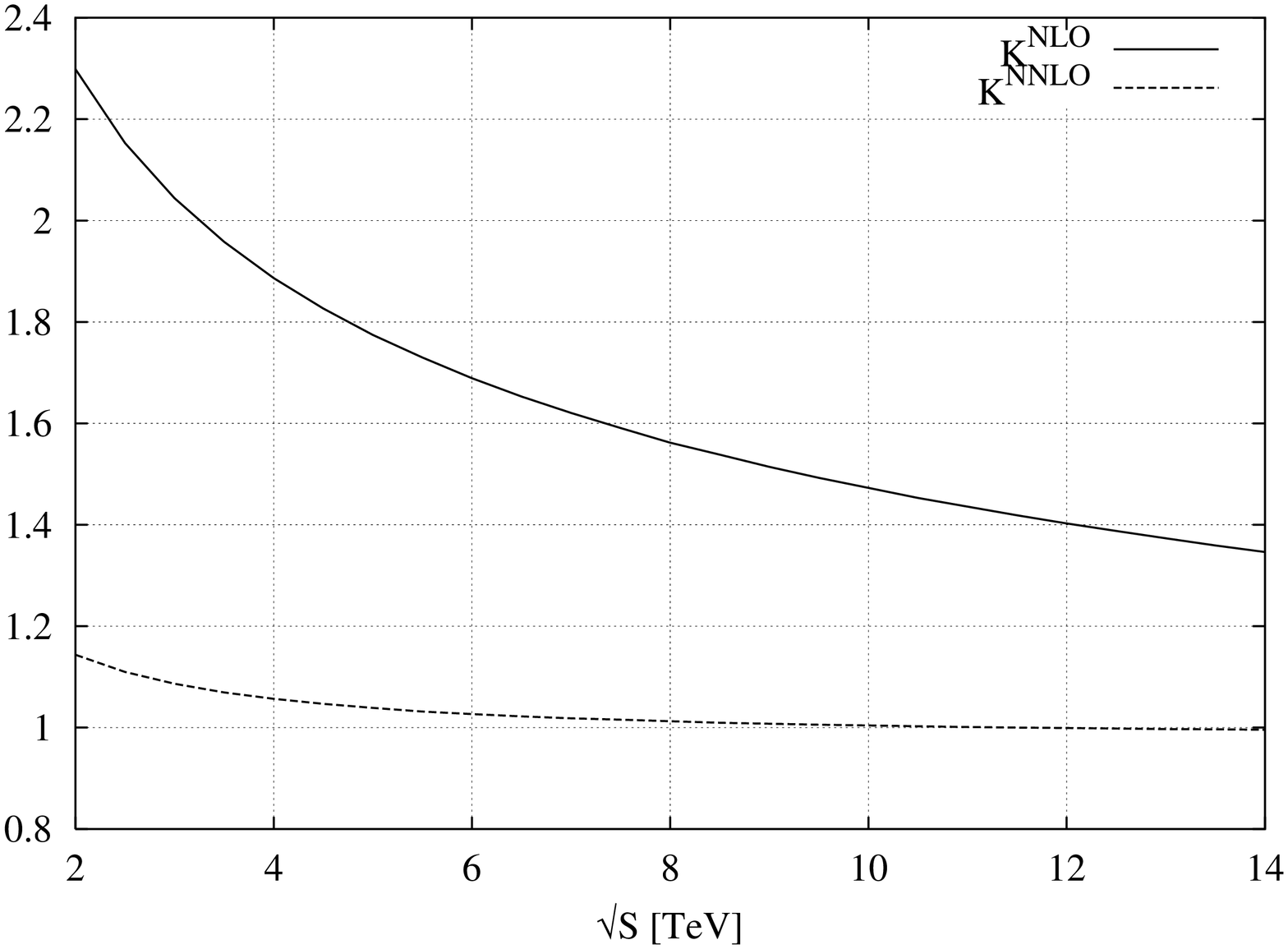}}
\caption{Cross sections and $K$-factors for the scalar Higgs
production at LHC as a function of $\sqrt{S}$ with $\mu_F=C m_H$,
with $\mu_F^2=k\mu_R^2$ and $m_H=114$ GeV. MRST inputs have been used.}
\label{ener3}
\end{figure}

The energy dependence of the NNLO predictions for the total cross sections 
and the corresponding K-factors at the LHC are shown in 
Figs.~\ref{ener1}-\ref{ener3}, where we have varied the ratio 
$C=\mu_F/m_H$ and $k=\mu_R^2/\mu_F^2$ in order to illustrate the variation 
of the results. The cross sections raises sharply with energy and the 
impact of the NNLO corrections is significant.  The K-factors, in most of 
the configurations chosen, vary between 1 and 2.2. We have chosen the 
MRST input. The behaviour of the K-factors is influenced significantly 
by the choice of the ratio $(k)$ between $\mu_R$ and $\mu_F$. 
In particular, in Figs.~\ref{ener2} the NNLO K-factors increase 
with $\sqrt{S}$ for $k=2$ , the center of mass energy, which is not 
found for other choices of scales. The case $k=1/2$ is close in behaviour 
to the coincident case $\mu_R^2=\mu_F^2$. The overall stability of the 
K-factors is clearly obtained with the choice $k=1$. We have finally 
included in a set of tables our numerical 
predictions in order to make them available to the experimental collaborations.

\section{Conclusions}
A study of the NNLO corrections to the cross section for Higgs production 
has been presented.  We have inplemented the exact three-loop splitting 
functions in our own parton evolution code. We used as 
initial conditions (at small scales) the boundary values of Martin, 
Roberts, Thorne and Stirling and of Alekhin.  This study shows that the 
impact of these corrections are important for the discovery of the 
Higgs and for a reconstruction of its mass.  The condition of stability 
of the perturbative expansion is also quite evident from these studies 
and suggests that the optimal choice to fix the arbitrary scales of the theory 
are near the coincidence point, with $\mu_R$ in the region of a plateau. 
The determination of the plateau has been performed by introducing in 
the perturbative expansion and in the evolution a new independent 
scale $(\mu_R)$, whose variation allows to accurately characterize 
the properties of the expansion in a direct way.

While this paper was being completed several authors have presented 
studies of the total Higgs cross section based on threshold 
resummation of soft or soft-plus-virtual logarithms, see \cite{mv}, \cite{lm},
\cite{ij}. Our work is based on exact NNLO partonic cross sections.
Another relevant paper which recently appeared is \cite{amp}.

\centerline{\bf Acknowledgements}
The work of A.C., C.C. and M.G. is partly supported by INFN and by MIUR. 
Numerical studies have been performed using the INFN-LE computational cluster. 
The work of J.S. was supported in part by the National Science Foundation 
grant PHY-0354776.


\begin{figure}
\subfigure[$m_H=110$GeV]{\includegraphics[%
  width=12cm,
  angle=-90]{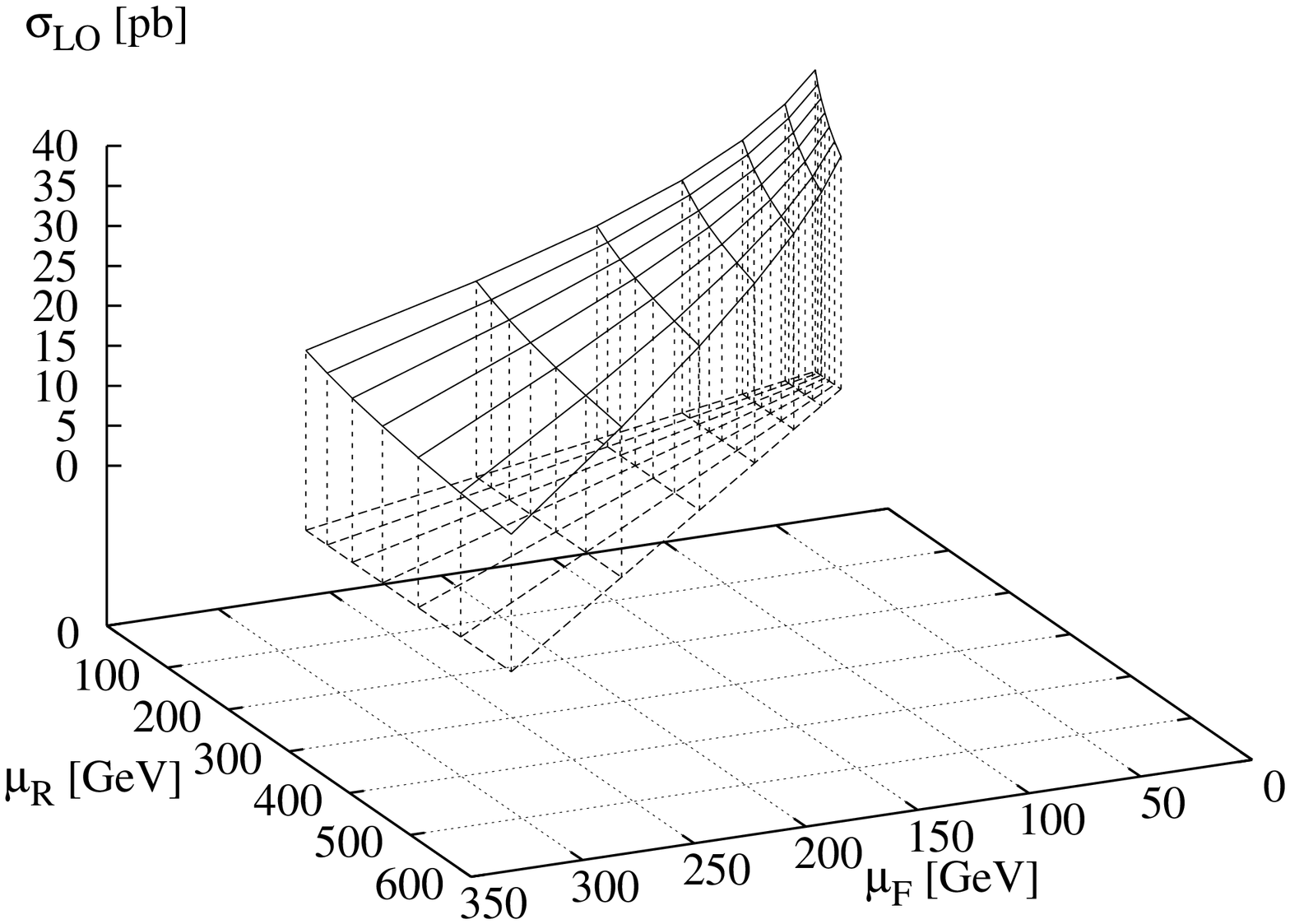}}
\caption{Three-dimensional graphs for the LO cross sections for scalar Higgs
production at LHC, as a function of $\mu_R$ and with $\mu_F$ with
a fixed value of $m_H$}
\label{3D1}
\end{figure}
\begin{figure}
\subfigure[$m_H=110$GeV]{\includegraphics[%
  width=12cm,
  angle=-90]{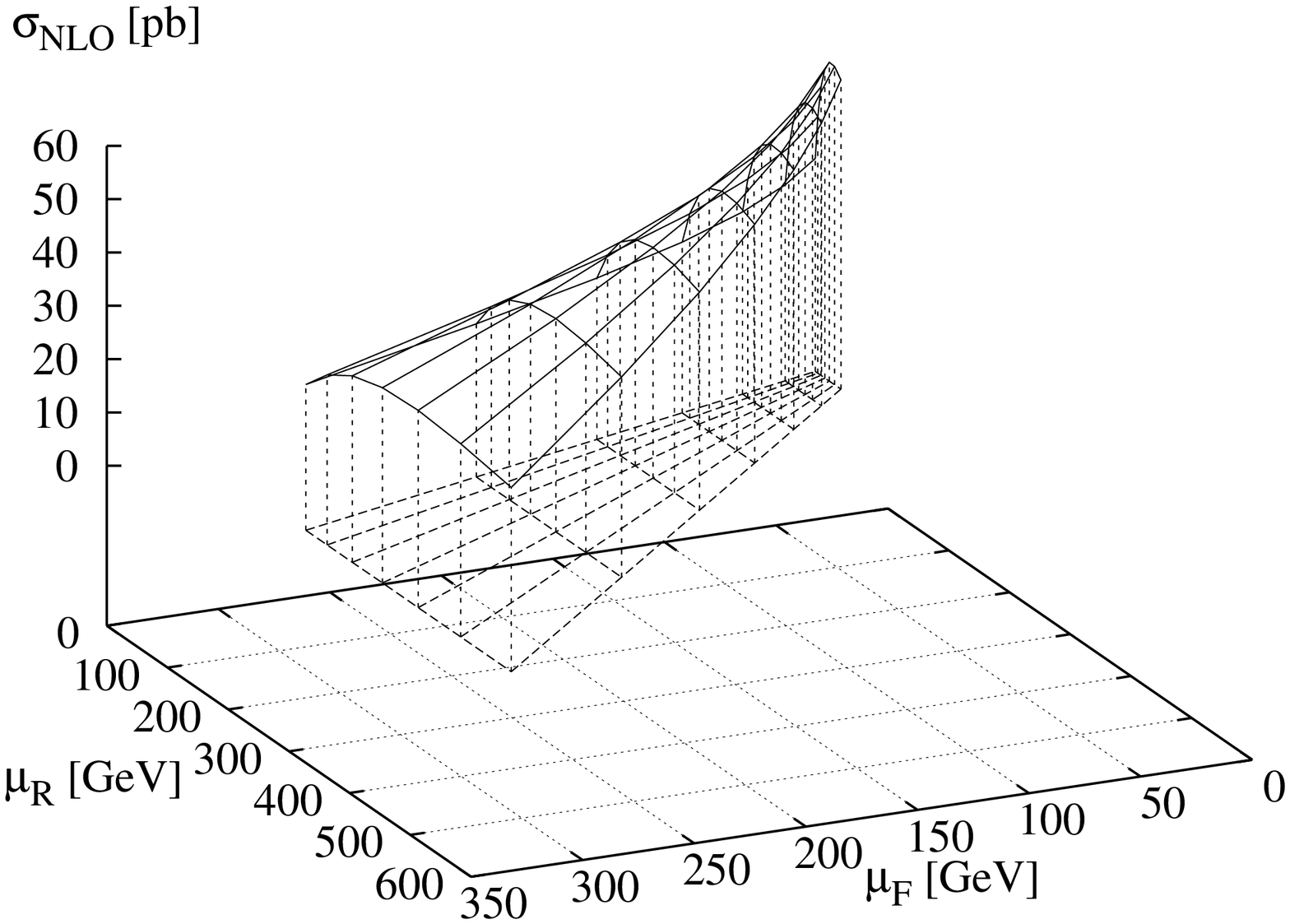}}
\caption{Three-dimensional graphs for the NLO cross sections for scalar Higgs
production at LHC, as a function of $\mu_R$ and with $\mu_F$ with
a fixed value of $m_H$}
\label{3D2}
\end{figure}
\begin{figure}
\subfigure[$m_H=110$GeV]{\includegraphics[%
  width=12cm,
  angle=-90]{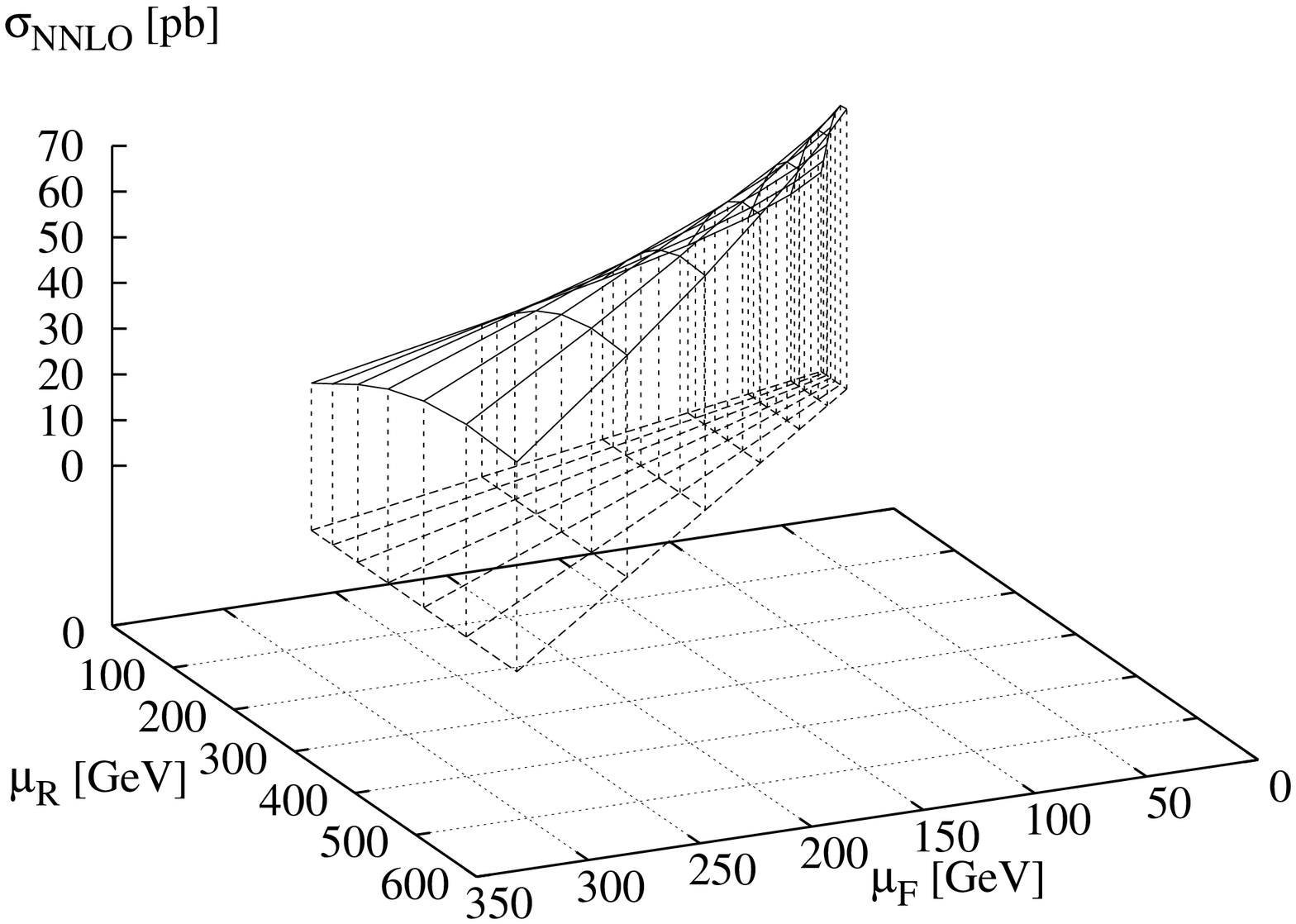}}
\caption{Three-dimensional graphs for the NNLO cross sections for scalar Higgs
production at LHC, as a function of $\mu_R$ and with $\mu_F$ with
a fixed value of $m_H$}
\label{3D3}
\end{figure}
\begin{figure}
\subfigure[$m_H=110$GeV]{\includegraphics[%
  width=12cm,
  angle=-90]{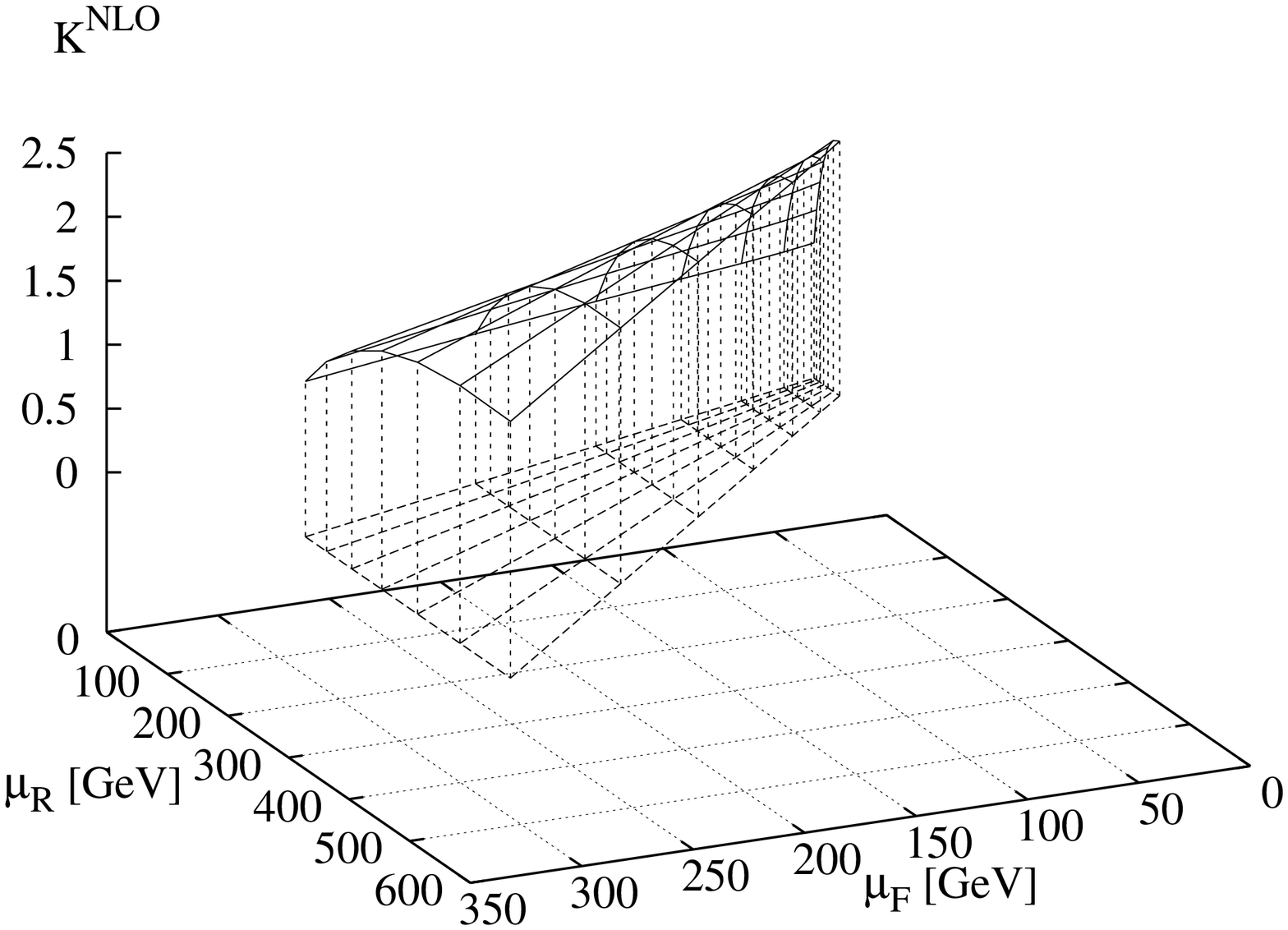}}
\caption{Three-dimensional graphs for the $K$-factor
$\sigma_{NLO}/\sigma_{LO}$ for scalar Higgs production at LHC,
as a function of $\mu_R$ and $\mu_F$ and at a
fixed value of $m_H$.}
\label{3D4}
\end{figure}
\begin{figure}
\subfigure[$m_H=110$GeV]{\includegraphics[%
  width=12cm,
  angle=-90]{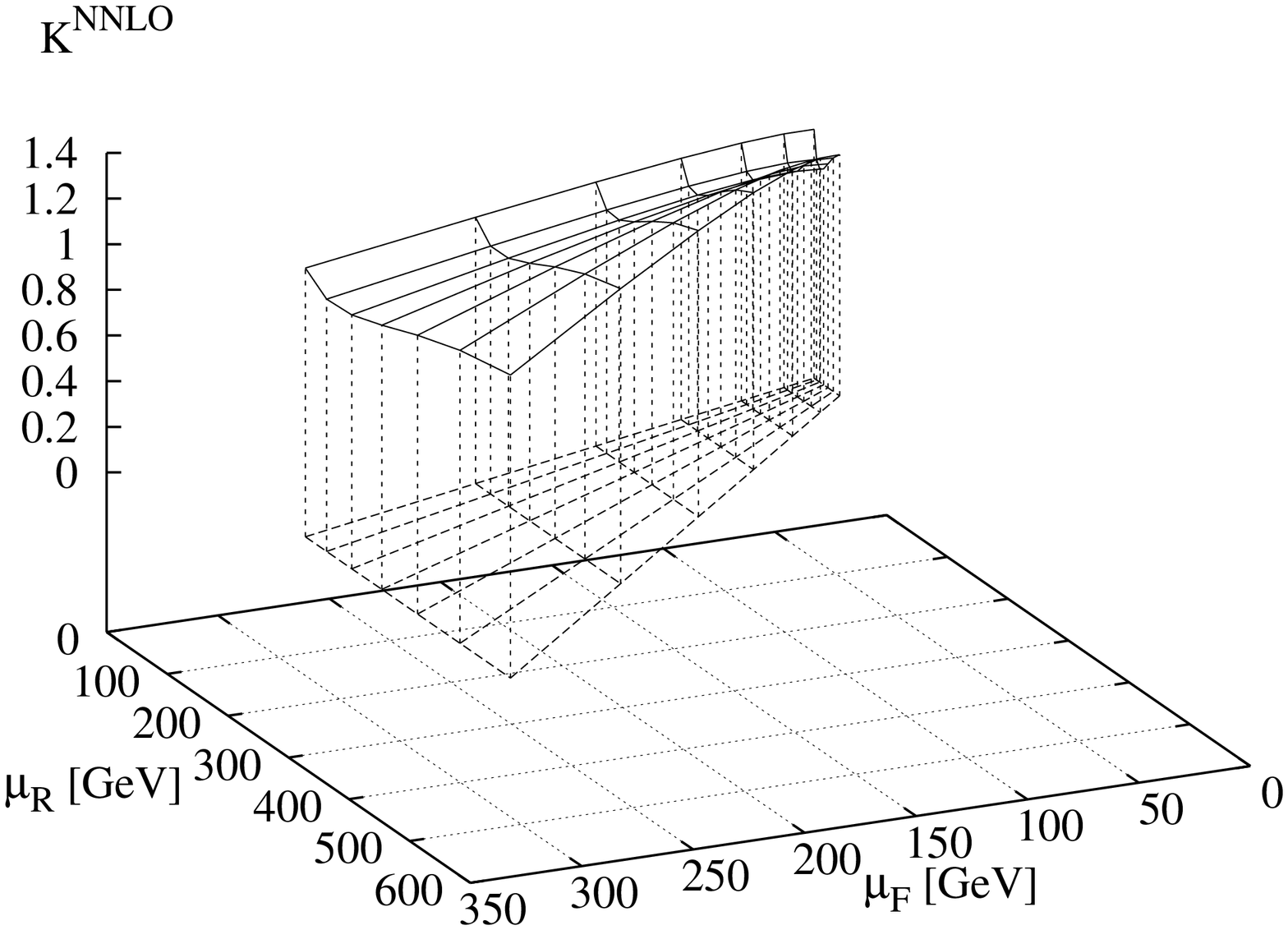}}
\caption{Three-dimensional graphs for the $K$-factor
$\sigma_{NNLO}/\sigma_{NLO}$ for scalar Higgs production at LHC,
as a function of $\mu_R$ and $\mu_F$ and at a
fixed value of $m_H$.}
\label{3D5}
\end{figure}

\newpage


\begin{table}
\begin{center}
\begin{tabular}{|c|ccccc|}
\hline
$\sqrt{S}$&
$\sigma_{LO}$&
$\sigma_{NLO}$&
$\sigma_{NNLO}$&
$K^{NLO}$&
$K^{NNLO}$\tabularnewline
\hline
$2.0$&
$0.4155$&
$0.8670$&
$1.242$&
$2.087$&
$1.433$\tabularnewline
$2.5$&
$0.7410$&
$1.521$&
$2.084$&
$2.053$&
$1.370$\tabularnewline
$3.0$&
$1.153$&
$2.335$&
$3.093$&
$2.025$&
$1.325$\tabularnewline
$3.5$&
$1.645$&
$3.292$&
$4.248$&
$2.001$&
$1.290$\tabularnewline
$4.0$&
$2.212$&
$4.380$&
$5.529$&
$1.980$&
$1.262$\tabularnewline
$4.5$&
$2.847$&
$5.587$&
$6.924$&
$1.962$&
$1.239$\tabularnewline
$5.0$&
$3.547$&
$6.903$&
$8.419$&
$1.946$&
$1.220$\tabularnewline
$5.5$&
$4.308$&
$8.318$&
$10.01$&
$1.931$&
$1.203$\tabularnewline
$6.0$&
$5.125$&
$9.826$&
$11.68$&
$1.917$&
$1.189$\tabularnewline
$6.5$&
$5.995$&
$11.42$&
$13.42$&
$1.905$&
$1.175$\tabularnewline
$7.0$&
$6.916$&
$13.09$&
$15.23$&
$1.893$&
$1.163$\tabularnewline
$7.5$&
$7.885$&
$14.84$&
$17.11$&
$1.882$&
$1.153$\tabularnewline
$8.0$&
$8.899$&
$16.66$&
$19.05$&
$1.872$&
$1.143$\tabularnewline
$8.5$&
$9.956$&
$18.55$&
$21.04$&
$1.863$&
$1.134$\tabularnewline
$9.0$&
$11.05$&
$20.49$&
$23.09$&
$1.854$&
$1.127$\tabularnewline
$9.5$&
$12.19$&
$22.50$&
$25.18$&
$1.846$&
$1.119$\tabularnewline
$10.0$&
$13.37$&
$24.56$&
$27.31$&
$1.837$&
$1.112$\tabularnewline
$10.5$&
$14.58$&
$26.68$&
$29.49$&
$1.830$&
$1.105$\tabularnewline
$11.0$&
$15.83$&
$28.85$&
$31.71$&
$1.822$&
$1.099$\tabularnewline
$11.5$&
$17.11$&
$31.06$&
$33.97$&
$1.815$&
$1.094$\tabularnewline
$12.0$&
$18.42$&
$33.32$&
$36.26$&
$1.809$&
$1.088$\tabularnewline
$12.5$&
$19.76$&
$35.62$&
$38.59$&
$1.803$&
$1.083$\tabularnewline
$13.0$&
$21.13$&
$37.97$&
$40.95$&
$1.797$&
$1.078$\tabularnewline
$13.5$&
$22.53$&
$40.36$&
$43.33$&
$1.791$&
$1.074$\tabularnewline
$14.0$&
$23.96$&
$42.78$&
$45.75$&
$1.785$&
$1.069$\tabularnewline
\hline
\end{tabular}
\end{center}
\caption{Values of the cross sections and $K$-factors for scalar Higgs
production at LHC as a function of $\sqrt{S}$ with $\mu_F=m_H$,
with $\mu_F^2=\mu_R^2$ and $m_H=114$ GeV. MRST inputs have been used.}
\end{table}


\begin{table}
\begin{center}
\begin{tabular}{|c|ccccc|}
\hline
$\sqrt{S}$&
$\sigma_{LO}$&
$\sigma_{NLO}$&
$\sigma_{NNLO}$&
$K^{NLO}$&
$K^{NNLO}$\tabularnewline
\hline
$2.0$&
$0.3029$&
$0.6871$&
$1.081$&
$2.268$&
$1.573$\tabularnewline
$2.5$&
$0.5508$&
$1.221$&
$1.830$&
$2.217$&
$1.499$\tabularnewline
$3.0$&
$0.8700$&
$1.893$&
$2.733$&
$2.176$&
$1.444$\tabularnewline
$3.5$&
$1.256$&
$2.690$&
$3.771$&
$2.142$&
$1.402$\tabularnewline
$4.0$&
$1.706$&
$3.602$&
$4.928$&
$2.111$&
$1.368$\tabularnewline
$4.5$&
$2.216$&
$4.619$&
$6.191$&
$2.084$&
$1.340$\tabularnewline
$5.0$&
$2.781$&
$5.733$&
$7.549$&
$2.061$&
$1.317$\tabularnewline
$5.5$&
$3.400$&
$6.937$&
$8.993$&
$2.040$&
$1.296$\tabularnewline
$6.0$&
$4.069$&
$8.225$&
$10.52$&
$2.021$&
$1.279$\tabularnewline
$6.5$&
$4.786$&
$9.590$&
$12.11$&
$2.004$&
$1.263$\tabularnewline
$7.0$&
$5.548$&
$11.03$&
$13.77$&
$1.988$&
$1.248$\tabularnewline
$7.5$&
$6.354$&
$12.53$&
$15.49$&
$1.972$&
$1.236$\tabularnewline
$8.0$&
$7.201$&
$14.11$&
$17.27$&
$1.959$&
$1.224$\tabularnewline
$8.5$&
$8.088$&
$15.74$&
$19.10$&
$1.946$&
$1.213$\tabularnewline
$9.0$&
$9.012$&
$17.43$&
$20.98$&
$1.934$&
$1.204$\tabularnewline
$9.5$&
$9.974$&
$19.17$&
$22.90$&
$1.922$&
$1.195$\tabularnewline
$10.0$&
$10.97$&
$20.97$&
$24.87$&
$1.912$&
$1.186$\tabularnewline
$10.5$&
$12.00$&
$22.82$&
$26.88$&
$1.902$&
$1.178$\tabularnewline
$11.0$&
$13.06$&
$24.71$&
$28.93$&
$1.892$&
$1.171$\tabularnewline
$11.5$&
$14.16$&
$26.65$&
$31.01$&
$1.882$&
$1.164$\tabularnewline
$12.0$&
$15.28$&
$28.63$&
$33.13$&
$1.874$&
$1.157$\tabularnewline
$12.5$&
$16.44$&
$30.65$&
$35.28$&
$1.864$&
$1.151$\tabularnewline
$13.0$&
$17.62$&
$32.71$&
$37.47$&
$1.856$&
$1.146$\tabularnewline
$13.5$&
$18.83$&
$34.81$&
$39.68$&
$1.849$&
$1.140$\tabularnewline
$14.0$&
$20.07$&
$36.95$&
$41.92$&
$1.841$&
$1.135$\tabularnewline
\hline
\end{tabular}
\end{center}
\caption{Values of the cross sections and $K$-factors for scalar Higgs
production at LHC as a function of $\sqrt{S}$ with $\mu_F=2 m_H$,
with $\mu_F^2=\mu_R^2$ and $m_H=114$ GeV. MRST inputs have been used.}
\end{table}

\begin{table}
\begin{center}
\begin{tabular}{|c|ccccc|}
\hline
$\sqrt{S}$&
$\sigma_{LO}$&
$\sigma_{NLO}$&
$\sigma_{NNLO}$&
$K^{NLO}$&
$K^{NNLO}$\tabularnewline
\hline
$2.0$&
$0.5817$&
$1.098$&
$1.396$&
$1.888$&
$1.271$\tabularnewline
$2.5$&
$1.014$&
$1.904$&
$2.328$&
$1.878$&
$1.223$\tabularnewline
$3.0$&
$1.551$&
$2.896$&
$3.440$&
$1.867$&
$1.188$\tabularnewline
$3.5$&
$2.182$&
$4.054$&
$4.707$&
$1.858$&
$1.161$\tabularnewline
$4.0$&
$2.899$&
$5.362$&
$6.111$&
$1.850$&
$1.140$\tabularnewline
$4.5$&
$3.695$&
$6.804$&
$7.635$&
$1.841$&
$1.122$\tabularnewline
$5.0$&
$4.563$&
$8.369$&
$9.266$&
$1.834$&
$1.107$\tabularnewline
$5.5$&
$5.498$&
$10.05$&
$10.99$&
$1.828$&
$1.094$\tabularnewline
$6.0$&
$6.496$&
$11.83$&
$12.81$&
$1.821$&
$1.083$\tabularnewline
$6.5$&
$7.552$&
$13.70$&
$14.71$&
$1.814$&
$1.074$\tabularnewline
$7.0$&
$8.662$&
$15.67$&
$16.67$&
$1.809$&
$1.064$\tabularnewline
$7.5$&
$9.824$&
$17.71$&
$18.71$&
$1.803$&
$1.056$\tabularnewline
$8.0$&
$11.03$&
$19.84$&
$20.81$&
$1.799$&
$1.049$\tabularnewline
$8.5$&
$12.29$&
$22.03$&
$22.97$&
$1.793$&
$1.043$\tabularnewline
$9.0$&
$13.59$&
$24.30$&
$25.18$&
$1.788$&
$1.036$\tabularnewline
$9.5$&
$14.93$&
$26.63$&
$27.44$&
$1.784$&
$1.030$\tabularnewline
$10.0$&
$16.31$&
$29.02$&
$29.75$&
$1.779$&
$1.025$\tabularnewline
$10.5$&
$17.72$&
$31.46$&
$32.10$&
$1.775$&
$1.020$\tabularnewline
$11.0$&
$19.18$&
$33.96$&
$34.50$&
$1.771$&
$1.016$\tabularnewline
$11.5$&
$20.66$&
$36.51$&
$36.93$&
$1.767$&
$1.012$\tabularnewline
$12.0$&
$22.18$&
$39.11$&
$39.40$&
$1.763$&
$1.007$\tabularnewline
$12.5$&
$23.73$&
$41.76$&
$41.91$&
$1.760$&
$1.004$\tabularnewline
$13.0$&
$25.31$&
$44.45$&
$44.45$&
$1.756$&
$1.000$\tabularnewline
$13.5$&
$26.92$&
$47.19$&
$47.02$&
$1.753$&
$0.9964$\tabularnewline
$14.0$&
$28.55$&
$49.96$&
$49.62$&
$1.750$&
$0.9932$\tabularnewline
\hline
\end{tabular}
\end{center}
\caption{Values of the cross sections and $K$-factors for scalar Higgs
production at LHC as a function of $\sqrt{S}$ with $\mu_F=(1/2) m_H$,
with $\mu_F^2=\mu_R^2$ and $m_H=114$ GeV. MRST inputs have been used.}
\end{table}

\begin{table}
\begin{center}
\begin{tabular}{|c|ccccc|}
\hline
$\sqrt{S}$&
$\sigma_{LO}$&
$\sigma_{NLO}$&
$\sigma_{NNLO}$&
$K^{NLO}$&
$K^{NNLO}$\tabularnewline
\hline
$2.0$&
$0.3546$&
$0.5343$&
$0.7766$&
$1.507$&
$1.453$\tabularnewline
$2.5$&
$0.6388$&
$1.004$&
$1.415$&
$1.572$&
$1.409$\tabularnewline
$3.0$&
$1.002$&
$1.626$&
$2.237$&
$1.623$&
$1.376$\tabularnewline
$3.5$&
$1.438$&
$2.393$&
$3.232$&
$1.664$&
$1.351$\tabularnewline
$4.0$&
$1.944$&
$3.300$&
$4.387$&
$1.698$&
$1.329$\tabularnewline
$4.5$&
$2.514$&
$4.340$&
$5.694$&
$1.726$&
$1.312$\tabularnewline
$5.0$&
$3.144$&
$5.507$&
$7.142$&
$1.752$&
$1.297$\tabularnewline
$5.5$&
$3.831$&
$6.795$&
$8.724$&
$1.774$&
$1.284$\tabularnewline
$6.0$&
$4.572$&
$8.199$&
$10.43$&
$1.793$&
$1.272$\tabularnewline
$6.5$&
$5.363$&
$9.714$&
$12.26$&
$1.811$&
$1.262$\tabularnewline
$7.0$&
$6.202$&
$11.33$&
$14.20$&
$1.827$&
$1.253$\tabularnewline
$7.5$&
$7.088$&
$13.06$&
$16.24$&
$1.843$&
$1.243$\tabularnewline
$8.0$&
$8.017$&
$14.87$&
$18.39$&
$1.855$&
$1.237$\tabularnewline
$8.5$&
$8.987$&
$16.79$&
$20.63$&
$1.868$&
$1.229$\tabularnewline
$9.0$&
$9.997$&
$18.79$&
$22.97$&
$1.880$&
$1.222$\tabularnewline
$9.5$&
$11.05$&
$20.88$&
$25.40$&
$1.890$&
$1.216$\tabularnewline
$10.0$&
$12.13$&
$23.05$&
$27.91$&
$1.900$&
$1.211$\tabularnewline
$10.5$&
$13.25$&
$25.31$&
$30.50$&
$1.910$&
$1.205$\tabularnewline
$11.0$&
$14.40$&
$27.64$&
$33.18$&
$1.919$&
$1.200$\tabularnewline
$11.5$&
$15.59$&
$30.05$&
$35.92$&
$1.928$&
$1.195$\tabularnewline
$12.0$&
$16.81$&
$32.53$&
$38.74$&
$1.935$&
$1.191$\tabularnewline
$12.5$&
$18.06$&
$35.08$&
$41.63$&
$1.942$&
$1.187$\tabularnewline
$13.0$&
$19.33$&
$37.70$&
$44.59$&
$1.950$&
$1.183$\tabularnewline
$13.5$&
$20.64$&
$40.39$&
$47.61$&
$1.957$&
$1.179$\tabularnewline
$14.0$&
$21.97$&
$43.14$&
$50.70$&
$1.964$&
$1.175$\tabularnewline
\hline
\end{tabular}
\end{center}
\caption{Values of the cross sections and $K$-factors for scalar Higgs
production at LHC as a function of $\sqrt{S}$ with $\mu_F=m_H$,
with $\mu_F^2=2\mu_R^2$ and $m_H=114$ GeV. MRST inputs have been used.}
\end{table}

\newpage

\begin{table}
\begin{center}
\begin{tabular}{|c|ccccc|}
\hline
$\sqrt{S}$&
$\sigma_{LO}$&
$\sigma_{NLO}$&
$\sigma_{NNLO}$&
$K^{NLO}$&
$K^{NNLO}$\tabularnewline
\hline
$2.0$&
$0.2597$&
$0.4159$&
$0.6614$&
$1.601$&
$1.590$\tabularnewline
$2.5$&
$0.4763$&
$0.7912$&
$1.216$&
$1.661$&
$1.537$\tabularnewline
$3.0$&
$0.7574$&
$1.293$&
$1.936$&
$1.707$&
$1.497$\tabularnewline
$3.5$&
$1.100$&
$1.917$&
$2.811$&
$1.743$&
$1.466$\tabularnewline
$4.0$&
$1.501$&
$2.660$&
$3.833$&
$1.772$&
$1.441$\tabularnewline
$4.5$&
$1.956$&
$3.517$&
$4.992$&
$1.798$&
$1.419$\tabularnewline
$5.0$&
$2.464$&
$4.482$&
$6.281$&
$1.819$&
$1.401$\tabularnewline
$5.5$&
$3.021$&
$5.553$&
$7.693$&
$1.838$&
$1.385$\tabularnewline
$6.0$&
$3.625$&
$6.724$&
$9.221$&
$1.855$&
$1.371$\tabularnewline
$6.5$&
$4.275$&
$7.991$&
$10.86$&
$1.869$&
$1.359$\tabularnewline
$7.0$&
$4.967$&
$9.351$&
$12.60$&
$1.883$&
$1.347$\tabularnewline
$7.5$&
$5.700$&
$10.80$&
$14.44$&
$1.895$&
$1.337$\tabularnewline
$8.0$&
$6.472$&
$12.33$&
$16.38$&
$1.905$&
$1.328$\tabularnewline
$8.5$&
$7.282$&
$13.95$&
$18.40$&
$1.916$&
$1.319$\tabularnewline
$9.0$&
$8.127$&
$15.65$&
$20.52$&
$1.926$&
$1.311$\tabularnewline
$9.5$&
$9.008$&
$17.42$&
$22.71$&
$1.934$&
$1.304$\tabularnewline
$10.0$&
$9.923$&
$19.27$&
$24.99$&
$1.942$&
$1.297$\tabularnewline
$10.5$&
$10.87$&
$21.19$&
$27.34$&
$1.949$&
$1.290$\tabularnewline
$11.0$&
$11.85$&
$23.18$&
$29.76$&
$1.956$&
$1.284$\tabularnewline
$11.5$&
$12.86$&
$25.24$&
$32.26$&
$1.963$&
$1.278$\tabularnewline
$12.0$&
$13.89$&
$27.36$&
$34.82$&
$1.970$&
$1.273$\tabularnewline
$12.5$&
$14.96$&
$29.55$&
$37.45$&
$1.975$&
$1.267$\tabularnewline
$13.0$&
$16.05$&
$31.80$&
$40.15$&
$1.981$&
$1.263$\tabularnewline
$13.5$&
$17.17$&
$34.11$&
$42.91$&
$1.987$&
$1.258$\tabularnewline
$14.0$&
$18.32$&
$36.48$&
$45.72$&
$1.991$&
$1.253$\tabularnewline
\hline
\end{tabular}
\end{center}
\caption{Values of the cross sections and $K$-factors for scalar Higgs
production at LHC as a function of $\sqrt{S}$ with $\mu_F=2 m_H$,
with $\mu_F^2=2\mu_R^2$ and $m_H=114$ GeV. MRST inputs have been used.}
\end{table}

\begin{table}
\begin{center}
\begin{tabular}{|c|ccccc|}
\hline
$\sqrt{S}$&
$\sigma_{LO}$&
$\sigma_{NLO}$&
$\sigma_{NNLO}$&
$K^{NLO}$&
$K^{NNLO}$\tabularnewline
\hline
$2.0$&
$0.4899$&
$0.6836$&
$0.8836$&
$1.395$&
$1.293$\tabularnewline
$2.5$&
$0.8642$&
$1.270$&
$1.598$&
$1.470$&
$1.258$\tabularnewline
$3.0$&
$1.333$&
$2.036$&
$2.512$&
$1.527$&
$1.234$\tabularnewline
$3.5$&
$1.890$&
$2.976$&
$3.612$&
$1.575$&
$1.214$\tabularnewline
$4.0$&
$2.526$&
$4.079$&
$4.885$&
$1.615$&
$1.198$\tabularnewline
$4.5$&
$3.237$&
$5.338$&
$6.321$&
$1.649$&
$1.184$\tabularnewline
$5.0$&
$4.016$&
$6.743$&
$7.908$&
$1.679$&
$1.173$\tabularnewline
$5.5$&
$4.858$&
$8.288$&
$9.638$&
$1.706$&
$1.163$\tabularnewline
$6.0$&
$5.761$&
$9.966$&
$11.50$&
$1.730$&
$1.154$\tabularnewline
$6.5$&
$6.719$&
$11.77$&
$13.49$&
$1.752$&
$1.146$\tabularnewline
$7.0$&
$7.730$&
$13.69$&
$15.60$&
$1.771$&
$1.140$\tabularnewline
$7.5$&
$8.790$&
$15.73$&
$17.82$&
$1.790$&
$1.133$\tabularnewline
$8.0$&
$9.898$&
$17.88$&
$20.15$&
$1.806$&
$1.127$\tabularnewline
$8.5$&
$11.05$&
$20.14$&
$22.58$&
$1.823$&
$1.121$\tabularnewline
$9.0$&
$12.24$&
$22.49$&
$25.11$&
$1.837$&
$1.116$\tabularnewline
$9.5$&
$13.48$&
$24.94$&
$27.74$&
$1.850$&
$1.112$\tabularnewline
$10.0$&
$14.75$&
$27.49$&
$30.45$&
$1.864$&
$1.108$\tabularnewline
$10.5$&
$16.06$&
$30.13$&
$33.25$&
$1.876$&
$1.104$\tabularnewline
$11.0$&
$17.41$&
$32.85$&
$36.13$&
$1.887$&
$1.100$\tabularnewline
$11.5$&
$18.79$&
$35.66$&
$39.09$&
$1.898$&
$1.096$\tabularnewline
$12.0$&
$20.20$&
$38.55$&
$42.12$&
$1.908$&
$1.093$\tabularnewline
$12.5$&
$21.64$&
$41.51$&
$45.24$&
$1.918$&
$1.090$\tabularnewline
$13.0$&
$23.11$&
$44.56$&
$48.42$&
$1.928$&
$1.087$\tabularnewline
$13.5$&
$24.62$&
$47.67$&
$51.66$&
$1.936$&
$1.084$\tabularnewline
$14.0$&
$26.15$&
$50.86$&
$54.98$&
$1.945$&
$1.081$\tabularnewline
\hline
\end{tabular}
\end{center}
\caption{Values of the cross sections and $K$-factors for scalar Higgs
production at LHC as a function of $\sqrt{S}$ with $\mu_F=(1/2) m_H$,
with $\mu_F^2=2\mu_R^2$ and $m_H=114$ GeV. MRST inputs have been used.}
\end{table}

\begin{table}
\begin{center}
\begin{tabular}{|c|ccccc|}
\hline
$\sqrt{S}$&
$\sigma_{LO}$&
$\sigma_{NLO}$&
$\sigma_{NNLO}$&
$K^{NLO}$&
$K^{NNLO}$\tabularnewline
\hline
$2.0$&
$0.4899$&
$1.276$&
$1.635$&
$2.605$&
$1.281$\tabularnewline
$2.5$&
$0.8641$&
$2.083$&
$2.567$&
$2.411$&
$1.232$\tabularnewline
$3.0$&
$1.333$&
$3.022$&
$3.622$&
$2.267$&
$1.199$\tabularnewline
$3.5$&
$1.890$&
$4.069$&
$4.775$&
$2.153$&
$1.174$\tabularnewline
$4.0$&
$2.526$&
$5.206$&
$6.009$&
$2.061$&
$1.154$\tabularnewline
$4.5$&
$3.237$&
$6.419$&
$7.312$&
$1.983$&
$1.139$\tabularnewline
$5.0$&
$4.016$&
$7.698$&
$8.673$&
$1.917$&
$1.127$\tabularnewline
$5.5$&
$4.858$&
$9.034$&
$10.08$&
$1.860$&
$1.116$\tabularnewline
$6.0$&
$5.761$&
$10.42$&
$11.54$&
$1.809$&
$1.107$\tabularnewline
$6.5$&
$6.719$&
$11.85$&
$13.03$&
$1.764$&
$1.100$\tabularnewline
$7.0$&
$7.730$&
$13.32$&
$14.55$&
$1.723$&
$1.092$\tabularnewline
$7.5$&
$8.790$&
$14.82$&
$16.11$&
$1.686$&
$1.087$\tabularnewline
$8.0$&
$9.898$&
$16.35$&
$17.69$&
$1.652$&
$1.082$\tabularnewline
$8.5$&
$11.05$&
$17.91$&
$19.30$&
$1.621$&
$1.078$\tabularnewline
$9.0$&
$12.24$&
$19.50$&
$20.93$&
$1.593$&
$1.073$\tabularnewline
$9.5$&
$13.48$&
$21.11$&
$22.57$&
$1.566$&
$1.069$\tabularnewline
$10.0$&
$14.75$&
$22.74$&
$24.24$&
$1.542$&
$1.066$\tabularnewline
$10.5$&
$16.06$&
$24.39$&
$25.92$&
$1.519$&
$1.063$\tabularnewline
$11.0$&
$17.41$&
$26.06$&
$27.61$&
$1.497$&
$1.059$\tabularnewline
$11.5$&
$18.79$&
$27.74$&
$29.32$&
$1.476$&
$1.057$\tabularnewline
$12.0$&
$20.20$&
$29.44$&
$31.05$&
$1.457$&
$1.055$\tabularnewline
$12.5$&
$21.64$&
$31.15$&
$32.78$&
$1.439$&
$1.052$\tabularnewline
$13.0$&
$23.11$&
$32.87$&
$34.53$&
$1.422$&
$1.051$\tabularnewline
$13.5$&
$24.62$&
$34.60$&
$36.28$&
$1.405$&
$1.049$\tabularnewline
$14.0$&
$26.15$&
$36.35$&
$38.04$&
$1.390$&
$1.046$\tabularnewline
\hline
\end{tabular}
\end{center}
\caption{Values of the cross sections and $K$-factors for scalar Higgs
production at LHC as a function of $\sqrt{S}$ with $\mu_F=m_H$,
with $\mu_F^2=(1/2)\mu_R^2$ and $m_H=114$ GeV. MRST inputs have been used.}
\end{table}

\begin{table}
\begin{center}
\begin{tabular}{|c|ccccc|}
\hline
$\sqrt{S}$&
$\sigma_{LO}$&
$\sigma_{NLO}$&
$\sigma_{NNLO}$&
$K^{NLO}$&
$K^{NNLO}$\tabularnewline
\hline
$2.0$&
$0.3549$&
$1.031$&
$1.446$&
$2.905$&
$1.403$\tabularnewline
$2.5$&
$0.6393$&
$1.707$&
$2.286$&
$2.670$&
$1.339$\tabularnewline
$3.0$&
$1.003$&
$2.503$&
$3.242$&
$2.496$&
$1.295$\tabularnewline
$3.5$&
$1.439$&
$3.398$&
$4.292$&
$2.361$&
$1.263$\tabularnewline
$4.0$&
$1.945$&
$4.377$&
$5.418$&
$2.250$&
$1.238$\tabularnewline
$4.5$&
$2.515$&
$5.428$&
$6.610$&
$2.158$&
$1.218$\tabularnewline
$5.0$&
$3.146$&
$6.542$&
$7.857$&
$2.079$&
$1.201$\tabularnewline
$5.5$&
$3.834$&
$7.711$&
$9.151$&
$2.011$&
$1.187$\tabularnewline
$6.0$&
$4.575$&
$8.927$&
$10.49$&
$1.951$&
$1.175$\tabularnewline
$6.5$&
$5.367$&
$10.19$&
$11.86$&
$1.899$&
$1.164$\tabularnewline
$7.0$&
$6.207$&
$11.48$&
$13.27$&
$1.850$&
$1.156$\tabularnewline
$7.5$&
$7.093$&
$12.82$&
$14.70$&
$1.807$&
$1.147$\tabularnewline
$8.0$&
$8.023$&
$14.18$&
$16.16$&
$1.767$&
$1.140$\tabularnewline
$8.5$&
$8.994$&
$15.57$&
$17.65$&
$1.731$&
$1.134$\tabularnewline
$9.0$&
$10.00$&
$16.99$&
$19.15$&
$1.699$&
$1.127$\tabularnewline
$9.5$&
$11.05$&
$18.43$&
$20.68$&
$1.668$&
$1.122$\tabularnewline
$10.0$&
$12.14$&
$19.89$&
$22.22$&
$1.638$&
$1.117$\tabularnewline
$10.5$&
$13.26$&
$21.37$&
$23.78$&
$1.612$&
$1.113$\tabularnewline
$11.0$&
$14.41$&
$22.87$&
$25.35$&
$1.587$&
$1.108$\tabularnewline
$11.5$&
$15.60$&
$24.39$&
$26.93$&
$1.563$&
$1.104$\tabularnewline
$12.0$&
$16.82$&
$25.92$&
$28.53$&
$1.541$&
$1.101$\tabularnewline
$12.5$&
$18.07$&
$27.46$&
$30.14$&
$1.520$&
$1.098$\tabularnewline
$13.0$&
$19.35$&
$29.02$&
$31.76$&
$1.500$&
$1.094$\tabularnewline
$13.5$&
$20.65$&
$30.59$&
$33.39$&
$1.481$&
$1.092$\tabularnewline
$14.0$&
$21.99$&
$32.18$&
$35.03$&
$1.463$&
$1.089$\tabularnewline
\hline
\end{tabular}
\end{center}
\caption{Values of the cross sections and $K$-factors for scalar Higgs
production at LHC as a function of $\sqrt{S}$ with $\mu_F=2 m_H$,
with $\mu_F^2=(1/2)\mu_R^2$ and $m_H=114$ GeV. MRST inputs have been used.}
\end{table}

\begin{table}
\begin{center}
\begin{tabular}{|c|ccccc|}
\hline
$\sqrt{S}$&
$\sigma_{LO}$&
$\sigma_{NLO}$&
$\sigma_{NNLO}$&
$K^{NLO}$&
$K^{NNLO}$\tabularnewline
\hline
$2.0$&
$0.6960$&
$1.600$&
$1.830$&
$2.299$&
$1.144$\tabularnewline
$2.5$&
$1.198$&
$2.579$&
$2.862$&
$2.153$&
$1.110$\tabularnewline
$3.0$&
$1.814$&
$3.708$&
$4.028$&
$2.044$&
$1.086$\tabularnewline
$3.5$&
$2.531$&
$4.956$&
$5.300$&
$1.958$&
$1.069$\tabularnewline
$4.0$&
$3.341$&
$6.303$&
$6.661$&
$1.887$&
$1.057$\tabularnewline
$4.5$&
$4.234$&
$7.734$&
$8.096$&
$1.827$&
$1.047$\tabularnewline
$5.0$&
$5.204$&
$9.234$&
$9.593$&
$1.774$&
$1.039$\tabularnewline
$5.5$&
$6.243$&
$10.80$&
$11.14$&
$1.730$&
$1.031$\tabularnewline
$6.0$&
$7.347$&
$12.41$&
$12.74$&
$1.689$&
$1.027$\tabularnewline
$6.5$&
$8.512$&
$14.07$&
$14.38$&
$1.653$&
$1.022$\tabularnewline
$7.0$&
$9.732$&
$15.77$&
$16.06$&
$1.620$&
$1.018$\tabularnewline
$7.5$&
$11.00$&
$17.50$&
$17.77$&
$1.591$&
$1.015$\tabularnewline
$8.0$&
$12.33$&
$19.26$&
$19.50$&
$1.562$&
$1.012$\tabularnewline
$8.5$&
$13.69$&
$21.06$&
$21.26$&
$1.538$&
$1.009$\tabularnewline
$9.0$&
$15.11$&
$22.88$&
$23.05$&
$1.514$&
$1.007$\tabularnewline
$9.5$&
$16.56$&
$24.72$&
$24.86$&
$1.493$&
$1.006$\tabularnewline
$10.0$&
$18.05$&
$26.58$&
$26.69$&
$1.473$&
$1.004$\tabularnewline
$10.5$&
$19.58$&
$28.45$&
$28.53$&
$1.453$&
$1.003$\tabularnewline
$11.0$&
$21.14$&
$30.35$&
$30.39$&
$1.436$&
$1.001$\tabularnewline
$11.5$&
$22.74$&
$32.26$&
$32.26$&
$1.419$&
$1.000$\tabularnewline
$12.0$&
$24.37$&
$34.18$&
$34.15$&
$1.403$&
$0.9991$\tabularnewline
$12.5$&
$26.03$&
$36.12$&
$36.05$&
$1.388$&
$0.9981$\tabularnewline
$13.0$&
$27.72$&
$38.07$&
$37.96$&
$1.373$&
$0.9971$\tabularnewline
$13.5$&
$29.44$&
$40.02$&
$39.89$&
$1.359$&
$0.9968$\tabularnewline
$14.0$&
$31.19$&
$41.99$&
$41.82$&
$1.346$&
$0.9960$\tabularnewline
\hline
\end{tabular}
\end{center}
\caption{Values of the cross sections and $K$-factors for scalar Higgs
production at LHC as a function of $\sqrt{S}$ with $\mu_F=(1/2) m_H$,
with $\mu_F^2=(1/2)\mu_R^2$ and $m_H=114$ GeV. MRST inputs have been used.}
\end{table}


\begin{thebibliography}{10}
\bibitem{vogt1}S. Moch, J. Vermaseren and A. Vogt, 
Nucl. Phys.\, {\bf B 688}, 101, (2004);
Nucl. Phys.\, {\bf B 691}, 129, (2004).
\bibitem{RSVN} V. Ravindran, J. Smith and W.L. van Neerven, 
Nucl. Phys. {\bf B665}, 325, (2003). 
\bibitem{VN1} W. Van Neerven and A. Vogt, 
Nucl.Phys.\, {\bf B 603}, 42, (2001); Nucl.Phys. {\bf B 588}, 345, (2000).
\bibitem{dawson} S. Dawson, hep-ph/9411325.
\bibitem{RSVN1}V. Ravindran , J. Smith, W.L. Van Neerven,
Nucl.Phys.\,{\bf B 704}, 332, (2005)
\bibitem{cafacor} A. Cafarella and C. \cc, 
Comput.Phys.Commun.\,{\bf 160}, 213, (2004).
\bibitem{CCG} A. Cafarella, C Corian\'o and M. Guzzi, in preparation.
\bibitem{remiddi1} T. Gehrmann, E. Remiddi, Comput. 
Phys. Commun. {\bf 141}, 296, (2001)
\bibitem{remiddi2} E. Remiddi, J.A.M. Vermaseren, 
Int. J. Mod. Phys. {\bf A 15} 725, (2000)
\bibitem{Vogt} A. Vogt, Comput.Phys.Commun. {\bf 170}, 65 (2005). 
\bibitem{bmsn}
M. Buza, Y. Matiounine, J. Smith and W.L. van Neerven,
Eur. Phys. J. {\bf C1}, 301 (1998).
\bibitem{cs}
A. Chuvakin and J. Smith, Comput. Phys. Commun. {\bf 143}, 257-286 (2002).
\bibitem{MRST}A.D.~Martin, R.G.~Roberts, W.J.~Stirling and R.S.~Thorne, 
Eur.Phys.J.C \textbf{23} (2002) 73;
Phys.Lett.B\textbf{531} (2002) 216.
\bibitem{Alekhin} S.I.~Alekhin, Phys.Rev.D \textbf{68} (2003) 014002.
\bibitem{mv}
S. Moch and A. Vogt, Higher order soft corrections to lepton pair production
and Higgs boson production, hep-ph/0508265.
\bibitem{lm}
E. Laenen and L. Magnea, Threshold resummation for electroweak 
annihilation from DIS data, hep-ph/0508284.
\bibitem{ij}
A. Idilbi, X. Ji, J-P. Ma and F. Yuan,
Threshold resummation for Higgs production in effective field theory,
hep-ph/0509294.
\bibitem{amp}
C. Anastasiou, K. Melnikov and F. Petriello, The gluon-fusion uncertainty
in Higgs coupling extractions, hep-ph/0509014.
\end{thebibliography}
\end{document}